# Effects of Data Resolution and Human Behavior on Large Scale Evacuation Simulations

A Dissertation Presented for the

Doctor of Philosophy

Degree

The University of Tennessee, Knoxville

Wei Lu

December 2013

*This dissertation is dedicated to my very supportive mother Yin Xu for her selfless love.*

# ACKNOWLEDGEMENTS

First, I would like to express my deepest gratitude to my academic advisor, Dr. Lee D. Han, for his mentorship during my PhD study. He really cares about my research interests and provides a lot of great ideas for my research projects. The dissertation would not exist without the inspiring conversations with him.

Also, I would like to thank Dr. Cheng Liu for his selfless help with the technical details in my research. Since June 2012, he has been my mentor in Oak Ridge National Laboratory. He provides many programs which boost my research to generate promising results in my academic papers.

I would like to acknowledge the contributions of Dr. Christopher R. Cherry in my research on vehicular communication and Dr. Joshua S. Fu in my minor degree in computational science. I am so grateful to have them on my committee and collaborate with them on research and publishing papers.

I would like to thank Oak Ridge National Laboratory for providing the LandScan (2011)$^{TM}$ USA Population Data, which is copyrighted by UT-Battelle, LLC, operator of Oak Ridge National Laboratory under Contract No. DE-AC05-00OR22725 with the US Department of Energy. I appreciate the help from Dr. Budhendra L. Bhaduri and Mark A. Tuttle for providing those data.

I enjoyed the time with all the graduate students in transportation program, Ryan Overton, Qiang Yang, Stephanie Hargrove, Jianjiang Yang, Hongtai Yang, Shuguang Ji, Casey Langford, Taekwan Yoon, Andrew Campbell, Amber Woodburn, Jun Liu, and so on. They make my life in Knoxville more colorful.

# ABSTRACT

Traffic Analysis Zones (TAZ) based macroscopic simulation studies are mostly applied in evacuation planning and operation areas. The large size in TAZ and aggregated information of macroscopic simulation underestimate the real evacuation performance. To take advantage of the high resolution demographic data LandScan USA (the zone size is much smaller than TAZ) and agent-based microscopic traffic simulation models, many new problems appeared and novel solutions are needed. A series of studies are conducted using LandScan USA Population Cells (LPC) data for evacuation assignments with different network configurations, travel demand models, and travelers' compliance behavior.

First, a new Multiple-Source-Nearest-Destination Shortest Path (MSNDSP) problem is defined for generating Origin-Destination (OD) matrix in evacuation assignments when using LandScan dataset. A new Super Node based Trip Generator (SNTG) algorithm is developed to significantly improve the computational performance in generating OD pairs. Second, a new agent-based traffic assignment framework using LandScan and TRANSIMS modules is proposed for evacuation planning and operation study. Impact analysis on traffic analysis area resolutions (TAZ vs. LPC), evacuation start times (daytime vs. nighttime), and departure time choice models (normal S-shape model vs. location based model) are studied. Third, based on the proposed framework, multi-scale network configurations (two levels of road networks and two scales of zone sizes) and three routing schemes (shortest network distance, highway biased, and shortest straight-line distance routes) are implemented for the evacuation performance comparison studies. Fourth, to study the impact of human behavior under evacuation operations, travelers' compliance behavior with compliance levels from total complied to total non-complied are analyzed. Finally, an experimental study of using vehicular communication techniques in a car sharing system is conducted, which can be used to provide real-time road information for evacuees to choose their own best evacuation routes.

# TABLE OF CONTENTS

# LIST OF TABLES

# LIST OF FIGURES

# CHAPTER 1
# INTRODUCTION

Evacuation studies from hurricane, wildfires, terrorist attack, and other severe events reveal that how to evacuate people based on different demographic information in affected area is one of the most important factors for a successful evacuation plan. As high resolution demographic data source, LandScan USA population data has become the community standard for national population distribution. A lot of population risk study based on LandScan USA dataset were conducted to predict better analysis and results (Bhaduri, Bright et al. 2002). LandScan USA population cell (LPC) data has a high resolution population distribution to provide 90m x 90m (3'×3') resolution with national population distribution data (Bhaduri, Bright et al. 2007). It is much more accurate than conventional TAZ because some TAZ zones are large in scale or dense in population (You, Nedović-Budić et al. 1998). Compared to TAZ-based traffic assignment that generates trips from large-scale zones, LPC-based methods allow small-scale, cell-to-cell trip generation. This increase the number of O-D pairs in trip generation and distribution stage. A fast and effective O-D assignment between all LPC origins and destination/shelters becomes a critical issue.

Due to the limitation of computing resources and high resolution demographic data availability, Traffic Analysis Zones (TAZ) based simulation researches dominated the evacuation simulation area. Conventional TAZ based models assign all the trips in a TAZ to the centroid point and then connected to the nearest node on the network. Even agent-based traffic simulation, used in Transportation Analysis and Simulation System (TRANSIMS) package, cannot represent the real-world situation with TAZ models because it evenly distributes travelers to all the activity locations in one TAZ. A more accurate agent-based traffic simulation model is needed to take advantage of high resolution demographic data to produce more accurate simulation results.

One of the challenges to develop a successful evacuation plan using high resolution demographic data is generating the OD matrix for the simulation tools, such as TRANSIMS. LandScan USA Population Cells (LPC) data has higher amount of origins and produces more OD pairs than TAZ zones. Using the shortest path algorithm to find the nearest destination for each LPC can be very time-consuming in large networks. This brings a new shortest path problem, Multiple-Source-Nearest-Destination Shortest Path (MSNDSP) problem, which bridges each LPC origin to its nearest destination from the pre-defined destinations pool. Using super node concept and backward star network structure, a new algorithm Super Node based Trip Generator (SNTG) is developed with promising computational performance to assign the OD pairs.

Evacuation planning and operation also needs simulation-based studies while real world evacuation data is hardly available. Han, Yuan et al. (2006) compared the static traffic assignment and dynamic traffic assignment for emergency evacuation scenarios and emphasized the advantage of using intelligent transportation systems to route evacuees. Chen and Zhan (2006) used an agent-based technique to model traffic flows at individual vehicle level and investigated the collective behaviors of evacuating vehicles in the city of San Marcos, Texas. It revealed that the road network structure and population density have impacts on evacuation strategies. Chen, Meaker et al. (2006) also used agent-based modeling to analyze the minimum clearance time and how many people would need to be accommodated if evacuation route got unavailable in the Florida Keys area. Some traditional microscopic traffic simulation can also be treated as agent-based simulation to some extent. Jha, Moore et al. (2004) used microscopic simulation model (MITSIM) to model the evacuation of Los Alamos National Laboratory. Cova and Johnson (2002) presented a method to develop neighborhood evacuation planning with microscopic traffic simulation in the urban - wildland interface. Henson and Goulias (2006) reviewed 46 activity-based models and their competency of homeland security applications. All these agent-based simulation and evacuation models are still based on existing TAZ-based traffic assignment during the traffic supply-demand modeling stage.

To take advantage of high resolution demographic data, a new agent-based evacuation assignment framework is proposed in this dissertation. A comparison study using road network and demographic data in Alexandria, Virginia is conducted to evaluate the framework accuracy and evacuation efficiency. Three aspects are considered, including traffic analysis area resolutions (TAZ vs. LPC), evacuation start times (daytime vs. nighttime), and departure time choice models (normal S-shape model vs. location-based model). Subsequent discussions on improving evacuation simulation accuracy and efficiency through high resolution demographic data are also presented.

Routing problem and network optimization are mainly concerned in emergency evacuation studies. Ng, Park et al. (2010) optimized the shelter locations through a hybrid bi-level model to improve the evacuation performance. Yue-ming and Hui (2010) modeled the shortest emergency evacuation time through Pontryagin minimum principle and dynamic traffic assignment. Xie and Turnquist (2011) integrated Lagrangian relaxation and Tabu search algorithm to optimize the lane-based evacuation system performance. Lu, George et al. (2005) simulated their evacuation plans with capacity constrained routing algorithms. Liu, Lai et al. (2006) studied the staged emergency evacuation planning with cell-based network optimization models. Kai-Fu and Wen-Long (2008) illustrated the emergency evacuation network statistics through genetic algorithm and kinematic wave models. Most of these researches focus on system-level analysis during emergency evacuation scenarios.

Most simulation studies only take the major road networks (without considering the local roads in the neighborhoods) as their inputs. This limits the travel time and driving behaviors on local roads, where might produce bottleneck during emergency evacuation scenarios. Different routing schemes, such as shortest network distance route, highway-biased route, and shortest straight-line distance route also have various impacts on evacuation performance. Comparison studies with multi-scale networks and three routing schemes are implemented to provide reasonable suggestions for evacuation planning and operations.

Traveler compliance behavior is a key factor in traffic simulation and evacuation assignments. Many researches focus on the compliance behavior resulting from Advanced Traveler Information Systems (ATIS) with real-time road conditions. Srinivasan and Mahmassani (2000) examined travelers' route choice decisions based on compliance and inertia mechanisms with simulation and empirical data comparison. The results indicate that information quality, network loading, level-of-service, and travelers' prior experience determine route choices. Al-Deek, Khattak et al. (1998) used traveler compliance behavior and traffic system performance to evaluate ATIS. Lu, Han et al. (2013) studied the impact of connected vehicle technology for travelers to exchange real-time traffic information in a car sharing system, which provided another approach to implement an ATIS framework. There are also some papers presenting the evacuee compliance behavior impact in evacuation performance. Pel et al. did a series of studies on evacuation modeling with traveler information and compliance behavior (Pel, Huibregtse et al. 2009, Pel, Hoogendoorn et al. 2010, Pel, Bliemer et al. 2011, Pel, Bliemer et al. 2012). Some results show that traveler compliance behavior affect evacuation efficiency and road capacities have no significant impact on evacuation. The reviewed dynamic traffic simulation models also indicate that some existing models still have some weakness with regard to the choice to evacuate, departure time choice, destination choice, and route choice. Yuan, Han et al. (2007) simulated a Nuclear Plant and Tennessee state with non-compliance route choices in evacuation assignments. The results suggest that there is no significant difference between full compliance and full non-compliance. Revised traffic analysis zones might help to reveal more realistic evacuation performance. All of these studies used TAZ for trip assignment. TAZ is good for planning purpose. But the biggest disadvantage of TAZ is that the data is not usually available to the general audience. Typically, city or county transportation planning agencies are responsible for their TAZ definition. And many of those data are out-of-date and do not represent the current road networks.

One-to-one origin/destination trip assignment is widely accepted for evacuation planning purposes. But it assumes all the travelers follow the instructions strictly (spatially and temporally) and do not change their destinations depending on the real-time road conditions. Evacuation operation surveys from several hurricane evacuation studies indicate that evacuees changed or are willing to change their route based on real-time traffic information. Evacuee compliance behavior is one of the key factors for evacuation planning. A detailed simulation study to examine whether the evacuees' spatial compliance behavior with population data in different resolutions would impact on the evacuation performance is conducted and some subsequent discussions on implementing travelers' spatial compliance behavior through high resolution data are also suggested.

Vehicular communication techniques can provide adequate real-time traffic information under evacuation scenarios. Yet even with an increase in car sharing programs worldwide, there has been little research on the application of Vehicular Ad hoc Network (VANET) in car sharing systems. To test the feasibility and system performance of vehicular communication networks in road traffic, a car sharing system is adapted. This study can help to implement the non-compliance travelers' behavior under evacuation scenarios. Some suggestions for future field deployment are provided.

The dissertation is organized in journal article formats since each chapter is either published, submitted, or to be submitted to academic journals. Following this chapter, the second chapter defines multiple-sources-nearest-destination shortest path problem in evacuation assignments and provides a computationally promising solution. The third chapter proposes a new evacuation assignment framework using high resolution demographic data and agent-based microscopic traffic simulation. The fourth chapter studies the emergency evacuation performance under multi-scale network configurations and routing schemes. The fifth chapter researches on evacuee compliance behavior in both TAZ and LPC based evacuation simulations at different compliance levels. The sixth chapter evaluates the vehicular communication networks in a car sharing system, which can provide an approach to broadcast real-time traffic information in evacuation scenarios. Finally, conclusions are drawn and future works are recommended in the sixth chapter.

# CHAPTER 2
# A MULTIPLE-SOURCE-NEAREST-DESTINATION SHORTEST PATH PROBLEM IN EVACUATION ASSIGNMENTS

**Abstract**

Efficient evacuation planning is important for local authorities to prepare for emergency events, such as terrorist attack or severe weather. Simulation based study is one of the major methods for evacuation planning. How to quickly build an origin-destination (OD) matrix to assign people in each source zone to their nearest destination for evacuation simulations becomes an issue. In this paper, we propose a new problem, Multiple-Source-Nearest-Destination Shortest Path (MSNDSP) problem, for generating OD matrix in evacuation assignments. Compared to a benchmark study using Dijkstra's algorithm with forward star network structure to solve the new problem, we propose a new algorithm Super Node based Trip Generator (SNTG) to improve the computational performance in generating OD pairs connecting multiple sources to their nearest destination respectively. The new algorithm transforms the OD generation from a MSNDSP problem to a normal single source shortest pat problem, which significantly reduces the computational time. Experimental studies using real world street networks and high resolution LandScan USA population data indicate that the SNTG algorithm can provides the identical OD output as the benchmark study but the computing time is about 500 to 45,000 times faster in different network sizes. An evacuation performance case study using the OD table from our algorithm is also conducted to prove its efficiency and accuracy.

*Keywords:* Shortest Path Problems; Multiple Sources; Nearest Destination; Routing with Super Node; Backward Star; Evacuation Assignment

## 2.1 Introduction

Taking advantage of the fast development of computational resources and high resolution demographic data, agent-based microscopic traffic simulation has been widely adapted in evacuation planning and operation area. The challenge to develop a successful evacuation plan is generating the origin-destination matrix for the simulation tools, such as Transportation Analysis and Simulation System (TRANSIMS). Under emergency evacuation situations, such severe weather or terrorist attack, evacuees should be evacuated to the nearest safe destination out of the affected area efficiently. Evacuees do not know the whole road conditions before leaving affected area. Their best destination choice is the nearest one among all assigned destinations based on their locations. Traffic Analysis Zones (TAZ) based OD assignment is commonly used for traffic simulations but it is not efficient to produce evacuation plans for each individuals. High resolution demographic data, LandScan USA Population Cells (LPC) data (Bhaduri, Bright et al. (2007), has higher amount of origins and produces more O-D pairs than TAZ zones. Using the shortest path algorithm to find the nearest destination to people in an LPC origin can be very time-consuming in large networks, because each LPC origin needs to be assigned with a destination through the shortest path calculation. This brings a new shortest path problem, Multiple-Sources-Nearest-Destination Shortest Path (MSNDSP) problem, which connects each LPC origin to its nearest destination from the pre-defined destinations pool.

The purpose of this paper is to define an efficient algorithm to generate O-D pairs in emergency evacuation scenarios, which can be used as the input of TRANSIMS simulation package. In our simulation, the number of generated trips is determined by the LPC population. The destinations are assigned to the exits to affected area, where evacuees are trying to move outward. All assigned destinations are equivalent to evacuees. People in each LPC are seeking the nearest destination for safety. This problem can be solved with Dijkstra's shortest path algorithm using forward star network structure. This method is implemented as a benchmark study. The computational time increases exponentially when the number of LPC zones, network nodes, and network links increase. To generate the OD matrix for the MSNDSP problem efficiently and effectively, we develop a new algorithm, Super Node based Trip Generator (SNTG), to connect multiple sources to their nearest destination respectively. Our algorithm converts the MSNDSP problem to a single source shortest path problem by adding a super node to a backward star represented network. It produces the same O-D table as the benchmark study but shorten the computational time dramatically. A case study using the O-D table generated with our algorithm from real world data is conducted to prove its accuracy and efficiency in evacuation application. The SNTG algorithm can also be applied to those cases where origins or destinations in a network are equal to each other.

The rest of this paper is organized as follows. In Section 2, we review the evacuation simulation models and shortest path algorithms. In Section 3, the MSNDSP problem is introduced and formally defined. In Section 4, we propose the SNTG algorithm and explain its implementation compared to the benchmark method. In Section 5, the algorithm is experimented on various evacuation areas by utilizing real-world road networks and population data. We summarize the computational results of SNTG algorithm and discuss the evacuation performance of a case study. Finally, important finding are concluded and future work are pointed in Section 6.

## 2.2 Related Work

Simulation studies are widely adapted as efficient tools in evacuation planning and operations while real world evacuation data is hardly available. It provides valuable reference for evacuation managers and general instructions for individual evacuees. Han, Yuan et al. (2006) compared the static traffic assignment and dynamic traffic assignment for emergency evacuation scenarios and emphasized the advantage of using intelligent transportation systems to route evacuees. Chen and Zhan (2006) used an agent-based technique to model traffic flows at individual vehicle level and investigated the collective behaviors of evacuating vehicles in the city of San Marcos, Texas. It revealed that the road network structure and population density have impacts on evacuation strategies. Chen, Meaker et al. (2006) also used agent-based modeling to analyze the minimum clearance time and how many people would need to be accommodated if evacuation route got unavailable in the Florida Keys area. Some traditional microscopic traffic simulation can also be treated as agent-based simulation to some extent. Jha, Moore et al. (2004) used microscopic simulation model (MITSIM) to model the evacuation of Los Alamos National Laboratory. Cova and Johnson (2002) presented a method to develop neighborhood evacuation planning with microscopic traffic simulation in the urban - wildland interface. Henson and Goulias (2006) reviewed 46 activity-based models and their competency of homeland security applications in TRANSIMS. All these agent-based simulation and evacuation models need pre-defined O-D matrix in the traffic supply-demand modeling stage to produce specific and detailed evacuation performance.

Shortest path based routing algorithms take an important place in trip assignment in evacuation plannings. Yin (2009) proposed a improved capacity constrained route planner algorithm for evacuation planning in large scale road network and provided aggregated performance analysis on evacuation time and computational time. Hobeika, Kangwalklai et al. (2003) discussed time-dependent label-constrained shortest path problems and its application in TRANSIMS with testing the Portland, Oregon, transportation system. Ng, Park et al. (2010) optimized the shelter locations through a hybrid bi-level model to improve the evacuation performance. Yue-ming and Hui (2010) modeled the shortest emergency evacuation time through Pontryagin minimum principle and dynamic traffic assignment. Xie and Turnquist (2011) integrated Lagrangian relaxation and Tabu search algorithm to optimize the lane-based evacuation system performance. Lu, George et al. (2005) simulated their evacuation plans with capacity constrained routing algorithms. Qiu, et al. (Kai-Fu and Wen-Long 2008) illustrated the emergency evacuation network statistics through genetic algorithm and kinematic wave models. Most of these researches focus on system-level analysis during emergency evacuation scenarios. Detailed evacuation simulation studies are appreciated to provide shortest path route for each individual evacuee.

Evacuation studies from hurricane, wildfires, terrorist attack, and other severe events reveal that how to evacuate people based on different demographic information in affected area is one of the most important factors for a successful evacuation plan. As high resolution demographic data source, LandScan USA population data has become the community standard for national population distribution. A lot of population risk study based on LandScan USA dataset were conducted to predict better analysis and results (Bhaduri, Bright et al. 2002). LandScan USA population cell (LPC) data has a high resolution population distribution to provide 90m x 90m (3'×3') resolution with national population distribution data (Bhaduri, Bright et al. 2007). It is much more accurate than conventional TAZ because some TAZ zones are large in scale or dense in population (You, Nedović-Budić et al. 1998). Compared to TAZ-based traffic assignment that generates trips from large-scale zones, LPC-based methods allow small-scale, cell-to-cell trip generation. This increase the number of O-D pairs in trip generation and distribution stage. A fast and effective O-D assignment between all LPC origins and destination/shelters becomes a critical issue.

## 2.3 Problem Definition

*2.3.1 Background*

The most important step for emergency evacuation planning is to define the evacuation area. This can be done by using either administrative boundaries or emergency event location based circles and polygons. Due to the randomness of emergency event locations, location based polygon works more efficient to evacuate affected people than administrative polygon. Thus, we define an evacuation scenario as described in Figure 2.1. The road network is cut from a location near Arlington, VA in Washington metropolitan area. We assume that some emergency happened near the centroid of the circle and the 3-mile-radius circle is defined as the affected area. All the destinations are connected to the roads cut by the circle. The LandScan USA Population Cells (LPC) in the circle are selected as the origins. The color represents cells with the number of population gradually, from red as high dense population to grey as no population. The total population in this area is 91,369. We also assume that all the population use personal cars as evacuation vehicles from their locations to the destinations. The vehicle per capita ratio is 0.8, which means every 125 people using 100 cars. People in each LPC cell access to the road network through the nearest network node. All the destinations are equally attractive to evacuees. They just look for the nearest destination based on their locations. This involves the shortest path algorithm to connect multiple LPC origins to their nearest destination respectively. It is different from most exiting shortest path problems, such as single source shortest path problem (which finds the shortest path from the source to every other nodes) and nearest neighbor problem (which finds the node from a set of nodes which is nearest to a given source node). Thus, we formally define this problem as Multiple-Sources-Nearest-Destination Shortest Path (MSNDSP) problem.

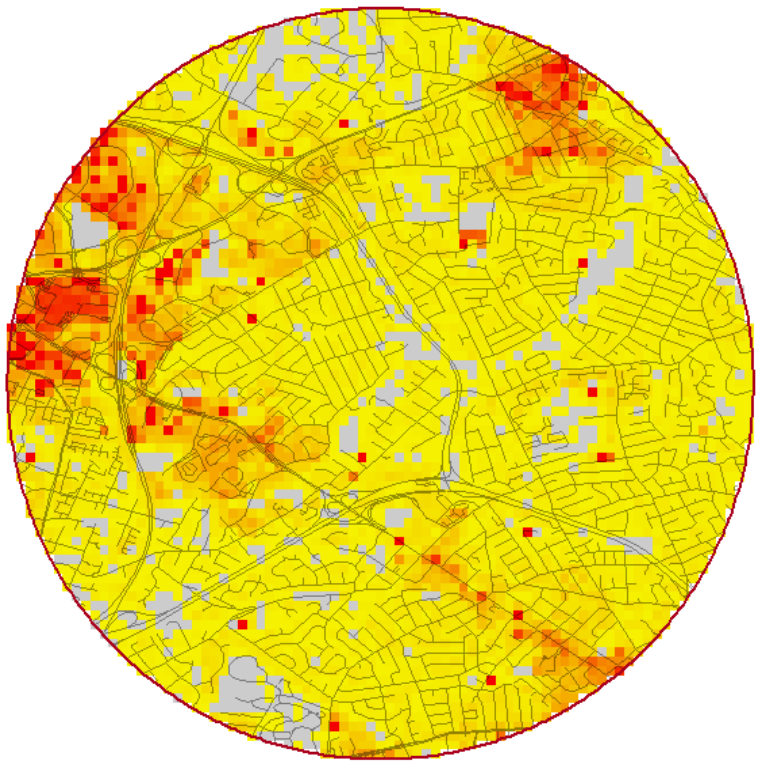

**Figure 2.1. The road network and LPC in evacuation area**

*2.3.2 The MSNDSP Problem*

**Definition:** Given a graph $G = (V, E)$ (where $V$ is the vertex and $E$ is the edge) and a group of centroid points of LandScan population cells $S$, find the shortest paths for each cell in $S$ to its nearest destination vertex $d$, where $d$ is from the defined destination pool $D$.

**Input:** Evacuation area is cut by a user-defined circle or polygon. A road network is defined with network nodes and links, as $G = (V, E)$. Destination nodes $D$ are built at the intersections of network links and the circle or polygon. A LPC map is matched to the road network by connecting the centroid of each LPC of $S$ to its nearest network node $v$.

**Output:** An O-D matrix connects each LPC origin $S$ to its nearest destination $d$. The purpose of our research is to generate the LPC-based input data for TRANSIMS simulation package. We use TRANSIMS to find the actual driving path using dynamic traffic assignment and microscopic traffic simulation.

**Goals:** 1) Minimize the initial travel cost for individual cell $s$: min $\{c(s)\}$
      2) Minimize the algorithm run-time

**Notes:** The travel cost can be travel distance, travel time, etc. Here we use travel time as travel cost in this algorithm. Since the evacuees do not know the road traffic conditions before their leaving, we assume that they use the route with shortest travel time based on an empty map. The destinations can also be assigned as shelters within the evacuation area. The evacuees can dynamically select the nearest shelter based on the real time travel conditions by solving the MSNDSP problem at each simulation step. This becomes an unknown destination problem. Also, instead of only finding the nearest one destination, we can find top 3 or 5 nearest destination by applying the MSNDSP problem multiple times to give the people optional destinations. This can be used for evacuees' compliance behavior analysis. Furthermore, This MSNDSP problem can be applied to other routing

problems, such as finding the nearest grocery store for each household among provided store locations.

## 2.4 Algorithm

*2.4.1 Overview*

As stated in previous sections, the MSNDSP problem has its unique characteristics to represent the OD assignment procedure for evacuation simulation with high resolution LPC data. First, the source nodes S are not from the network nodes N. They are consisted of the centroid points of LPC cells and connected to the nearest network nodes with edge cost (travel time) as 0. Second, the destination nodes D are determined by the intersections of circle and network links. We built a new node at the intersection as the destination node in selected area, as illustrated in Figure 2. The evacuees in each LPC cell choose the nearest destination out of D based on travel time on links. Here we use the ratio of link length to speed limit as the cost function. The objective function is to find the minimal initial travel time for people in each LPC cell. If using Dijkstria's shortest path algorithm with Forward Star (FS) network representation method, we have to run S times shortest path algorithm to find the O-D pair for each source node and destination node, where S is the number of LPC source nodes. If we connect all the destinations to a dummy super node and reverse the network with Backward Star (BS) representation method, as shown in Figure 2.2, we can have the same O-D matrix output as the FS method but with just one time run of Dijkstria's algorithm. In this paper, we implement the FS method as a benchmark study and build a new algorithm using BS method, which is called Super Node Trip Generator (SNTG).

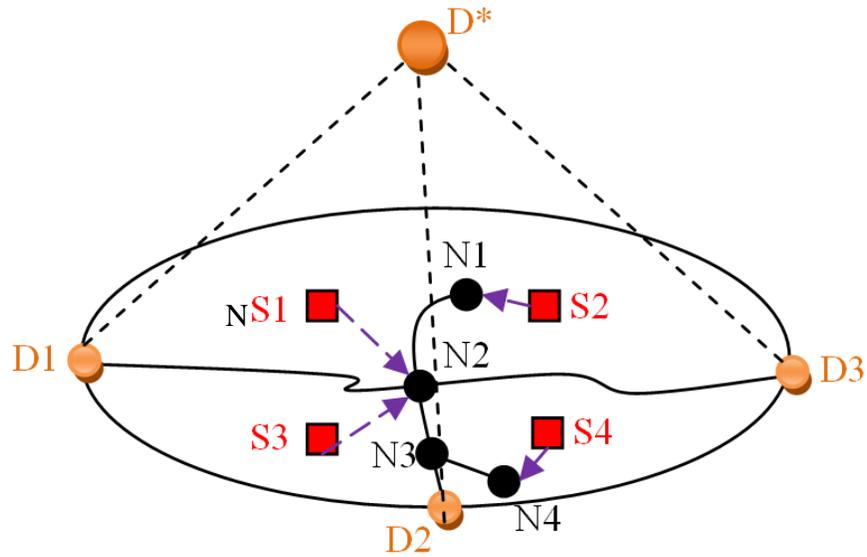

**Figure 2.2. Demonstration of MSNDSP problem with four source nodes and three destinations, including the super node concept**

### 2.4.2 Implementation with Forward Star

The MSNDSP problem can be tackled by solving the single source shortest problem for each LPC source node. All the shortest paths connecting the pre-defined destination nodes are saved for the next step. The final shortest path between a source node and its nearest destination are saved as a record in the O-D matrix file. An example of a small network and forward start structure is illustrated in Figure 2.3.

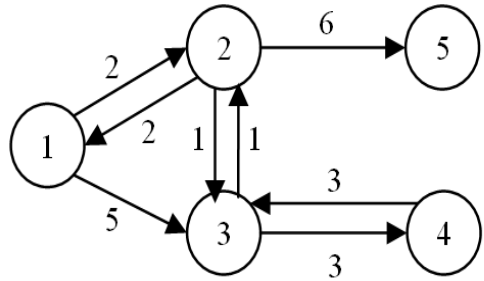

| Node | Pointer |
|---|---|
| 1 | 0 |
| 2 | 2 |
| 3 | 5 |
| 4 | 7 |
| 5 | 8 |
| (6) | 8 |

| Link ID | From Node | To Node | Cost |
|---|---|---|---|
| 0 | 1 | 2 | 2 |
| 1 | 1 | 3 | 5 |
| 2 | 2 | 1 | 2 |
| 3 | 2 | 3 | 1 |
| 4 | 2 | 5 | 6 |
| 5 | 3 | 2 | 1 |
| 6 | 3 | 4 | 3 |
| 7 | 4 | 3 | 3 |

**Figure 2.3. Network example with Forward Star (FS) structure**

The nodes N1, N2, N3 are network nodes and the nodes N4, N5 are destination nodes. We assume some LPC centroid points are connected to N1 with straight lines. The shortest paths between these LPC cells and destination nodes are transferred to the shortest paths between N1 and N4, N5. With simple calculation based on Dijkstria's algorithm, the shortest path between N1 and N4 is 1→2→3→4 with the sum link value as 6. The shortest path connecting N1 and N5 is 1→2→5 with the sum link value as 8. Thus, the solution for this special MSNDSP problem is the first path. All the source nodes connecting to this network node use this path for the O-D assignment.

*2.4.3 SNTG with Backward Star*

The method with FS structure can provide reasonable solution for our MSNDSP problem. But every source node needs a shortest path calculation to find all the shortest paths and then select the nearest one. This is very time consuming, especially in LPC data format. We define a new algorithm to solve the MSNDSP problem with adding a super node and reverse all the link directions, shown as Figure 2.4. The network is constructed with Backward Star (BS) structures. We call this new algorithm as Super Node Trip Generator (SNTG).

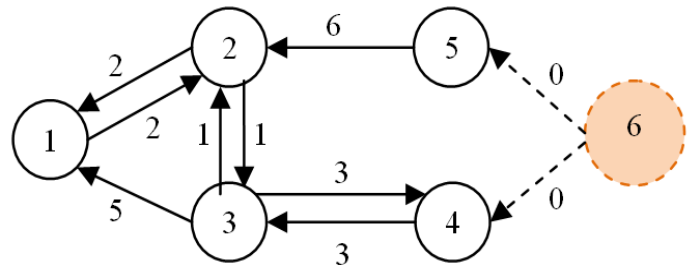

| Node | Pointer |
|------|---------|
| 1    | 0       |
| 2    | 1       |
| 3    | 3       |
| 4    | 6       |
| 5    | 7       |
| 6    | 8       |
| (7)  | 10      |

| Link ID | From Node | To Node | Cost |
|---------|-----------|---------|------|
| 0       | 1         | 2       | 2    |
| 1       | 2         | 1       | 2    |
| 2       | 2         | 3       | 1    |
| 3       | 3         | 1       | 5    |
| 4       | 3         | 2       | 1    |
| 5       | 3         | 4       | 3    |
| 6       | 4         | 3       | 3    |
| 7       | 5         | 2       | 6    |
| 8       | 6         | 4       | 0    |
| 9       | 6         | 5       | 0    |

**Figure 2.4. Network example for SNTG with Backward Star (BS) structure**

We solve the same MSNDSP problem using node 1, 2, 3 as the network nodes and node 4,5 as the destination nodes. Instead of finding the shortest paths between each source node to the destinations node, we begin our search from the destination nodes end with a super node 6. Two dummy links are also added to node 4 and node 5 with the link value 0. The problem is transferred to find the shortest path between node 6 and node 1. After using the Dijkstria's shortest path calculation, the shortest path is 6→4→3→2→1 with the sum link value as 6. Based on the shortest path algorithm's characteristics, the subset of the origin shortest path is also the shortest one. After deducting the link between node 6 and node 4, the shortest path between all the destinations and the source node is 4→3→2→1. It is exactly the same path with reverse direction in the previous section. Also, if there is another source node connecting to a different network node, the SNTG algorithm can use the same BS structure without building a new FS structure. This converts the MSNDSP problem to a normal single source shortest path problem with only using the Dijkstria's algorithm once, which can improve the computation performance significantly.

*2.4.4 Time Complexity Comparison*

To compare the algorithm performance, a benchmark study using Dijkstria's shortest path with forward star network structure is also implemented. The time complexity for the benchmark method is $\mathbf{O}(S * (E + V) * \log V)$, where S is the number of source nodes; E is the number of network links; V is the number of network nodes. Basically, it run the Dijkstria's algorithm S times. Our SNTG algorithm uses a super node to connect all the equivalent destinations and reverse the network structure, as represented with backward star structure. This transforms the MSNDSP problem to a single source shortest path problem. The time complexity of proposed SNTG algorithm is $\mathbf{O}(((E + D) + (V + 1)) * \log(V + 1))$, where D is the number of destination nodes. After adding the super node and connecting it to all destination nodes, the total number of links is E+D and the total number of nodes is V+1. This is almost equivalent to the time complexity $\mathbf{O}((E + V) * \log V)$, which only use the Dijkstria's algorithm once to find all the OD pairs. With the network size increasing, the SNTG algorithm works more efficient than the benchmark method, which almost follows an exponential relationship.

## 2.5 Algorithm

To demonstrate the feasibility and benefits of the proposed SNTG algorithm for improving evacuation O-D assignment efficiency, we developed several case studies of real-world evacuation operations at different scales and ran one full evacuation simulation with an O-D matrix generated by our algorithm. The experiment results are based on a Ubuntu 12.04 32-bit Linux workstation. The configuration of this workstation is 4GB RAM, Intel (R) Core (TM) 2 Quad CPU Q6700 @ 2.66GHz *4, and 500GB hard disk.

### *2.5.1 Data Preparation*

To test the robustness of our SNTG algorithms, we select some road networks near Arlington, VA, in Washington metropolitan area. They are cut by circles with different radius values, including 1 mile, 3 miles, 5 miles, 7 miles, and 10 miles, as shown in Figure 2.5. The detailed network information is summarized in Table 2.1. Also, we use the same circles to cut the LandScan USA daytime population data to get the LPCs respectively. Similar to Figure 2.1, all these LPC cells are defined as origins. The population numbers are also presented in Table 1. The number of destinations is determined by the intersections of the circle and the road networks. Once evacuees get out of the circle, they are safe.

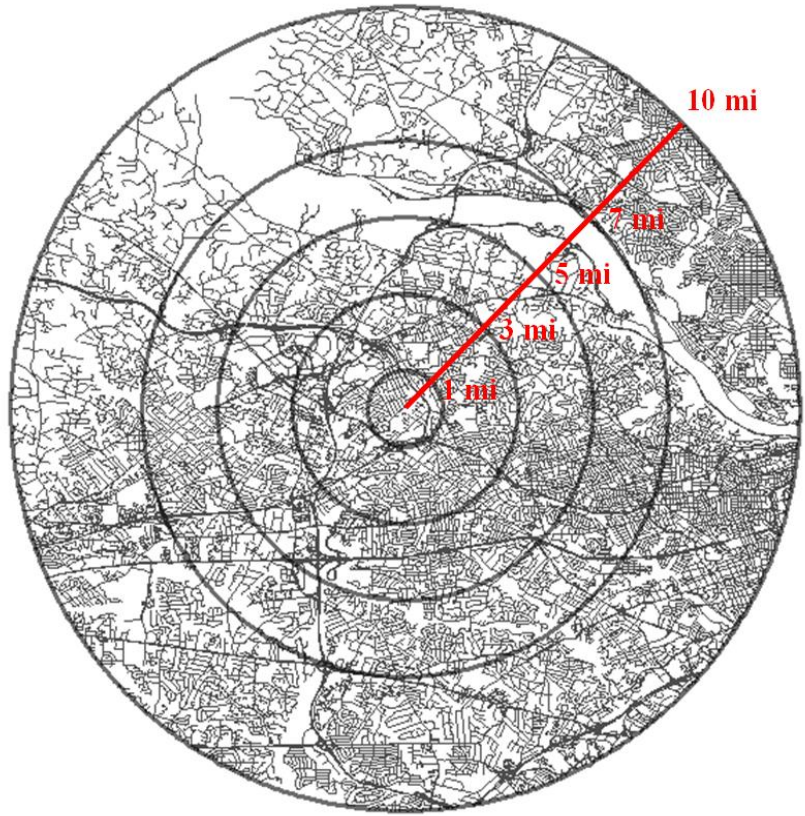

**Figure 2.5. Evacuation area with different radii**

**Table 2.1. Network and population scales**

| Radius (mile) | 1 | 3 | 5 | 7 | 10 |
|---|---|---|---|---|---|
| Area (mile$^2$) | 3.14 | 28.26 | 78.5 | 153.86 | 314 |
| # Nodes | 297 | 2854 | 6852 | 11574 | 24435 |
| # Links | 369 | 3599 | 8759 | 14549 | 31051 |
| # LPCs | 521 | 4659 | 12222 | 21396 | 46807 |
| # Dest | 48 | 151 | 176 | 226 | 395 |
| # Pop | 3,664 | 91,369 | 234,077 | 338,127 | 758,204 |

*2.5.2 Experiment Results in MSNDSP problem*

We use both Dijkstria's shortest path algorithm with Forward Star (FS) network structure and our backward star based SNTG algorithm to solve the MSNDSP problem. For each network size, the two methods generate the same output for O-D tables. The computing time in different network sizes with these two methods are summarized in Table 2.2 and Figure 2.6a. The SNTG algorithm produces very promising results compared to the FS method. Even with a large network containing 24,435 nodes, 31,051 links and 46,807 LPCs, the SNTG algorithm just need 0.0131 second to generate the O-D table, but the FS method need almost 10 minutes to get the same output.

Table 2.2. Computing time for FS and SNTG (in second)

| R (mi) | 1 | 3 | 5 | 7 | 10 |
|---|---|---|---|---|---|
| **SNTG** | 0.00007 | 0.00107 | 0.00327 | 0.00581 | 0.01310 |
| **FS** | 0.0340 | 3.3105 | 27.2254 | 107.7086 | 592.6513 |

Figure 2.6b shows the ratio of computing time for these two methods. It increases exponentially according to the network size. The SNTG algorithm works very efficient in large networks. For the 1 mile circle, it is about 500 times faster than the FS method. In 3-miles case, the computing time of SNTG is about 45,000 times shorter than the FS method.

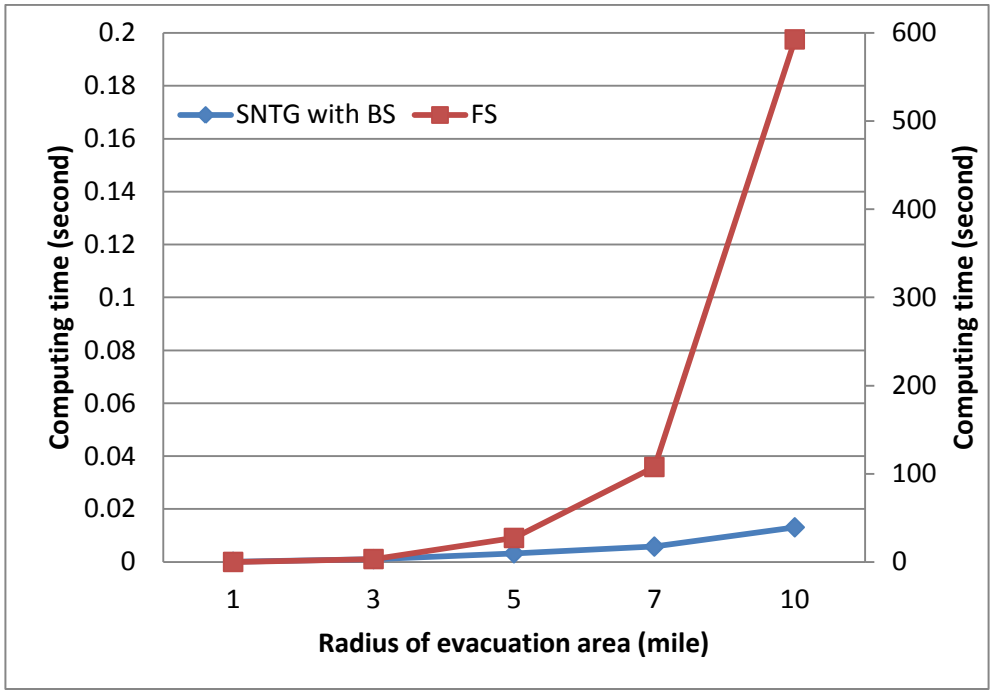

**Figure 2.6a. The computing time for FS structure and BS based SNTG algorithm**

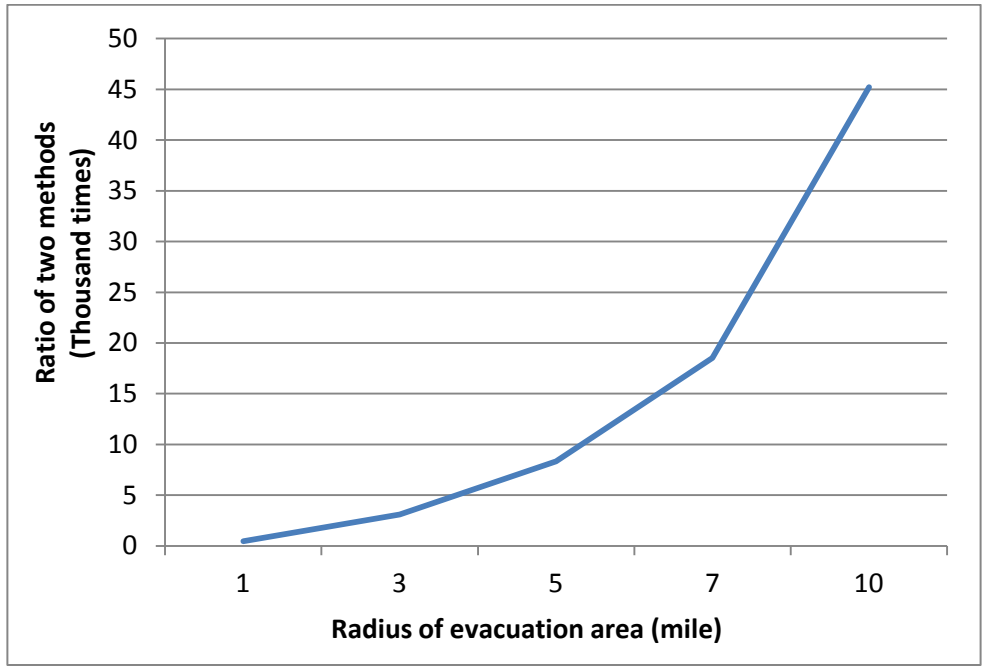

**Figure 2.6b. The ratio of computing time for FS structure and BS based SNTG algorithm**

*2.5.3 Evacuation Simulation Results*

To test the effectiveness of the SNTG algorithm for evacuation assignments, we simulate a 3-mile evacuation case study with the network and population data in section 3.1. Since the vehicle per capita ratio is 0.8, the total number of trips in this case study is 73,095 (91,369 * 0.8). All the vehicles are seeking safe destinations with a one-hour loading curve. TRANSIMS 4.0.8 package on Linux is used for the agent-based microscopic traffic simulation. The evacuation simulation time is set as 5 hours. To evaluate the evacuation efficiency, various measures of effectiveness (MOEs) are introduced (Han, Yuan et al. 2007). Evacuation clearance time is the most significant MOE to be considered in this study.

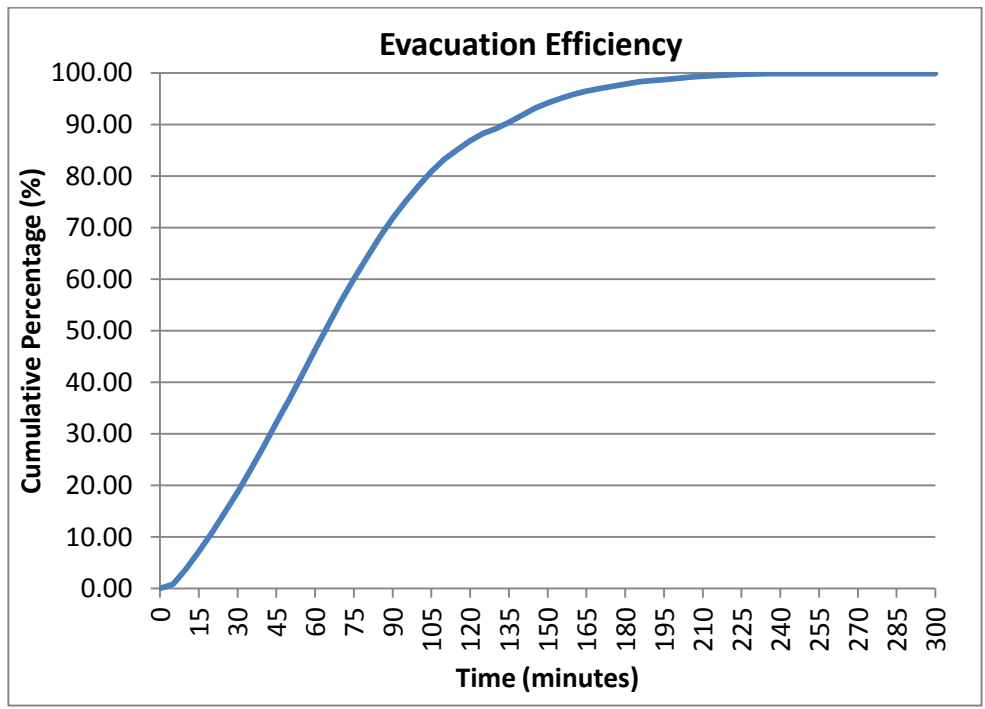

**Figure 2.7. Evacuation curve for the case study area**

The evacuation performance is plotted in Figure 2.7. All evacuees arrive at the pre-assigned destinations after about four hours. The evacuation is very efficient at the beginning stage due to the almost empty network, but the evacuation curve gets flat after three hours because of the congestions near some intersections. The average travel time for evacuees is 18.2 minutes. The O-D matrix, which is generated by the SNTG algorithm, works well in TRANSIMS simulation based evacuation assignments.

## 2.6 Discussions and conclusions

There are two major contributions in this paper. First, we define a new MSNDSP problem to represent our research problem in generating the high resolution LPC-based OD matrix for TRANSIMS evacuation simulation. Then we propose a new SNTG algorithm to solve our research problem efficiently and effectively. With the network size increasing, the SNTG algorithm works more efficient than the benchmark method, which almost follows an exponential relationship. A full evacuation performance case study using the OD table from our algorithm is also conducted and produces reasonable results.

The SNTG algorithm is proved to be accurate through computational analysis by comparing whether its output is identical to the benchmark study. The super node idea works because a part of the shortest path is also the shortest by Dijkstria's algorithm. A theoretical proof of SNTG algorithm generating identical OD pairs as benchmark study will be provided in the near future.

As mentioned in Section 3.2, the MSNDSP problem can be used in other routing problems besides emergency evacuation assignments. In the future, we would like to formulize this problem to apply to other research fields, such as nearest grocery store problem. In addition, we will apply our SNTG algorithm in dynamic destination assignment problem. In this case, people in the origins will be updated their nearest destinations in every simulation time step based on the real time network conditions. This can represent the real situation when people are seeking safety. They will change their destination depending on the road conditions.

# CHAPTER 3
# USING HIGH RESOLUTION DEMOGRAPHIC DATA IN AGENT-BASED EVACUATION ASSIGNMNETS

# Abstract

The successful evacuation planning in a transportation system depends chiefly on the input data accuracy and simulation model effectiveness. This paper proposes a new traffic modeling framework using high spatiotemporal resolution demographic data and microscopic simulation models for evacuation planning and operation. A comparison study using real-world road network and demographic data in Alexandria, Virginia is conducted to evaluate the framework accuracy and evacuation efficiency. Three aspects that impact evacuation performance are analyzed, including spatial population resolutions in traffic analysis area (Traffic Analysis Zones vs. LandScan USA Population Cells), temporal population resolutions in evacuation start times (daytime vs. nighttime) and combined spatiotemporal resolutions in departure time models (normal S-shape model vs. location-based model). TAZ-based traffic assignment underestimates evacuation travel time and simulation computational time. The two evacuation start time results suggest it takes more time for evacuees in Alexandria to travel to shelters in daytime under normal traffic conditions than the case in nighttime because of the different temporal demographic patterns. The studies of two departure time models also indicate that location-based model improves the evacuation efficiency at the first several hours when utilizing the road network sufficiently. But the average travel time for individual vehicles decreases because severe congestion would happen if all people are trying to access the road in a short period. The new framework shows flexibility in implementing different evacuation strategies and accuracy in evacuation performance. The use of this framework can be explored to day-to-day traffic assignment to support daily traffic operations.

*Keywords:* agent-based model, LandScan, high resolution demographic data, evacuation management, evacuation traffic assignment

## 3.1 Introduction

Traffic assignment with macroscopic or microscopic simulation is used to develop evacuation models with evacuation performance analysis at multiple levels. Due to the limitation of computing resources and high resolution demographic data availability, Traffic Analysis Zones (TAZ) based simulation researches dominated the evacuation simulation area. Conventional TAZ based models assign all the trips in a TAZ to the centroid and then connected to the nearest node on the network. Even agent-based traffic simulation, used in Transportation Analysis and Simulation System (TRANSIMS) package, cannot represent the real-world situation with TAZ models because of it evenly distribute travelers to all the activity locations in one TAZ. A more accurate agent-based traffic model is needed to take advantage of high resolution demographic data to produce more accurate simulation results.

Agent-based micro-simulation models have been well explored in traffic and transportation planning and operation domain. Bo and Cheng (2010) reviewed the applications of agent technology in traffic and transportation systems from various aspects, including agent-based systems architecture and platforms, as well as their applications in roadway transportation, railway transportation, and aviation systems. Agent-based behavior models are developed to describe the individual driving behaviors from system level to intersection level (Dia 2002, Doniec, Mandiau et al. 2008, Nagel and Flötteröd 2009). These papers provide some methods to implement and simulate travel demand models with multi-scale choice dimensions and constraints. Zhang and Levinson (2004) developed an agent-based travel demand model considering the interactions of three types of agents in the transportation system: node, arc, and traveler. Lin, Eluru et al. (2008) combined activity-based modeling and dynamic traffic assignment with two different software packages to provide a conceptual framework and explored practical integration issues for this combination. Khalesian and Delavar (2008) developed a prototype of spatio-temporal multi-agent system for microscopic traffic simulation of highway traffic at peak hours, which integrating the geospatial information systems for transportation systems. Besides these agent-based modeling studies for transportation simulation in relative small areas, large-scale agent-based microscopic traffic simulation based on queuing theory is explored for multiple applications and empirical case studies in Europe (Cetin, Burri et al. 2003, Balmer, Cetin et al. 2004, Balmer, Axhausen et al. 2006, Meister, Balmer et al. 2010). Agent-based approach can provide detailed driving information and interaction between individual vehicles, which helps evacuation operation managers to produce detailed evacuation strategies.

Evacuation planning and operation also needs simulation-based studies while real world evacuation data is hardly available. Han, Yuan et al. (2006) compared the static traffic assignment and dynamic traffic assignment for emergency evacuation scenarios and emphasized the advantage of using intelligent transportation systems to route evacuees. Chen and Zhan (2006) used an agent-based technique to model traffic flows at individual vehicle level and investigated the collective behaviors of evacuating vehicles in the city of San Marcos, Texas. It revealed that the road network structure and population density have impacts on evacuation strategies. Chen, Meaker et al. (2006) also used agent-based modeling to analyze the minimum clearance time and how many people would need to be accommodated if evacuation route got unavailable in the Florida Keys area. Some traditional microscopic traffic simulation can also be treated as agent-based simulation to some extent. Jha, Moore et al. (2004) used microscopic simulation model (MITSIM) to model the evacuation of Los Alamos National Laboratory. Cova and Johnson (2002) presented a method to develop neighborhood evacuation planning with microscopic traffic simulation in the urban - wildland interface. Henson and Goulias (2006) reviewed 46 activity-based models and their competency of homeland security applications. All these agent-based simulation and evacuation models are still based on existing TAZ-based traffic assignment during the traffic supply-demand modeling stage.

Evacuation planning and operation studies from hurricane, wildfires, terrorist attack, and other severe events reveal that how to evacuate people based on different demographic information in affected area is one of the most important factors for a successful evacuation plan. As high resolution demographic data source, LandScan USA population data has become the community standard for national population distribution. A lot of population risk study based on LandScan USA dataset were conducted to predict better analysis and results (Dobson, Bright et al. 2000, Bhaduri, Bright et al. 2002). LandScan USA population cell (LPC) data has a high resolution population distribution to provide 90m x 90m (3'×3') resolution with national population distribution data (Bhaduri, Bright et al. 2007). It is much more accurate than conventional TAZ because some TAZ zones are large in scale or dense in population (You, Nedović‐Budić et al. 1998). Compared to TAZ-based traffic assignment that generates trips from large-scale zones, LPC-based methods allow small-scale, cell-to-cell trip generation. In addition, LPC provides both day-time and night-time population distributions. Traffic assignment based on different time pattern may influence the accuracy of evacuation simulation.

To take advantage of high resolution demographic data, a new agent-based evacuation assignment framework is proposed in this paper. A comparison study using road network and demographic data in Alexandria, Virginia is conducted to evaluate the framework accuracy and evacuation efficiency. Three aspects are considered, including traffic analysis area resolutions (TAZ vs. LPC), evacuation start times (daytime vs. nighttime), and departure time choice models (normal S-shape model vs. location-based model). The simulation results indicate that high resolution demographic data increases evacuation accuracy from both spatial and temporal perspectives. Also, the location-based departure time choice model provides a method to maximize the evacuation performance through location-based strategies. Subsequent discussions on improving evacuation simulation accuracy and efficiency through high resolution demographic data are also presented.

## 3.2 Framework Description

LandScan population data and open-source agent-based traffic simulation package TRANSIMS provide valuable data and programs to build an efficient and accurate evacuation assignment and simulation platform. LPC-based traffic assignment brings many new issues to integrate modules in TRANSIMS package. A new evacuation framework built on several programs in TRANSIMS is proposed, as shown in Figure 3.1.

There are nine modules in this agent-based evacuation modeling framework, which convert the raw input data at the beginning to detailed simulation performance results at the end. It can provide both aggregated traffic information and individual vehicle tracking analysis.

1. Selection. The original input data are LandScan USA population (daytime/nighttime) and Navteq (a data company) national road network data. The users choose predefined evacuation area by circle, polygon, or county boundary through selection module. All the affected population and origins/shelters locations are summarized in two new files, as selected population and evacuation area. LandScan Global population data and OpenStreetMap (open source map data) can be implemented for simulating areas outside United States.

2. Conversion. To take advantage of TRANSIMS simulation tools, the selected Navteq road network is converted to TRANSIMS-based format. The population is converted to agents based on demographic information.

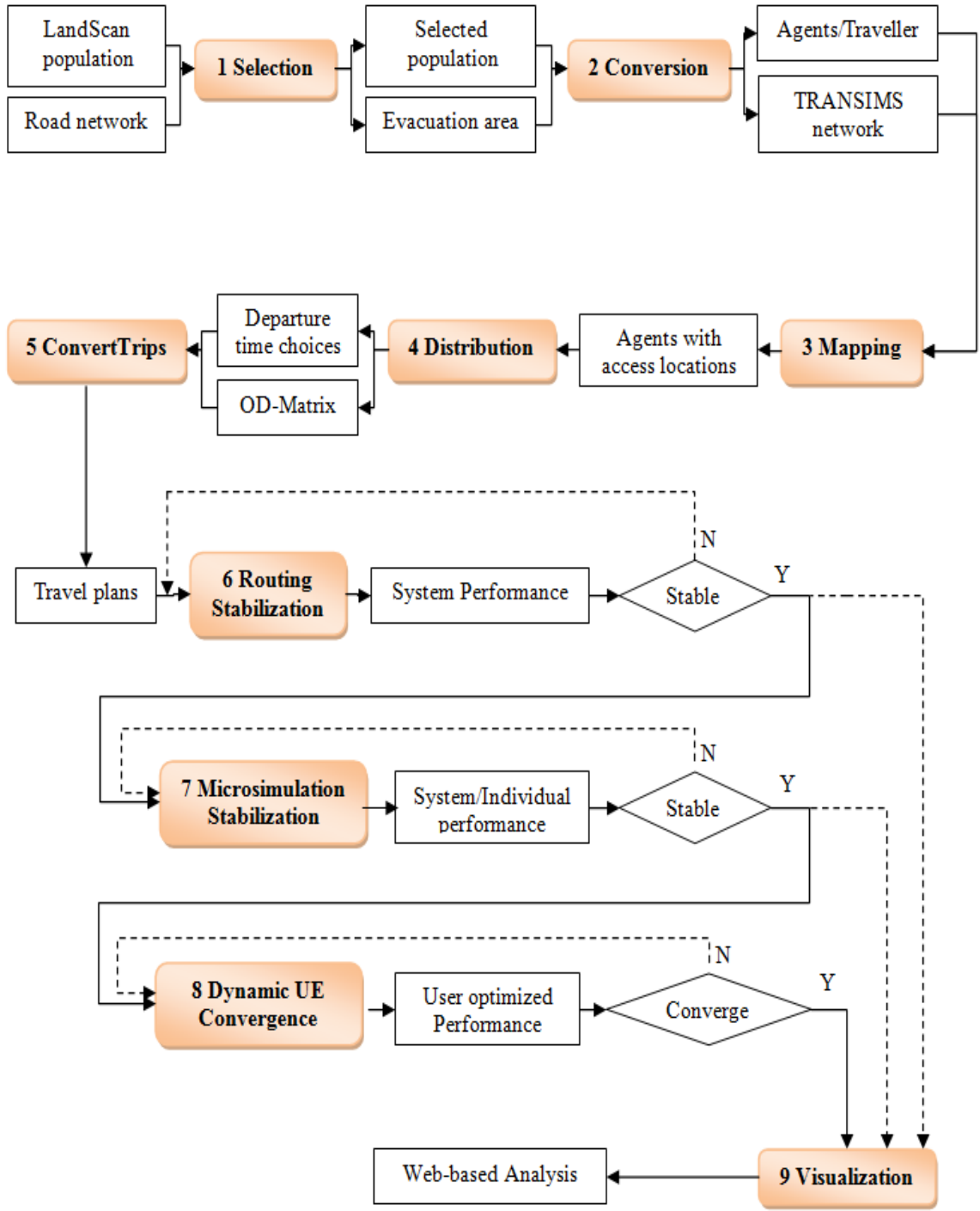

**Figure 3.1. Process structure of LandScan agent-based evacuation modeling framework**

3. Mapping. The mapping module assigns the travelers in each LandScan cell to their nearest activity locations. Vehicles access to the network through activity locations along the road. This module is used to resolve the limitation of activity-based traffic simulation in TRANSIMS. The original program assumes that there are at least two activity locations in a TAZ for inner-zone trips and at least one activity location for inter-zone trips. This is understandable in TAZ since the TAZ is usually large enough to include several road segments. But LPC size is much smaller than TAZ and there are many cells having no roads crossing through their areas. In this case, the trips in that zone cannot be assigned to the activity locations. Also, the original TRANSIMS algorithm evenly assigns all the trips in one TAZ to all the activity locations in that zone. In the real world, travelers prefer to access roads at the nearest spots to their locations. These two limits are resolved in this mapping module.

4. Distribution. Each LandScan USA population cell (LPC) is assigned as one origin zone. All the exit nodes generated in the first module are assigned as destinations/shelters. The OD matrix can be generated as formula (1) with minimum travel cost to each LPC. The travel cost can be calculated as equation (2), where impedance can be summarized with different shortest-path based routing algorithms, including shortest network distance routing, highway-based routing, and straight-line distance routing. In this evacuation model, only the shelters out of the evacuation area are considered.

$$Min \{Travel\ Cost_i, i\ is\ the\ i^{th}\ LPC\} \qquad (1)$$

$$Travel\ Cost_i = \sum_j Impedance_j, j\ is\ the\ j^{th}\ link\ in\ a\ route \qquad (2)$$

5. ConvertTrips. This is a program provided by TRANSIMS to generate travel plans for every individual traveler with OD-matrix and used-defined departure time choice model (or loading curve). Trip chain can also be implemented. But in this paper, we assume that all the vehicles travel to shelters directly after evacuation order is placed. TRANSIMS only consider 1000 TAZ zones as maximum for trip assignment. There are more than 1000 LPC cells in Alexandria. Technically, each LPC is equivalent to one TAZ zone. This limitation has been adjusted to accommodate LPC-based traffic assignment method.

6. Routing Stabilization. To save the computational time, link-based trip assignment is implemented to generate macroscopic system performance results. There is a loop to optimize the results to achieve stable status. Volume/Capacity ratio, travel time change ratio, et al. can be used as the condition statement. After the macroscopic simulation is stable, the results can be pulled to the last visualization module for analysis.

7. Microsimulation Stabilization. Microscopic simulation is also implemented if the users want to see detailed simulation results of each vehicle for operation management. Similar to the module 6, users can use those ratios to decide the number of iterations to achieve stable condition. The simulation results are also able to be plotted for intermediate analysis.

8. Dynamic UE Convergence. To improve the simulation accuracy, dynamic traffic assignment is also integrated to simulate user optimized evacuation dynamically. The convergence conditions are defined by users, such as less than 2% changes in travel time for each vehicle. The final outputs with both system level information and individual vehicle movement are summarized.

9. Visualization. Beyond the research analysis with output data, the simulation results file is processed to display on a web-based application front-end with both macroscopic and microscopic information.

This framework provides a one-stop application tool for both researchers and practitioners. From a scientific research perspective, we focus on the comparison study of TAZ and LPC based traffic assignment, evacuation happening times, and loading curve impact on assignment performance in this paper. Therefore, a corresponding TAZ-based simulation is also implemented with the same modules from five to night.

## 3.3 Evacuation Case Study Design

To assess evacuation efficiency with this proposed LPC-based high resolution traffic assignment framework in different scenarios, an evacuation case study using data in Alexandria, Virginia is conducted through comparing with TAZ-based traffic assignment.

### *3.3.1 Data Resources*

As stated in the LandScan agent-based evacuation modeling framework, population distribution and road networks are two major input data for the evacuation assignment. The road network of Alexandria is shown as red lines in Figure 3.2, which consists of 3634 links and 2608 nodes. The 62 TAZ zones are also displayed in Figure 1. Most of TAZ boundary uses real-world road segment, which is widely used in traffic planning field. The background map source is from OpenStreetMap in Esri ArcGIS 10.1 package.

The definition of high resolution data in this paper does not only include spatial demographic distribution, but also refers to temporal distribution. LandScan USA population cells (LPC) data, in both daytime and nighttime formats, are implemented. The daytime is defined from 6:00am to 6:00pm and the nighttime is the left 12 hours. The LPC daytime data consists of 5657 non-zero cells (6131 in total) and locates at downtown (the right side) with high dense population, as shown in Figure 3.3a. The color represents cells with the number of population gradually, from red as high dense population to grey as no population. Technically, TAZ and LPC have the same zone definition, but LPC size is much smaller. The total daytime population in Alexandria is 161,519 with a cell of maximum population as 2641. The LPC nighttime data is consisted of 4522 non-zero cells (still 6131 in total) and distributes more high dense area on the left residential area, as displayed in Figure 3.3b. The total nighttime population is 141,529 with a cell of maximum population as 419. The standard division values for daytime and nighttime population distribution data are 33.613 and 62.469, which means the nighttime population data distributes more evenly. The total amounts of population and distribution patterns may impact the evacuation performance in Alexandria. The TAZ population data are aggregated from LPC data, in both population amount and temporal distribution, to make the inputs consistent for comparison studies.

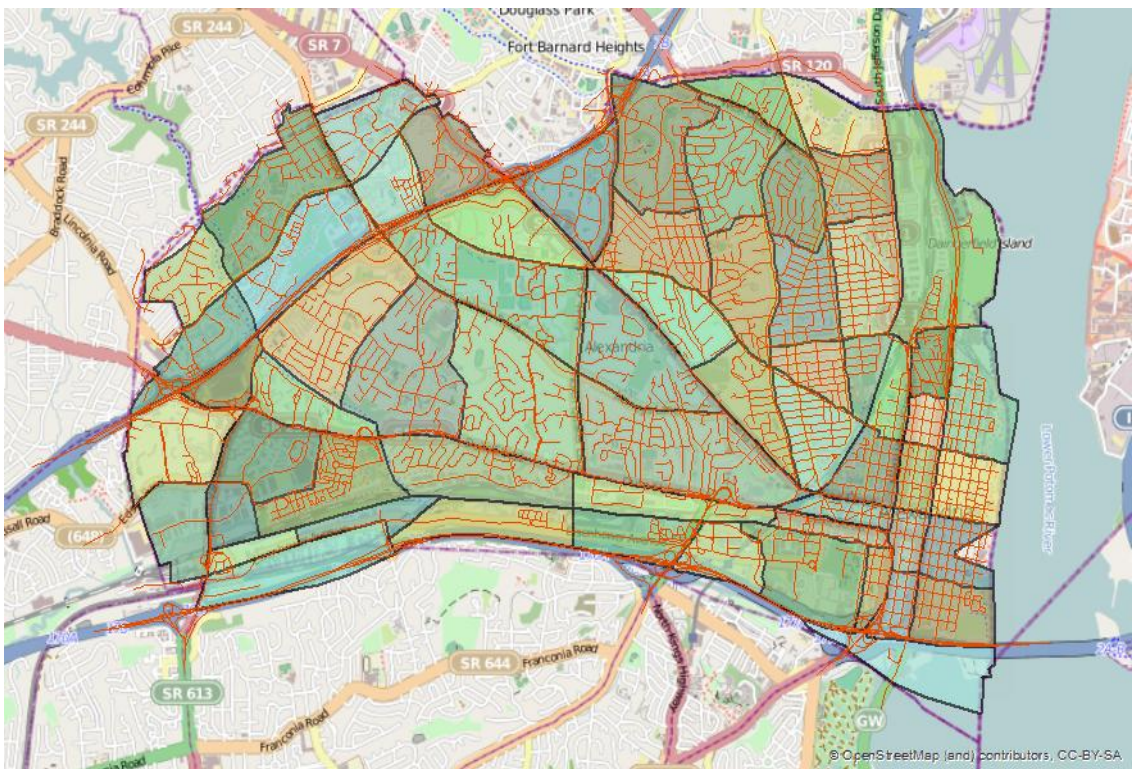

**Figure 3.2. Road network (red lines) and Traffic Analysis Zones in Alexandria (colored polygons)**

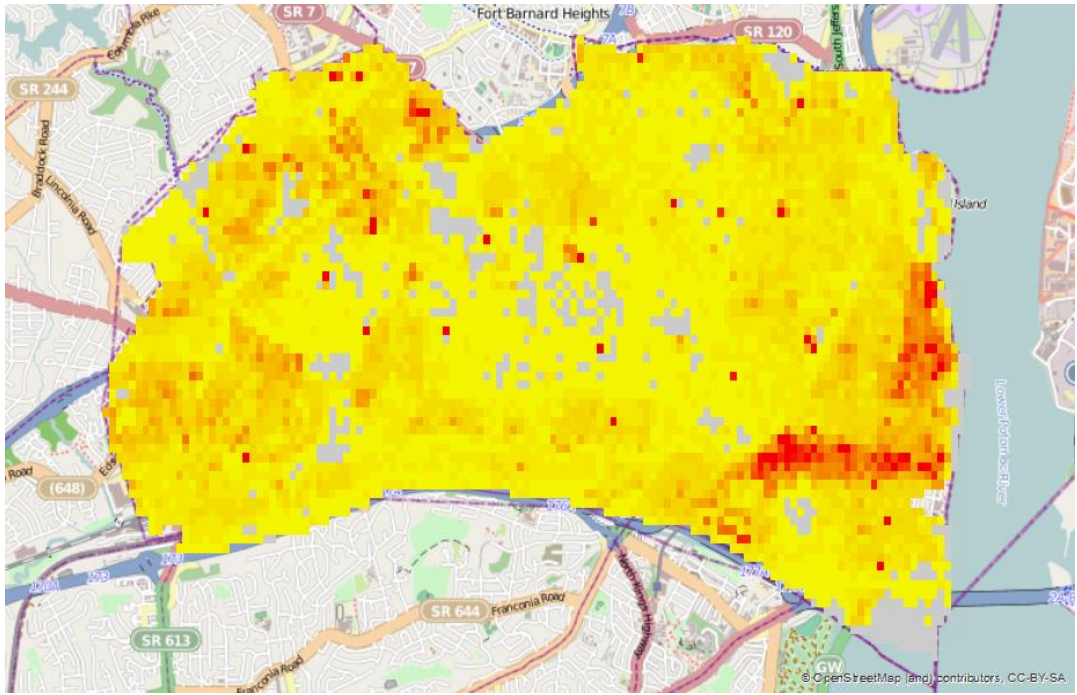

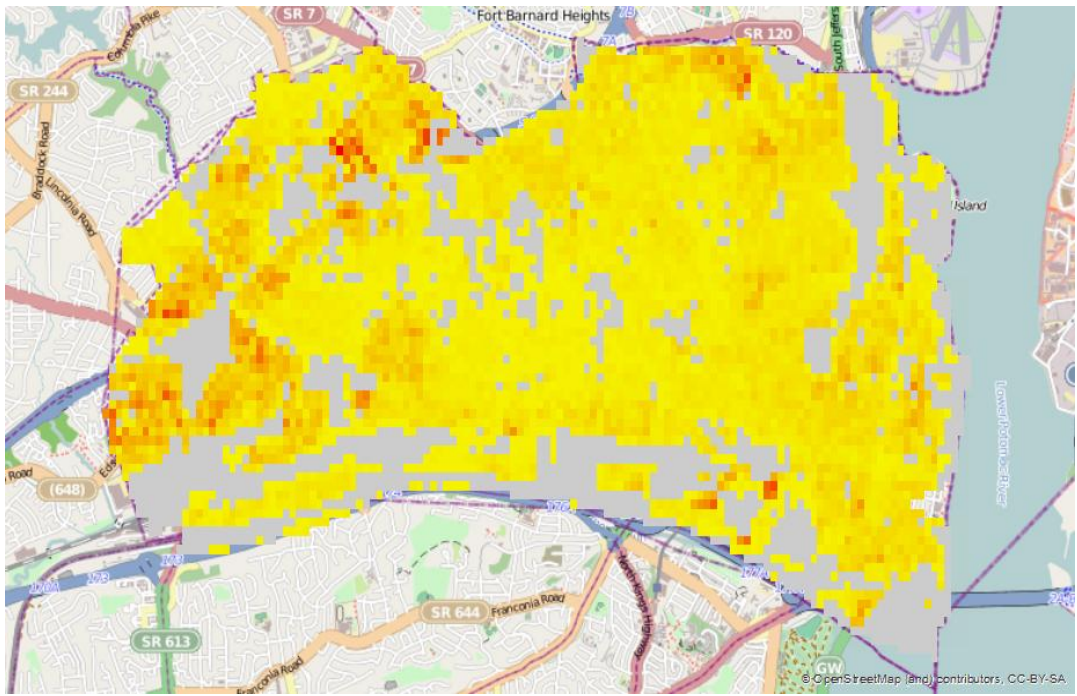

**Figure 3.3a. LandScan USA 2011 daytime population data in Alexandria (red: high density; yellow: low density; grey: zero)**
**Figure 3.3b. LandScan USA 2011 nighttime population data in Alexandria (red: high density; yellow: low density; grey: zero)**

*3.3.2 Site Modeling*

Based on the Alexandria road network map in Figure 1, a detailed network configuration is generated for the agent-based microscopic simulation. This includes 2608 nodes, 3634 links, 7718 activity locations, 1062 stop/yield signs, and 261 traffic signals. The Alexandria area is about 15.2 sq. mi. In our case study, only private vehicle mode is considered. The traffic mode, such as transit, can be implemented later. The trip chain for each traveler is "walk-drive-walk", which means walking to the car from origin, driving to the nearest parking lot at destination, and then walking to assigned shelters. Since the vehicle per capita ratio is about 0.8 in the United States, the total trips for evacuation in Alexandria is 129,215 (equal to 161,519 * 0.8) in daytime and 113,223 (equal to 141,529) in nighttime. In addition, 21 shelters are assigned and connected one-to-one with 21 exits in the road network. Evacuation trips generally move outward as evacuees leave an evacuation area and seek safety.

The evacuation departure time choice model for each zone or cell is converted to 15-munites interval volumes using loading curve. Florida Department of Transportation researched various response curves for evacuation conditions (Radwan, Mollaghasemi et al. 2005). A normal S-shape loading curve is implemented here for benchmark study, as the purple line in Figure 3.4. To adjust the evacuation performance for this area size, a 6-hours loading curve is adapted. To test the loading curve impacts on evacuation, a location-based departure time choice model is also modeled, as the red line in Figure 4. The basic idea is to evacuate people near the boundary with fast loading curve and people inner the city with a normal loading curve. Compared to a single loading curve for all evacuees, this helps people near the exits to evacuate to shelters fast. Figure 5 helps to explain the location-based departure time choice model. The blue points are the TAZ central points. The red points are the safe shelters as destinations. The LPC cells are not drawn here due to the large amount, but the concept is similar to TAZ. The location-based departure time choice model chooses the TAZ zones in shadow to evacuate with normal S-shape loading curve and the surrounding TAZ zones to evacuate with fast loading curve. To be clear, we still use TAZ boundaries, rather than the drawing line as boundary in Figure 3.5. The total number of trips in the shadow area is 52,978 in daytime and 50,950 in nighttime. In this case study, the corresponding LPC cells are selected based on TAZ zones.

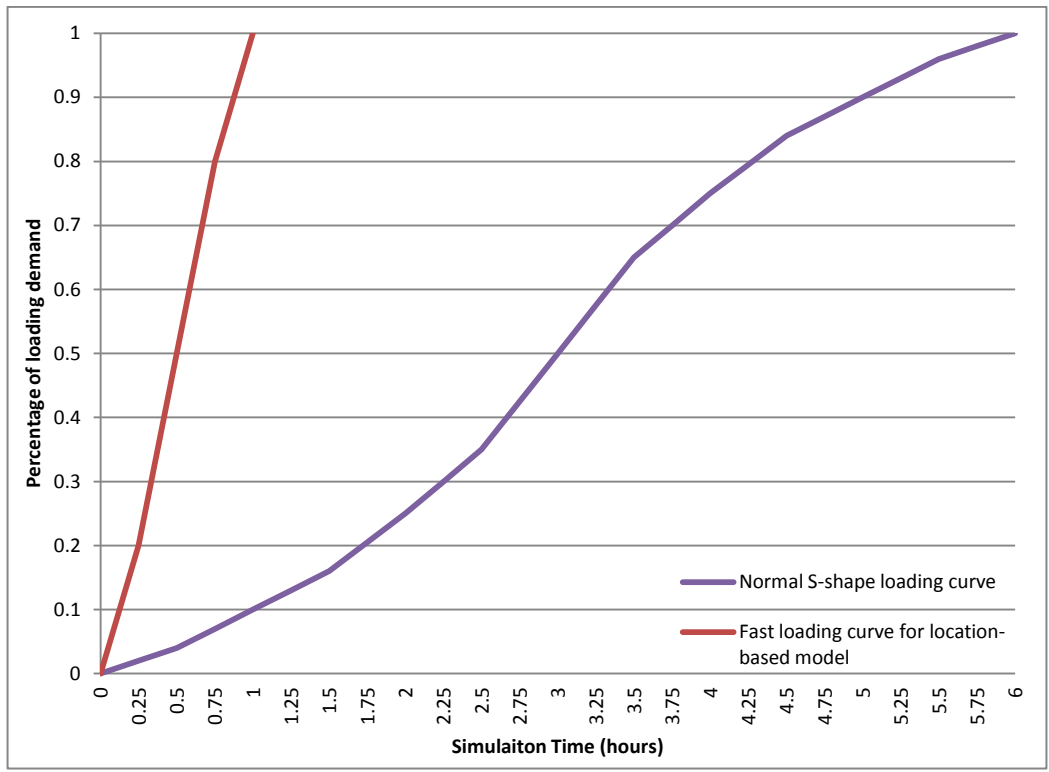

**Figure 3.4. Loading curves for different evacuation scenarios**

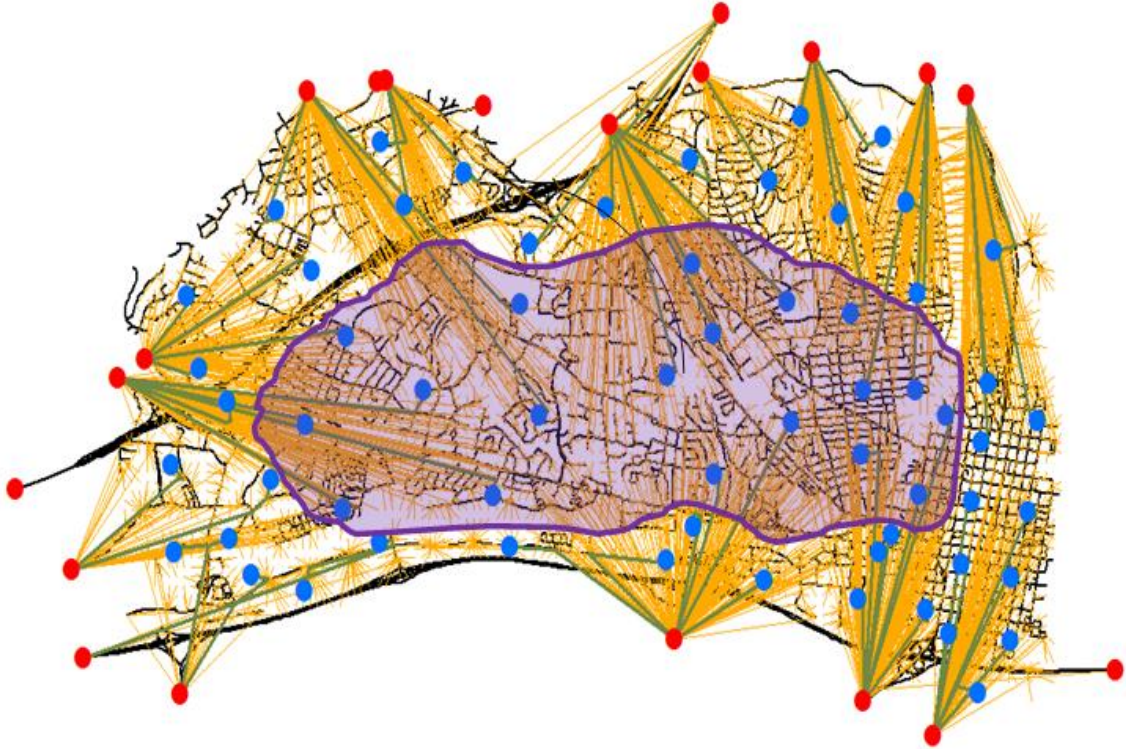

**Figure 3.5. Location-based trip assignment in TAZ and LPC (red points: shelters; blue points: TAZ centroids; black lines: road network; green lines: O-D links for TAZ; orange lines: O-D links for LPC; shadow polygon: areas that only use normal S-shape loading curve)**

How to generate an optimal evacuation plan for every evacuee with destination and route assignment is crucial for a city level evacuation planning. Route based on shortest path algorithm is appreciated from both planners' and evacuees' perspective. The conventional highway-biased shortest path assignment is implemented in Figure 5 for both TAZ and LPC scenarios. The highway-biased shortest path is calculated by the ratio of link distances and speed limits. The bold green links connects the TAZ central points to their shelters. The orange links bridges the LPC central points to their shelters. The population data in Figure 5 is daytime data. The nighttime data has the same O-D links for TAZ assignment and similar links for LPC assignment (with less orange connection links because of less non-zero LPC cells). Each TAZ or LPC is assigned to the destination based on the geometric location, rather than the amount of population. Thus, the shortest path routing algorithm for both daytime and nighttime data should be the same if we consider cells with value as zero.

*3.3.3 Simulation Scenarios*

Conventionally, transportation planners use traffic analysis zones for trip assignments. This study focuses on analyzing the agent-based modeling evacuation performance with high resolution data in both spatial (TAZ and LPC) and temporal (daytime and nighttime) manner. Also, to maximize the evacuation performance, two departure time choice models are simulated, including normal S-shape model and location-based model. In total, 8 evacuation scenarios are studied in this paper, combing two spatial models, two temporal models, and two departure time choice models. They are TAZ-Daytime-1, TAZ-Nighttime-1, LPC-Daytime-1, LPC-Nighttime-1, TAZ-Daytime-2, TAZ-Nighttime-2, LPC-Daytime-2, and LPC-Nighttime-2. Here, "1" means the normal S-shape model. "2" means the location-based model. The proposed framework was run on a Window 7 64-bit laptop computer. The configuration of this laptop is 16GB RAM, 2.6GHz Intel(R) Core(TM) i5-3320 CPU, and 500GB hard disk. A 10-hour simulation time is used and the other parameters are set as stated in the preceding sections. All the simulation results are collected after reaching user equilibrium status. To adjust the impact of random number in the framework, all the simulation results are based on the average of 30 independent runs. To evaluate the evacuation efficiency of these scenarios, various measures of effectiveness (MOEs) are developed. Han et al. (Han, Yuan et al. 2007) suggested a four-tier MOE framework for evacuation, including evacuation time, individual travel time and exposure time, time-based risk and evacuation exposure, and time-space-based risk and evacuation exposure. They can be used in various scenarios in evacuation modeling stage to satisfy different evacuation planning purpose. Evacuation efficiency is the most significant MOE to be considered in this study. Also, average travel time for evacuees and computational time are summarized to provide a system level reference.

## 3.4 Simulation Results and Discussions

Evacuation simulation results from comparison studies of traffic analysis area resolutions, evacuation start times, and departure time choice models are summarized in Table 3.1. In general, LPC based evacuation performance is not as efficient as TAZ based evacuation. Since LPC based traffic assignment can represent how people access the road network in real world, TAZ based assignment underestimates the real travel time in evacuation scenarios. The daytime and nighttime cases also have slight difference because of the population distribution patterns and the total number of evacuees. The location-based departure time choice model significantly improves the evacuation efficiency but causes longer average travel time for each individual car. Detailed plot and table analysis are also presented below to reveal the impact of these three aspects.

**Table 3.1. Evacuation time for different scenarios**

| Departure Time Choice Model | Evacuation Event Times | Population Resolutions | Percent Evacuated | | | | | | | | |
|---|---|---|---|---|---|---|---|---|---|---|---|
| | | | 20% | 30% | 40% | 50% | 60% | 70% | 80% | 90% | 100% |
| | | | Evacuation Times in Minutes | | | | | | | | |
| Normal S-shape Model | Daytime | TAZ | 95 | 127 | 151 | 174 | 195 | 222 | 256 | 295 | 375 |
| | | LPC | 98 | 130 | 158 | 185 | 213 | 245 | 285 | 335 | 435 |
| | Nighttime | TAZ | 96 | 127 | 153 | 175 | 199 | 226 | 258 | 299 | 355 |
| | | LPC | 99 | 130 | 156 | 181 | 209 | 240 | 276 | 319 | 395 |
| Location-based Model | Daytime | TAZ | 27 | 39 | 52 | 65 | 83 | 103 | 130 | 170 | 370 |
| | | LPC | 31 | 45 | 62 | 81 | 102 | 131 | 175 | 235 | 420 |
| | Nighttime | TAZ | 31 | 47 | 65 | 83 | 102 | 125 | 163 | 206 | 400 |
| | | LPC | 32 | 47 | 65 | 85 | 105 | 135 | 175 | 241 | 490 |

To analyze the impacts of spatial high resolution data, the simulation results for TAZ and LPC based traffic assignment under different evacuation event times and departure time choice models are extracted as Figure 3.6. In all these four scenarios, TAZ based traffic assignments have better performance than LPC based assignments. In original TRANSIMS package, it assigned all the trips in one TAZ zone evenly to all the activity locations in that zone. But our LPC based assignment assigned each population cell to their nearest activity locations based on corresponding demographic information, which is how people in real-world access the road network. The high resolution LPC data might cause some activity locations having more assigned vehicles to produce potential congestions, which slow down the evacuation efficiency.

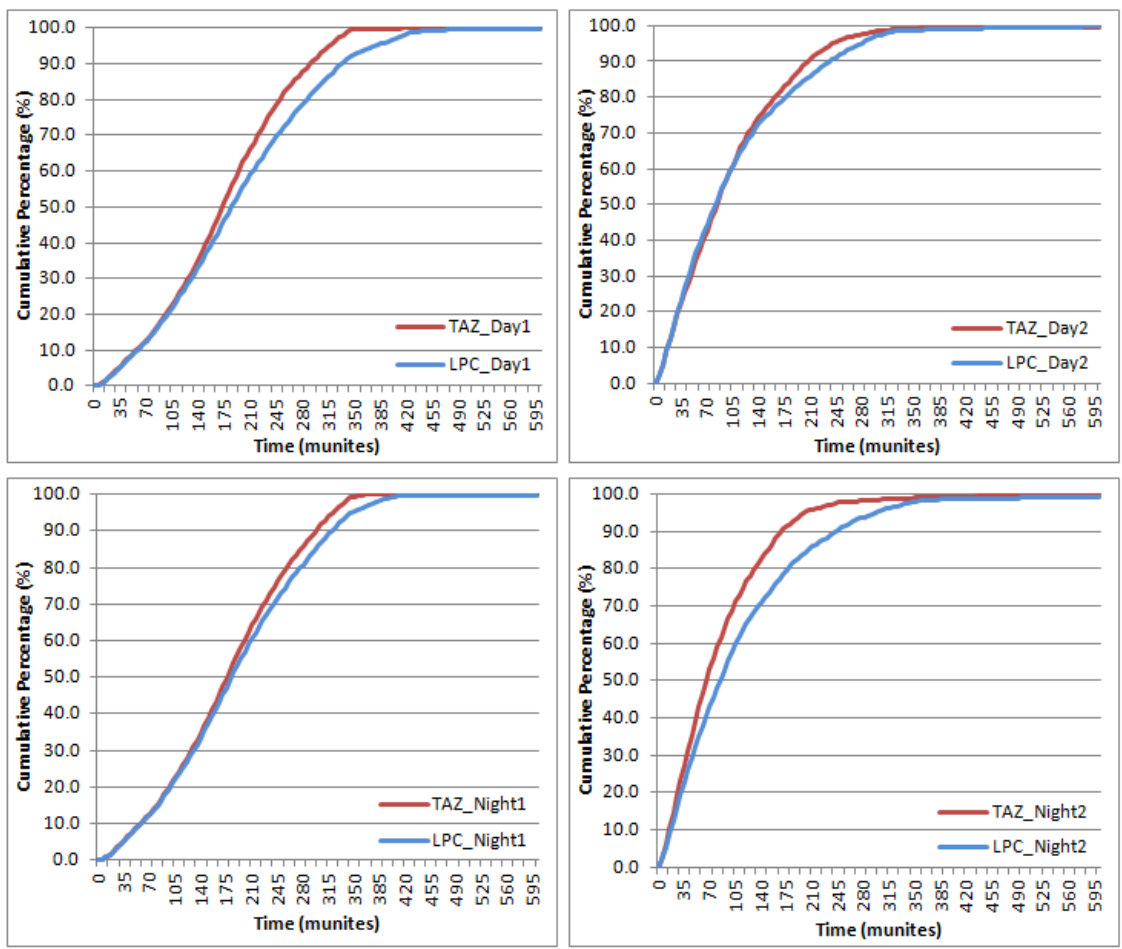

**Figure 3.6. Evacuation curves for TAZ and LPC comparison study**

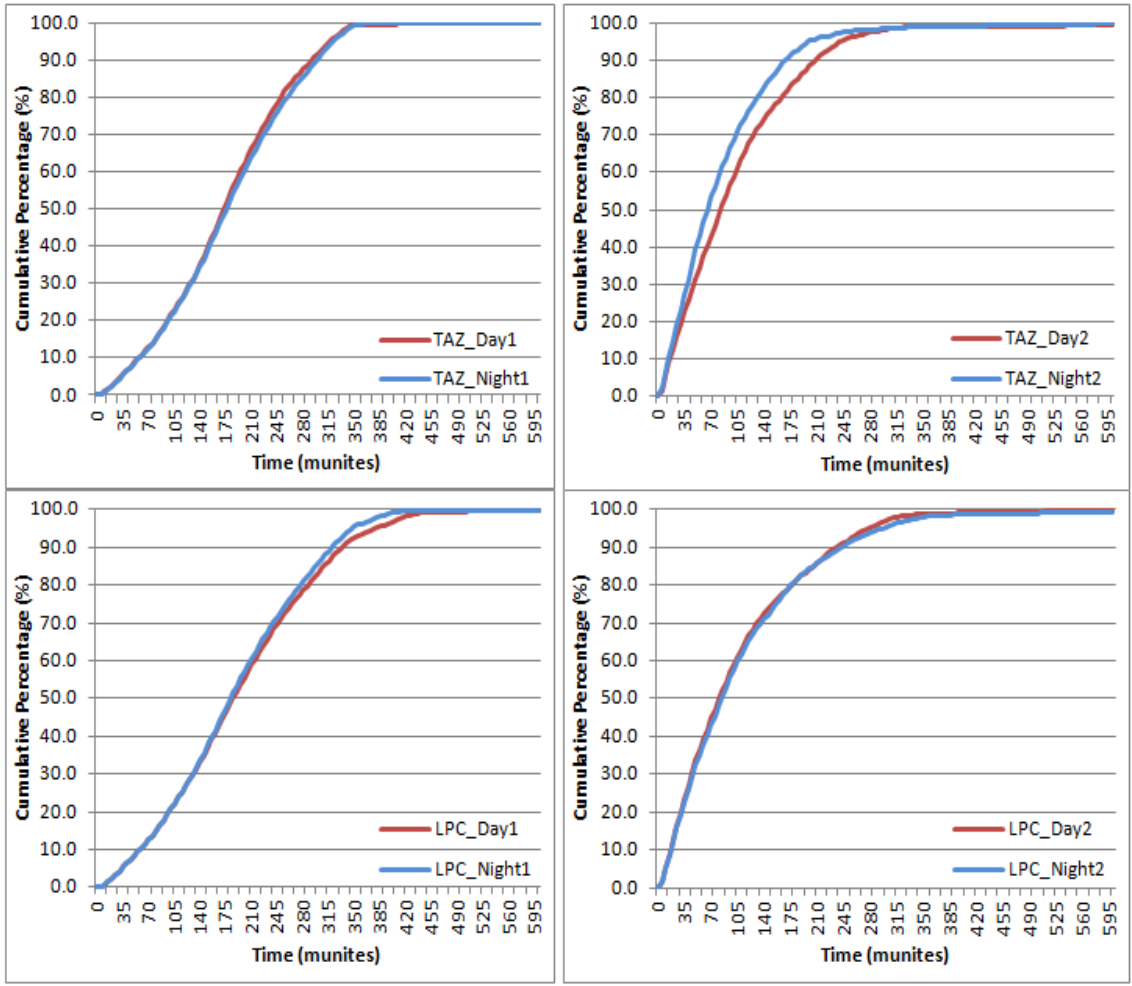

**Figure 3.7. Evacuation curves for daytime and nighttime comparison study**

The impacts of temporal high resolution data with daytime and nighttime formats under four scenarios, combining two population resolutions and two departure time choice models, are summarized in Figure 3.7. With the normal S-shape departure time model, there is no significant different between daytime and nighttime evacuation performance under TAZ and LPC because there is no significant large amount of vehicles assigned on the network. The evacuation efficiency in this case is more determined by the loading curve. But the nighttime evacuation performance with TAZ and location-based departure time choice model has relatively better efficiency because of the smaller amount of evacuees in the shadow area, which reliefs the potential congestions on the network with TRANSIMS original even activity location assignment within a TAZ zone. The efficiency does not happen on LPC based assignment because of the benefit of even trip distribution disappears in LPC based assignment. In a word, there is no significant difference between different temporal data from the system evacuation efficiency level. But there is difference in individual travel time and computational time with daytime and nighttime data. This will be discussed later.

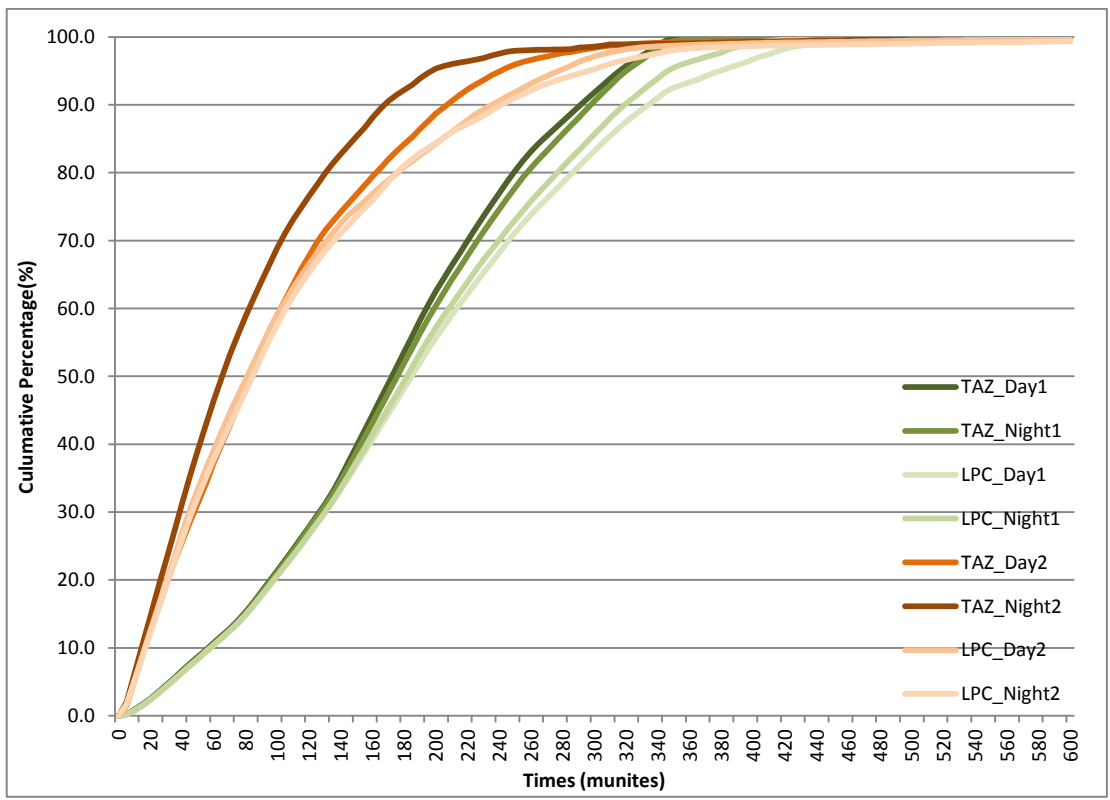

**Figure 3.8. Evacuation curves for departure time choice models comparison study**

Figure 3.8 illustrates the impact of departure time choice models under different spatial and temporal resolution data. The pattern is easy to observe. Therefore, one single plot of all the simulation results is displayed instead of four sub-plots. The location-based departure time model has better system efficiency at the first several hours evacuation. The fast loading curve helps people near the exits to access the road network fast and arrive at shelters in a relative short time. But this also causes some congestion on the network, especially near the shadow boundary zones/cells. On contrast, normal S-shape delays the vehicle to access the road network at the beginning but it also control the number of accessing vehicles on the network, which can control the congestion situations. The evacuation efficiency decreases after 6 to 7 hours with the location based departure time choice model. At this time, the system is trying to dissolve congestions on the network.

**Table 3.2. Operational and computational performance for different evacuation scenarios**

| Population Resolutions | Evacuation Time | Loading Curve Models | Average Travel Time (minutes) | Computation Time (minutes) |
|---|---|---|---|---|
| LPC | Daytime | Normal S-shape | 25.37 | 2.75 |
| | | Location-based model | 42.23 | 4.6 |
| | Nighttime | Normal S-shape | 19.57 | 2.33 |
| | | Location-based model | 43.15 | 4.72 |

Because of this, the average travel time for arriving evacuees with location-based model is higher than normal S-shape model, which is discussed in Table 3.2. Here, we only consider scenarios with LPC-based assignment to reflect the real world situation. As mentioned in the previous daytime and nighttime comparison study, the average travel time for nighttime scenario is better than daytime scenario under normal S-shape model due to the relative small amount of evacuees at night. But the average travel times under location-based model are almost the same because the congestions reduce the benefit of relative less people. Microscopic traffic simulation is time-consuming compared to macroscopic simulation. The computation time for the last run of microscopic simulation module is also provided in table 3.2. It is almost linear to average travel time. The computation time is determined by the number of vehicles calculated in one simulation step. Congestion significantly increases the simulation time.

## 3.5 Conclusions

In this paper, an agent-based traffic assignment framework with LandScan USA high resolution demographic data is presented for evacuation planning and simulation. To test the effectiveness and efficiency of this framework, comparison studies using population data in different spatial and temporal resolutions and driver departure time choice models are conducted. Through simulation studies of those eight scenarios in Alexandria, VA, three major findings are concluded here.

- TAZ-based traffic assignment underestimates evacuation travel time and simulation computational time due to its even distribution assignment in a TAZ zone. The proposed LPC-based framework provides a more accurate environment to represent how travelers access the road network.

- It takes more time for evacuees in Alexandria to travel to shelters in daytime under normal traffic conditions than the case in nighttime because of the different temporal demographic patterns. But the travel times with location-based model in both daytime and nighttime are almost the same due to congestions.

- Location-based departure time choice model improves the evacuation efficiency at the beginning stage of evacuation. But it also causes congestion, which delays the average travel time.

These findings should inspire evacuation managers and traffic operation engineers to reconsider the definition of traffic analysis zones concept. The conventional TAZs might be not effective in real-world traffic operation, especially in evacuation operation cases, where every individual is treated as an agent and should be studied independently. Also, evacuation plans for both daytime and nighttime are suggested due to the different demographic pattern. Although the evacuees' driving behavior is hard to predict, location-based evacuation plans can improve the system efficiency and travelers' travel time if the congestions can be well controlled.

This agent-based traffic assignment framework can be easily expanded to simulate any area in the world since LandScan Global has the whole world demographic data and OpenStreetMap provides the worldwide road network for free. The potential applications of this framework are not limited to evacuation studies. With daily demand-modeling framework(Balmer, Axhausen et al. 2006), the normal daily traffic operation and prediction can be conducted to provide various application purposes for researchers and practitioners. Though implementation of intelligent transportation systems with modern communication technologies, such as connected vehicle technology (Lu, Han et al. 2013), to provide real-time travel information, a real agent-based dynamic traffic assignment can be developed for various application purposes.

# CHAPTER 4
# STUDY OF EMERGENCY EVACUATION PERFORMANCE AS A FUNCTION OF MULTI-SCALE NETWORK CONFIGURATIONS AND ROUTING SCHEMES

## Abstract

The emergency situations, such as natural disasters, terror attacks, and major traffic accidents, have drawn great attention on evacuation planning and modeling. Most of these studies are based on the highway network and Traffic Analysis Zones (TAZ) concepts. Very few studies have documented the impacts of the data resolution of road networks and TAZ on evacuation performance. In this paper, a revised framework for emergency evacuation planning built on LandScan USA population cells (LPC) data and activity-based microscopic traffic simulation is proposed. Multi-scale networks are modeled, including two levels of real-world road networks and two scales of trip generation zones. Also, three routing schemes are implemented, including shortest network distance route, highway biased route, and shortest straight-line distance route. The purpose of this paper is to compare the evacuation travel times with respect to data in different resolutions and routing algorithms. The results indicate that conventional emergency evacuation plans with TAZ and major road network underestimated the travel time during evacuation. The evacuation time of 80% of trips in the evacuation area with the most complicated LPC_Full network are 30%, 36%, and 37% more than the conventional TAZ_Major network in those three trip distribution methods. The performance of highway biased trip routing algorithm beat the other two methods. More importantly, this research provides an easy implemented framework for emergency planners to take advantage of high resolution traffic and population data in today's big data era.

*Key words:* emergency evacuation, high resolution data, activity-based traffic assignment, microscopic simulation, special event operations, disasters, traffic management

## 4.1 Introduction

Emergency situations of recent times such as the massive wildfire in Colorado and terrorist attack in the Boston marathon have drawn great attention on effective evacuation planning and operations. The emergency evacuation planning, or no-notice evacuation planning, is different from the conventional transportation planning models. Emergency evacuation emphasizes on estimating evacuation time and identifying potential traffic flow bottlenecks with detailed road network traffic flow information to save more life and property. For this reason, plenty of studies were performed in the past years to evaluate the performance of different evacuation models, and to explore potential problems that can be improved.

Due to the limitation of computing resources and high resolution data availability, Traffic Analysis Zones (TAZ) based simulation studies dominated the evacuation simulation area.The conventional TAZ based models assigned all the trips in the TAZ to the centroid point and then connected it to the nearest node (intersection) on the network. Even activity based traffic simulation, used in Transportation Analysis and Simulation System (TRANSIMS), cannot represent the real-world situation with TAZ models. TRANSIMS assigned all the trips in one TAZ area evenly to all the activity locations in that area. But the trip distributions depend on the population density in a certain small neighborhood. Also, most simulation studies only take the major road networks (without considering the local roads in the neighborhoods) as their inputs. This limits the travel time and driving behaviors on local roads, which might produce bottleneck during emergency evacuation scenarios.

In this paper, a revised framework for emergency evacuation planning built on LandScan USA population cells (LPC) data and activity-based microscopic traffic simulation is proposed. The framework takes advantage of detailed population distribution data and road network in full level. The purpose of this paper is to compare the evacuation performance with respect to data in different resolutions and routing algorithms. The simulation results of Alexandria city in Virginia indicate that conventional emergency evacuation plans with TAZ and major road network underestimated the travel time during evacuation. Some subsequent discussions on improving evacuation simulation accuracy and efficiency through high resolution data are also presented.

## 4.2 Literature Reviews

A large amount of transportation literature exists on evacuation planning and operations, especially in notice evacuation scenarios, like hurricanes. Different from the prepared evacuation planning, emergency evacuation, such as no-notice evacuation, has its own characteristics. This section provides some relevant recent research on system-level and microscopic approaches on emergency evacuation simulation studies.

Alsnih and Stopher (2004) provided a detailed review of various emergency evacuation models and pointed out the essential procedures to devise emergency evacuation plans. Evacuees' behavior analysis, transportation engineering analysis (including transportation component analysis, microscopic traffic simulation, and dynamic traffic assignment), and role of government in emergency situations are the three main procedures for evacuation planning. Sisiopiku, Jones et al. (2004) also provide similar guidance from the regional system level. Ozbay et al., Han et al., and Liu et al. analyzed the real-world case studies of New Jersey, Washington D.C. and Tennessee area to evaluate the evacuation time and the system efficiency issues (Han, Yuan et al. 2006, Liu, Chang et al. 2008, Ozbay, Yazici et al. 2012).

Routing problem and network optimization are mainly concerned in emergency evacuation studies. Ng, Park et al. (2010) optimized the shelter locations through a hybrid bi-level model to improve the evacuation performance. Yue-ming and Hui (2010) modeled the shortest emergency evacuation time through Pontryagin minimum principle and dynamic traffic assignment. Xie and Turnquist (2011) integrated Lagrangian relaxation and Tabu search algorithm to optimize the lane-based evacuation system performance. Lu, George et al. (2005) simulated their evacuation plans with capacity constrained routing algorithms. Liu, Lai et al. (2006) studied the staged emergency evacuation planning with cell-based network optimization models. Kai-Fu and Wen-Long (2008) illustrated the emergency evacuation network statistics through genetic algorithm and kinematic wave models. Most of these researches focus on system-level analysis during emergency evacuation scenarios.

With the improvement of computing resources in transportation field, many researchers explore the emergency evacuation modeling through microscopic traffic simulations. Jha, Moore et al. (2004) took advantage of microscopic simulation model (MITSIM) to model the evacuation of Los Alamos National Laboratory. Cova and Johnson (2002) presented a method to develop neighborhood evacuation planning with microscopic traffic simulation in the urban - wildland interface. Household-level evacuation planning is implemented in various scenarios. Henson and Goulias (2006) reviewed 46 activity-based models and their competency of homeland security applications. Transportation Analysis and Simulation System (TRANSIMS) tool was discussed and demonstrated to those applications with publicly available data sources.

Beyond the traditional research on traffic assignment and traffic simulation analysis, intelligent transportation systems with advanced technologies were also introduced in emergency evacuation studies. Michael Robinson (2012) examined the impacts of Advanced Traveler Information Systems (ATIS) used to guide evacuees through integrating a mesoscopic traffic simulation and evacuees' route choice decision model for southeastern Virginia. This study indicated that the effectiveness of ATIS was increased through route changes before the evacuation and evacuees diverting to alternate routes may worsen road network congested situation. Han, Yuan et al. (2006) compared the static traffic assignment and dynamic traffic assignment for emergency evacuation scenarios and emphasized the advantage of using intelligent transportation systems to route evacuees. Lu, Han et al. (2013) analyzed the connected vehicle performance in an crowded urban area, which could be used to build the intelligent transportation systems for emergency evacuation models. Hamza-Lup, Hua et al. (2005) developed a smart traffic evacuation management system by leveraging real-time information obtained from sensors or other surveillance technologies.

Despite the number of evacuation studies in both academic and practical areas, these studies are still limited to the traditional data format and computational constraints. There is an interesting phenomenon that most large-scale evacuation simulation studies were conducted based on highway road network macroscopic or mesoscopic traffic simulation packages. Microscopic simulation based evacuation studies were more focusing on small regions because of the limitation of commercial traffic simulation software packages. Besides these, all the existing evacuation simulations are using traditional Traffic Analysis Zones (TAZ) to generate the trips. TAZ is good for planning purpose. But they are usually defined by city or county transportation planning agencies. Many of those data are out-of-date and cannot represent the updated road networks.

The road networks data and population distribution data are becoming more detailed and accurate. LandScan population data, developed by Oak Ridge National Laboratory in United States, has become the community standard for global population distribution (LandScan 2005). It has approximately 1 km resolution with global population distribution data available and represents an ambient population. A lot of population risk study based on LandScan dataset were conducted to predict better analysis and results (Dobson, Bright et al. 2000, Bhaduri, Bright et al. 2002). LandScan USA population cell (LPC) data is a higher resolution population data based on LandScan to provide more granular 90m x 90m (3'×3') resolution with national population distribution data (Bhaduri, Bright et al. 2007). It is much more accurate than conventional TAZ because some TAZ zones are unusual big in scale or dense in population. Compared to TAZs, LandScan challenged the traditional trip generation method. The trips are generated from cell to cell in very small scale, instead of relatively large scale zone to zone method. Besides that, LPC provides different population distribution between day-time and night-time. Trip assignment based on different time pattern could increase the accuracy of emergency evacuation planning.

## 4.3 Simulation Model

To assess evacuation efficiency with different population distribution data and road networks, a real-world evacuation operation case study with 12 different scenarios was conducted using the same microscopic simulation software package.

### *4.3.1 Alexandria City Network and Population Data*

A city level emergency evacuation operation is simulated with different data inputs. Alexandria city in Virginia is selected as a case study because of the well coded network data that is freely available with TRANSIMS. It has been widely used in many research areas. This helps us to focus more on the simulation and analysis using different data resolution, rather than spending time on geocoding the road networks. The road network data is show in Figure 4.1. The background map is from ESRI World Image data. Two levels of real-world road networks are built in the simulation: the major road network (the red links) and the full road network including the local roads (the blue links). The major network consists of 1278 links and 1061 nodes, while the full network is composed with 3634 links and 2608 nodes.

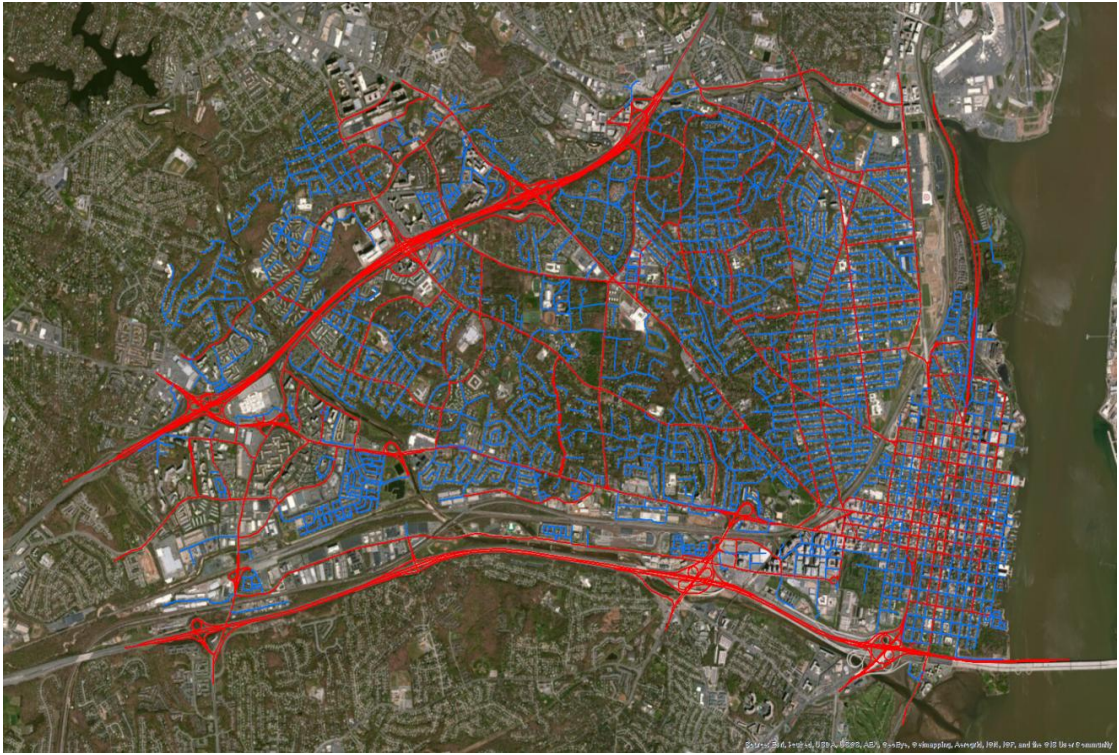

**Figure 4.1.** Alexandria City Road Network

To compare the two resolution population datasets, TAZ and LPC, LandScan USA 2011 population data is utilized. The TAZ population data is aggregated from LPC data to make the comparison consistent. The total population in Alexandria is 161,519 based on 2011 LandScan data. The TAZ distribution map of Alexandria is shown in Figure 4.2. The background map is from OpenStreetMap. Usually, TAZ uses real-world road segment to cut the boundary. Alexandria contains 62 TAZ zones.

In contrast, LPC consists of 6131 cells (5657 non-zero cells), as shown in Figure 4.3. The color indicates the density of population in that cell. Red means higher population and grey means no population. Each cell has a value for population number in that cell. Technically, TAZ and LPC can be considered as the same zone definition, but LPC size is much smaller. Since the vehicle per capita ratio is about 0.8 in the United States, the total trips for evacuation in Alexandria city is 129,215 (that is, 161,519 * 0.8). Only integer is used in defining number of trips for both TAZ and LPC. For those LPC cells with only 1 person, 1 trip is generated to ensure every person in the cell has the access to evacuate to the safe shelters. Adjustment is amended to make sure TAZ and LPC have the same number of trips. The objective of the evacuation scenarios is to evacuate all the people out of Alexandria city.

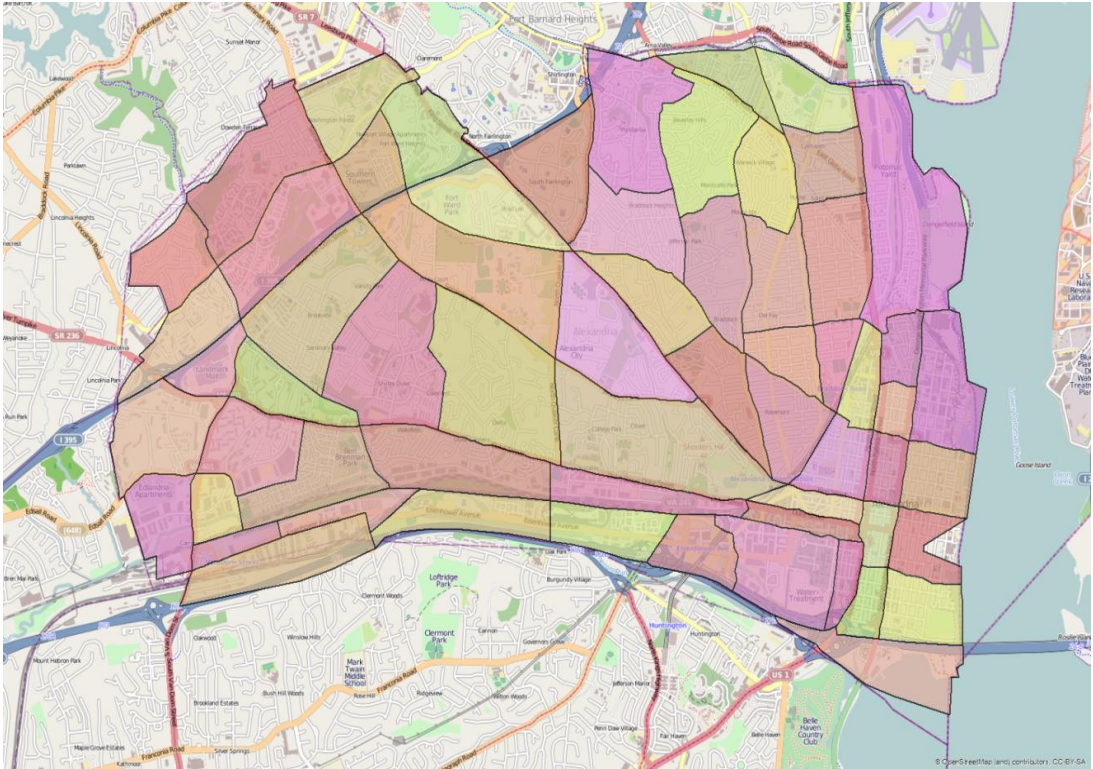

**Figure 4.2. Traffic Analysis Zones in Alexandria.**

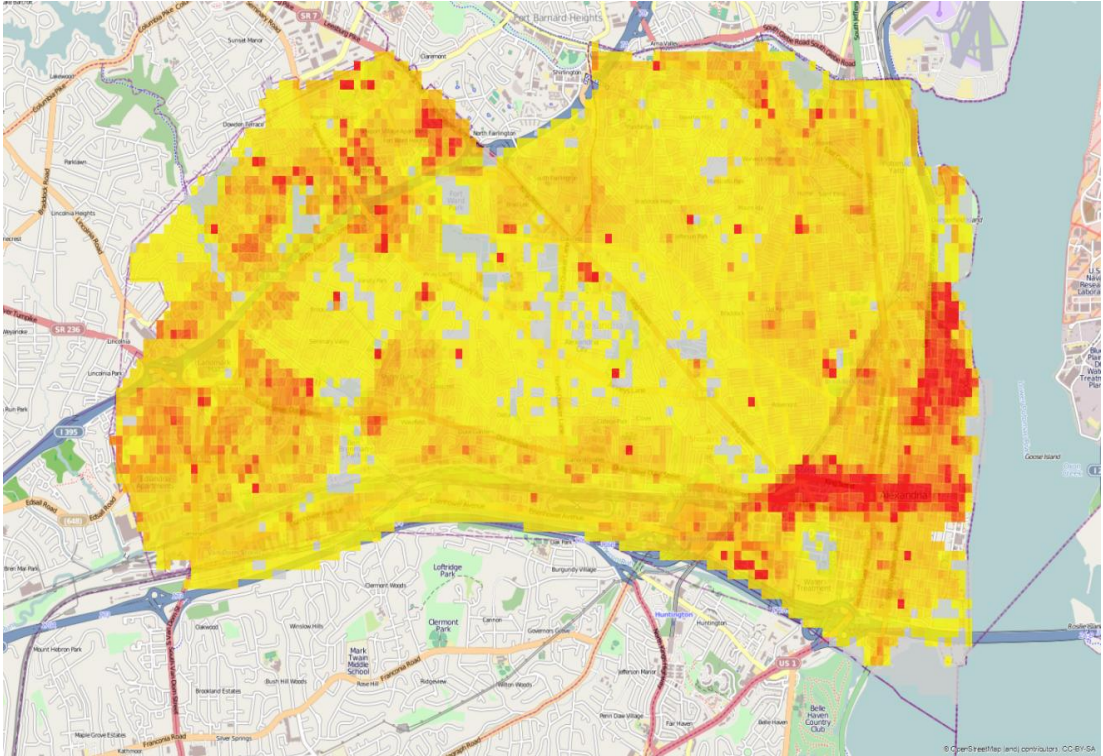

**Figure 4.3. LandScan USA 2011 population data in Alexandria**

*4.3.2 Simulation Tool*

TRANSIMS is well known as an open-source traffic modeling tool for transportation planning and operation analysis(Nagel, Beckman et al. 1999). Because the novel traffic assignment model based on LandScan population data breaks the traditional four-step traffic assignment method, all existing commercial traffic simulation software is not capable to deal with the large dataset such as LandScan. TRANSIMS provides a good platform for traffic model developers to modify the source code to fit their specific needs.

TRANSIMS was developed for traditional TAZ based traffic modeling and simulation; therefore, there are two major limits to overcome for applying it to LPC based models. The first aspect is the number of TAZs allowed in trip assignment. TRANSIMS only consider 1000 zones as maximum for TAZ. There are 6131 LPC cells, technically as the same as 6131 TAZ zones. The second limit is still related to the TAZ concept. TRANSIMS implemented activity-based traffic assignment algorithm to achieve more realistic simulation. It assumes that there are at least two activity locations in a TAZ for trip generation allowing inner-zone trips and at least one activity location for only inter-zone trips. This is understandable in traditional TAZ concept since the TAZ is usual large enough to include several road segments. But LPC size is much smaller than TAZ and there are many cells having no roads crossing through their areas. In this case, the trips in that zone cannot be assigned to the activity locations. To fix this problem, a self-developed program was written to assign each LPC to their nearest activity location. This meets the real-world situation where travelers prefer to access the roads through the nearest spots.

### *4.3.3 Site Modeling*

The Alexandria city was configured in TRANSIMS with two levels road networks and two resolution population data, which forms four combination scenarios: TAZ-Major, TAZ-Full, LPC-Major, and LPC-Full. The comparison of major and full level road networks is presented in Table 4.1. There are 62 TAZ zones and 5657 non-zero LPC cells with the same number of trips, 129,215 trips, are modeled. In addition, 21 safe shelters are assigned and connected one-to-one with 21 exits in the road network respectively. Evacuation trips generally move outward as evacuees leave an evacuation area and seek safety.

**Table 4.1. Major road network vs. Full road network**

| Road network type | Major | Full |
|---|---|---|
| Number of nodes | 1061 | 2608 |
| Number of links | 1278 | 3634 |
| Number of activity locations | 1644 | 7718 |
| Number of un-signalized nodes | 110 | 1062 |
| Number of signalized nodes | 182 | 261 |

The number of activity locations is determined by TRANSIMS. Two parameters, maximum access points and minimum split lengths, are set to different combination to represent the real-world scenarios. For comparison, five different sets with the real Alexandria image map, three access points and 100 meters split interval are modeled in our simulation. The traffic control is set as default by TRANSIMS.

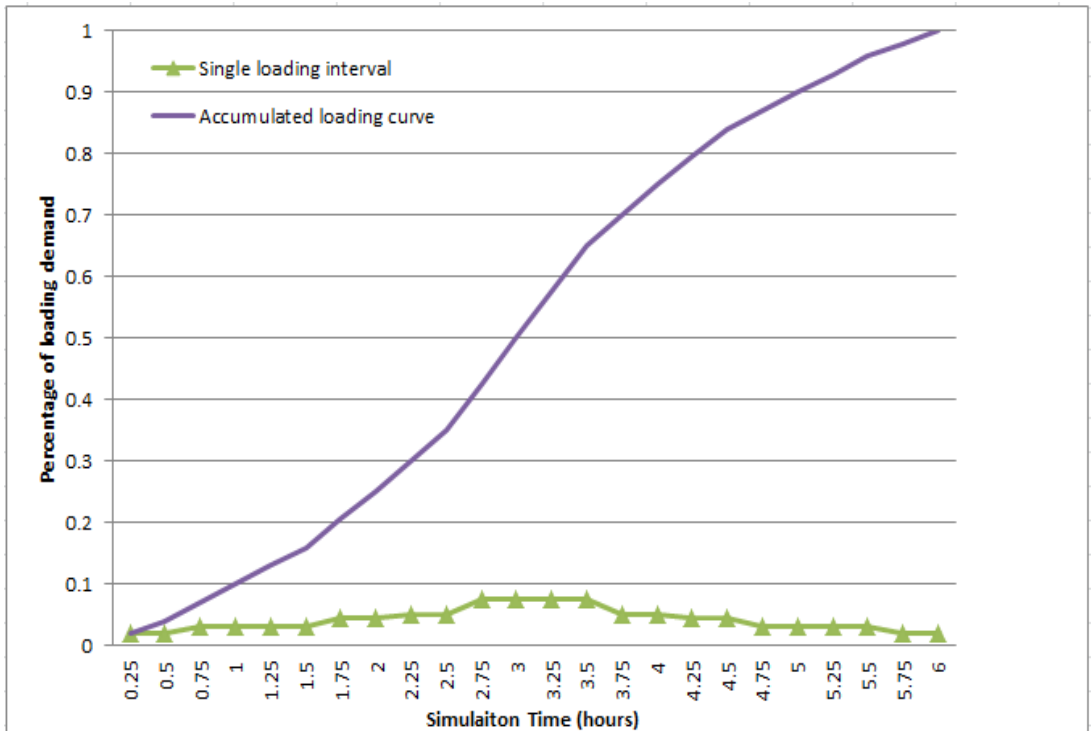

**Figure 4.4. Loading curve for evacuation demand**

The evacuation demand for each zone or cell was converted to 15-minutes interval volumes using loading curve. Florida Department of Transportation researched various response curves for evacuation conditions. The S-shape loading curve is used in this research. The whole Alexandria city area is about 15.2 sq. mi. To maximize the evacuation performance with this evacuation area size, the evacuation demand was loaded over 6 hours and evacuation simulation time is 6 hours. Figure 4 shows the trip loading curve implemented in this study. The purple line is the accumulated loading curve during the 6 hours. The green line shows the percentage of loading demand in each 15-minutes interval. This is how TRANSIMS implements the loading curve.

### *4.3.4 Simulation Scenarios*

For a city level evacuation planning in this study, it is very challenging for local officials to generate an optimal evacuation plan that includes destination and route assignment for every evacuee. Route based on shortest path algorithm is well appreciated from both planners' and evacuees' perspective. Conventionally, the shortest network distance assignment and highway-biased (shortest time) assignments are two typical routing strategies for evacuation purpose, as shown in Figures 4.5a and 4.5b. Meanwhile, from the naïve users' perspective, the shortest straight-line distance assignment is easiest routing finding algorithm, as shown in Figure 4.5c. This is also implemented in this study for comparison with the other two assignment algorithms. Figure 4.5 also illustrates the assignment patterns for both TAZ and LPC scenarios. The bold green links in Figures 4.5 connected the TAZs central points to their shelters with three kinds of trip routing algorithm. The blue points are the TAZs central points. The red points are the safe shelters as destinations. The LPC cells is not drawn here due to the large amount, but the concept is similar to TAZ. The orange links in these figures bridged the LPC central points to their shelters with the same three routing algorithms. These three routing algorithms are defined as three categories: shortest network distance assignment (SND), highway biased assignment (HWB), and shortest straight-line distance assignment (SLD). The pattern for each category is well compared in Figures 4.5. The shortest network distance assignment searches the nearest exits surrounding each TAZ or LPC area. They might use more local roads. The highway biased assignment will take more advantage of the freeways and major roads as they have higher capacity and speed limit. The shortest straight-line distance assignment is the easiest to understand for naïve users based on the calculation of geometrical distance between origins and destinations. These three figures only represent the full network scenarios. The major road network scenarios have similar pattern and the details will be implemented in the simulation studies.

In total, we implemented 12 simulated scenarios in this paper, combing the four resolution levels of road network and population data with three categories of routing algorithms. They are TAZ-Major-SND, TAZ-Major-HWB, TAZ-Major-SLD, TAZ-Full-SND, TAZ-Full-HWB, TAZ-Full-SLD, LPC-Major-SND, LPC-Major-HWB, LPC-Major-SLD, LPC-Full-SND, LPC-Full-HWB, and LPC-Full-SLD. TRANSIMS 4.0.8 version was adapted on a Window 7 64-bit laptop computer. The configuration of this laptop is 16GB RAM, 2.6GHz Intel(R) Core(TM) i5-3320 CPU, and 500GB hard disk. An 8-hour simulation time was used and the other parameters are set as stated in the preceding sections. After multiple testing of the TRANSIMS simulation, the traffic assignment achieve user equilibrium status after 10 runs of Router and 15 runs of Microsimulator for this case study, where Router and Microsimulator are two major procedures for trip assignment and microscopic simulation. To adjust the effect of random number in the Microsimulation procedure, all the simulation results are based on the average of 30 independent runs.

## 4.4 Results and Discussions

To evaluate the effectiveness of the evacuation operations, various measures of effectiveness (MOEs) are developed. Han, Yuan et al. (2007) suggested a four-tier MOE framework for evacuation, including evacuation time, individual travel time and exposure time, time-based risk and evacuation exposure, and time-space-based risk and evacuation exposure. They can be used in different scenarios in evacuation modeling stage to satisfy different evacuation planning purpose. In this evacuation study, the evacuation efficiency is the most significant MOE to be considered. Comparison studies of four resolution data and three traffic assignment categories were analyzed.

For the SND traffic assignment scenario, in which evacuees were assigned to the nearest of the 21 exits based on shortest network distance, the four resolution datasets are

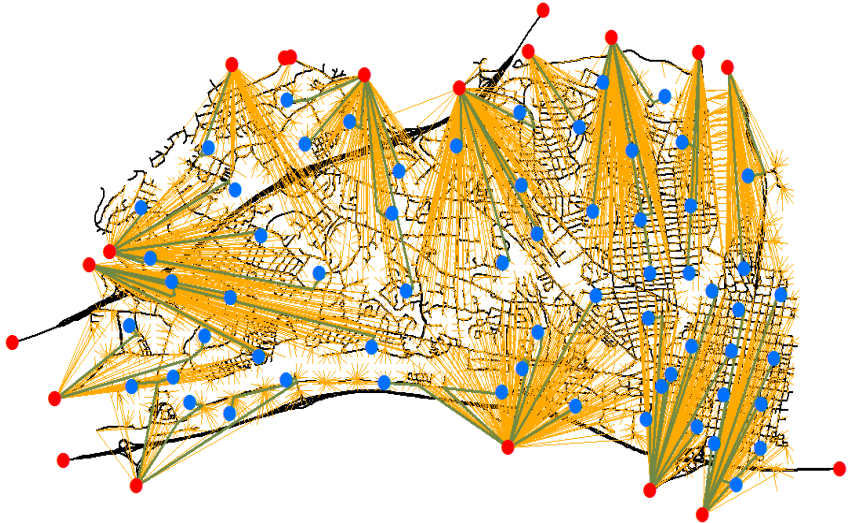

Figure 4.5 (a) TAZ and LPC trip assignment with shortest network distance routing algorithm

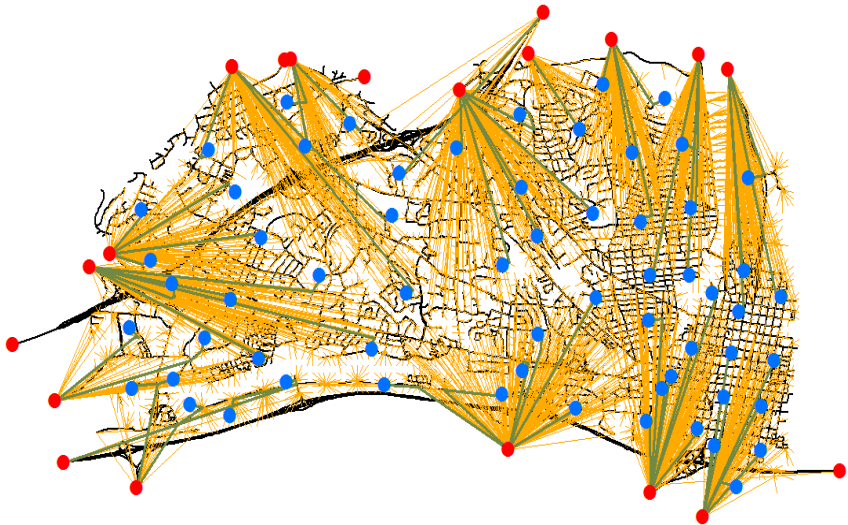

Figure 4.5 (b) TAZ and LPC trip assignment with highway biased routing algorithm.

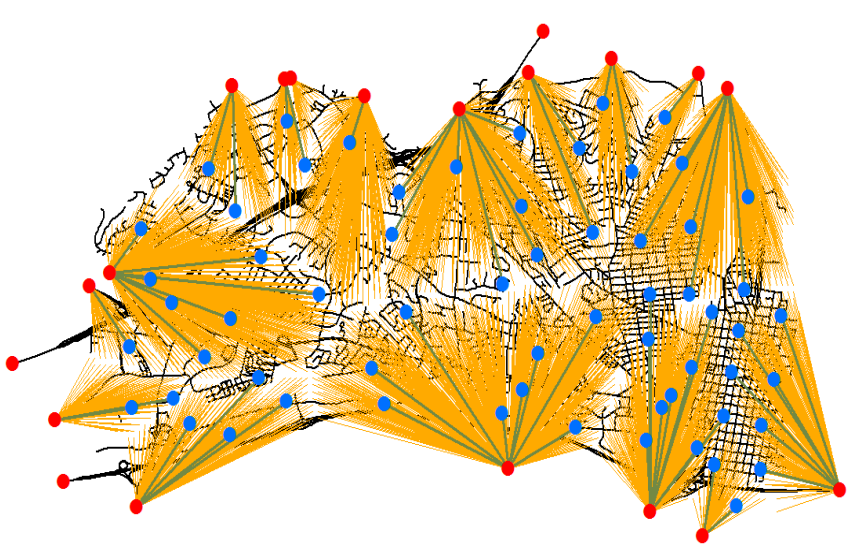

Figure 4.5 (c) TAZ and LPC trip assignment with shortest straight-line distance routing algorithm.

compared in Figure 4.6. The resolution level increases from TAZ_Major to LPC_Full. Under the same network structure, the performance of TAZ assignment is always better than LPC method because TRANSIMS evenly assign the trips to all the activity locations in each zone. But LPC method does not use the default TRANSIMS method. Instead, it assigns each LPC cell to the nearest activity location, which is more realistic, comparing to the real-world scenario. Some LPC cells might have much more trips than others, like downtown area, which formulates traffic congestion when people are trying to evacuate at the same time. For the same population resolution scales, the evacuation performance of major road network is better in both TAZ and LPC situations. This is because the major network did not count the travel time on those local roads. After 6 hours, between 81.6% and 96.5% of the population were evacuated to safe shelters in this routing method.

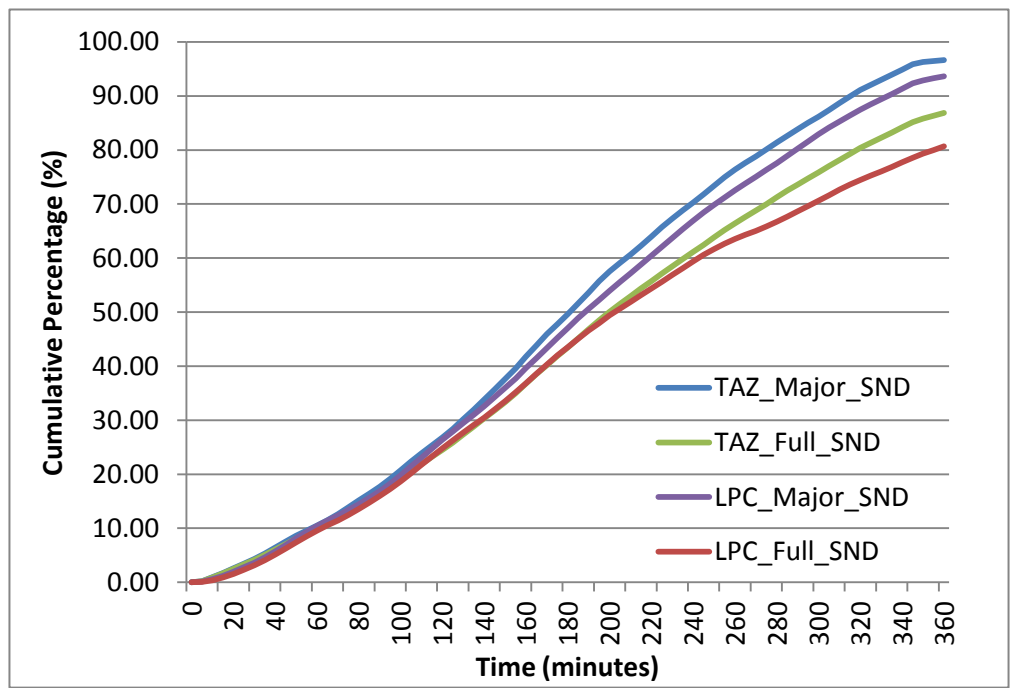

**Figure 4.6. Evacuation curve under shortest network distance routing.**

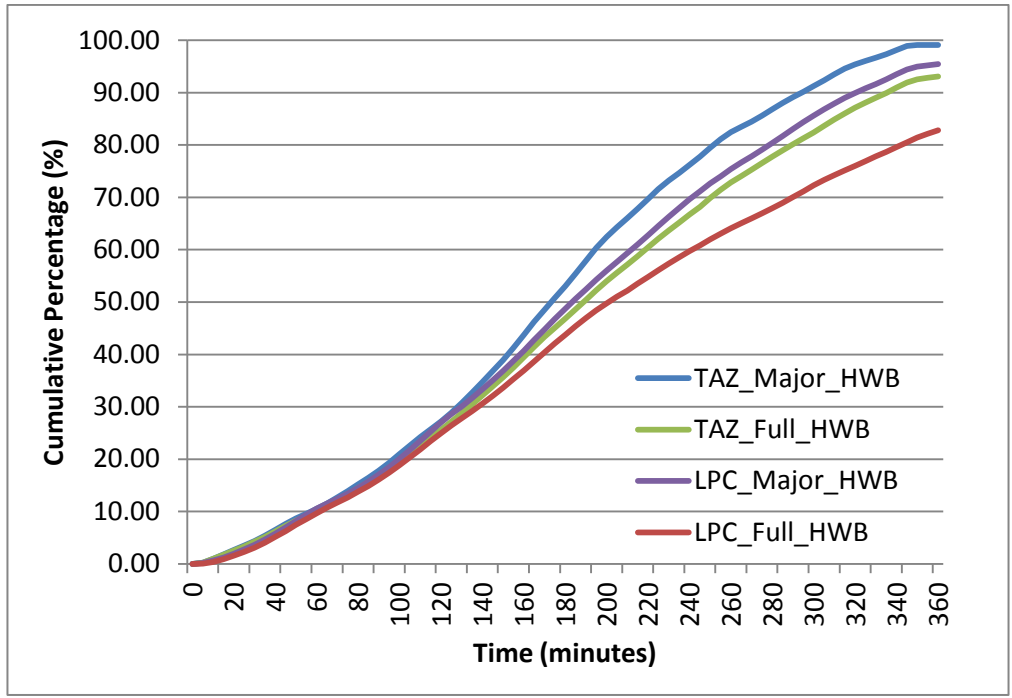

**Figure 4.7. Evacuation curve under highway-biased routing.**

Similar to SND routing, the percentage of evacuated people of TAZ scenario is better than the LPC case under the same network structure for highway-biased (HWB) routing method, as shown in Figure 4.7. However, the evacuation performance of four different datasets in HWB routing is relatively better than the SND cases. The highway network provides faster moving ability but access to highway also causes some traffic delay when exceeding the capacity. LPC with full network is still worse than the major network when we considered the detailed population distribution. At the end of the simulation, between 82.3% and 99.3% of the population were evacuated in HWB routing algorithm.

The evacuation results for shortest straight-line distance (SLD) routing the same patterns but with the worst performance, as shown in figure 4.8. This is because the straight-line distance method does not use the road network characteristics. The naïve user method is easy to the general audience but has less efficiency on evacuation performance. Between 71.7% and 87.8% of the population were evacuated to their destinations in SLD assignment method.

Through the comparison among Figures 4.6 to 4.8, the highest resolution data with LPC and full network has the worst evacuation performance from simulation results, but it is the closest scenario to the real-world case when we consider evacuating each individual person under realistic conditions. Most existing studies ignored the evacuation time difference in local road networks and population distribution in a building-to-building level.

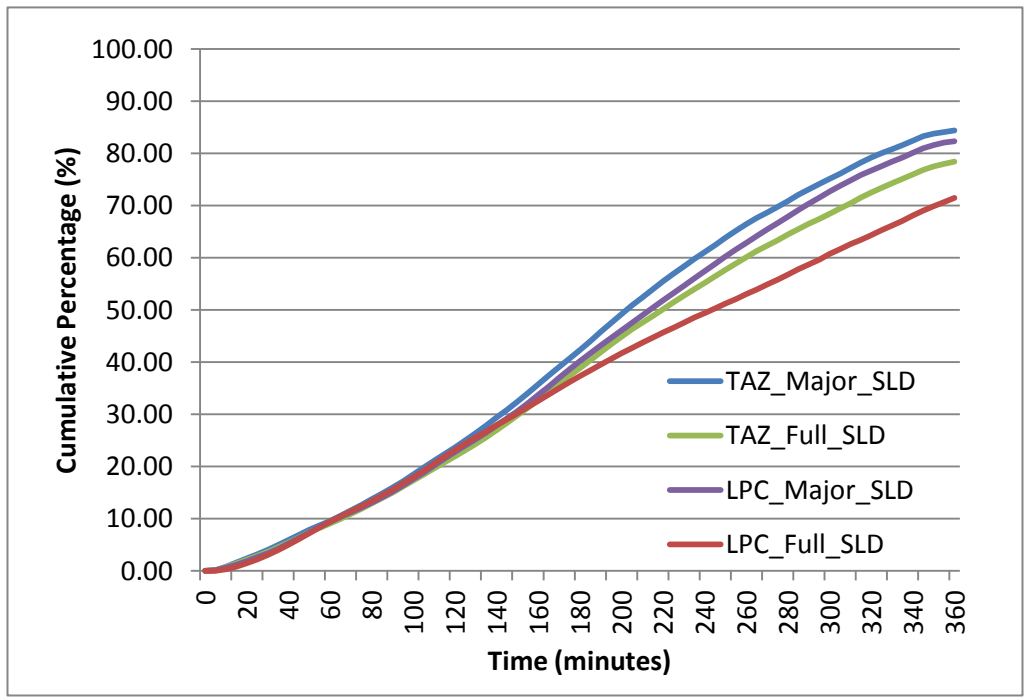

**Figure 4.8. Evacuation curve under shortest straight-line distance routing**

To compare the three assignment methods, the time needed to evacuate certain number of population are compared. Figure 4.9 illustrates the time to evacuate 80% of the total population in Alexandria under those 12 different scenarios. HWB method always performs relatively better than the other two. The SLD method produced the worst results. Also, the trends for these three routing schemes clearly represent the impact of data resolution in different network configurations. The evacuation time of 80% of trips in the evacuation area with the most complicated LPC_Full network are 30%, 36%, and 37% more than the conventional TAZ_Major network in those three routing schemes. This provides guidance for evacuation planners to propose a reasonable evacuation strategy.

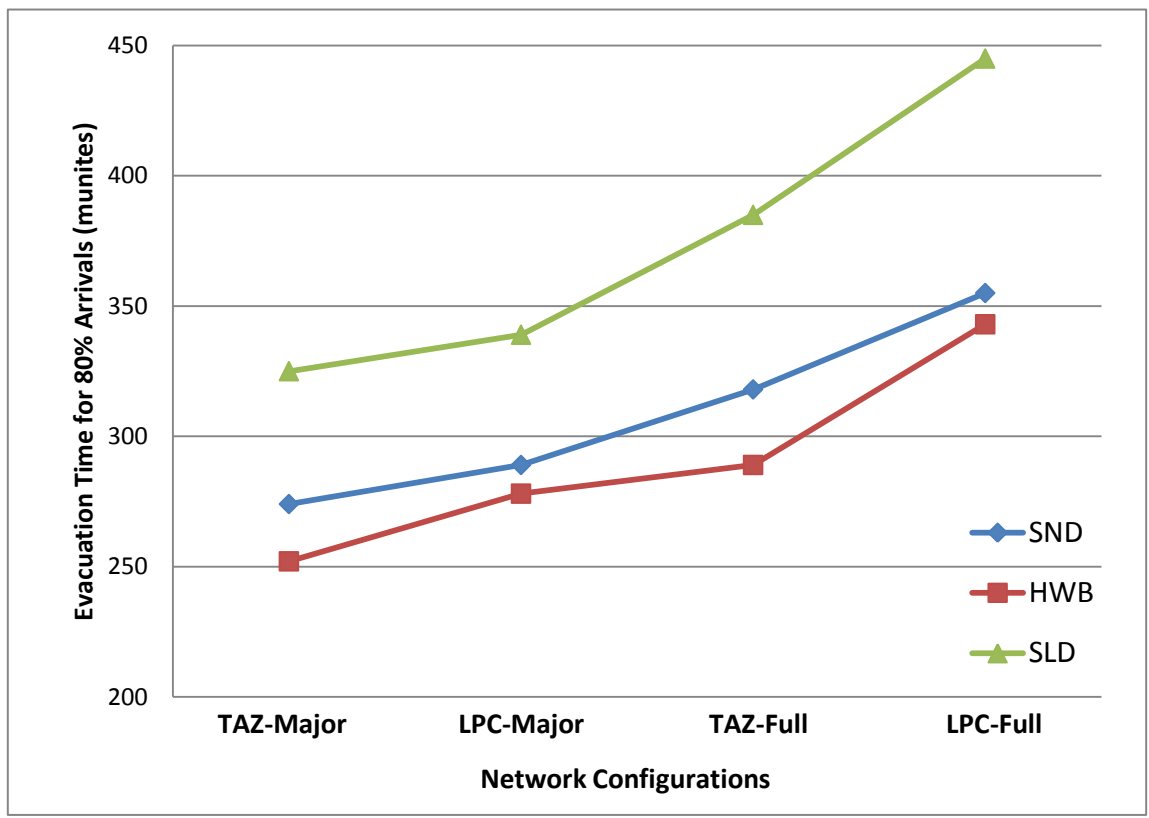

**Figure 4.9. Time for evacuating 80% population in 12 scenarios.**

## 4.5 Conclusions

In this paper, two levels of real-world road networks (major and full) and two scales of trip generation (TAZ and LPC) are modeled and compared using three different route finding algorithms. Two major findings are concluded from the simulation results of 12 scenarios in data resolution and routing algorithm aspects. For the data resolution analysis, the highest resolution data with LPC and full network, which is also the more realistic one, need more travel time for emergency evacuation. The traditional activity-based TAZ trip assignment underestimated the vehicle travel time and congestion time on the network due to its aggregated data characteristics and evenly distributed trips on each activity location in a zone. For the routing algorithm analysis, the highway-biased trip assignment has relatively better performance on evacuation time than the other two methods. But the performance is not significantly better than HWB method, especially when the network is very congested. The shortest straight-line distance assignment, the naïve user preferred intuitive method, produced the worst results. As discussed in this paper, state and local agencies in charge of modeling detailed emergency evacuation plans should consider using more high resolution population distribution data and detailed road network when testing their evacuation plans. Also, the shelter locations for each effected area should be determined by the road conditions in the neighborhood. If the shelter is close to the neighborhood, a shortest network distance might be better. If the evacuees are in the middle of evacuation area, a highway-biased method might work better. In conclusion, this research provides an easily implementable framework for emergency planners to take advantage of high resolution traffic and population data in today's big data era.

The simulations in this paper were implemented with distance-based traffic assignment. A time-based traffic assignment, like dynamic traffic assignment, could improve the performance when we implement more intelligent transportation systems (ITS) devices in the city to provide real-time travel information. Also, we assume all the travelers will follow the origin-destination assignment, which is not usually the case in the real world (Yuan, Han et al. 2007, Fang 2013). A non-compliance traffic assignment study will be conducted with high resolution data to evaluate the evacuation efficiency.

# CHAPTER 5
# EVACUEE COMPLIANCE BEHAVIOR ANALYSIS USING HIGH RESOLUTION DEMOGRAPHIC INFORMATION

## Abstract

The purpose of this study is to examine whether evacuee compliance behavior with route assignments from different resolutions of demographic data would impact the evacuation performance. Most existing evacuation strategies assume that travelers will follow evacuation instructions, while in reality a certain percent of evacuees do not comply with prescribed instructions. In this paper, a comparison study of evacuation assignment based on Traffic Analysis Zones (TAZ) and high resolution LandScan USA Population Cells (LPC) were conducted for the detailed road network representing Alexandria, Virginia. A revised platform for evacuation modeling built on high resolution demographic data and activity-based microscopic traffic simulation is proposed. The results indicate that evacuee compliance behavior affects evacuation efficiency with traditional TAZ assignment, but it does not significantly compromise the efficiency with high resolution LPC assignment. The TAZ assignment also underestimates the real travel time during evacuation, especially for high compliance simulations. This suggests that conventional evacuation studies based on TAZ assignment might not be effective at providing efficient guidance to evacuees. From the high resolution data perspective, traveler compliance behavior is an important factor but it does not impact the system performance significantly. The highlight of evacuee compliance behavior analysis should be emphasized on individual evacuee level route/shelter assignments, rather than the whole system performance.

*Key words:* evacuation assignment, traveler compliance behavior, high resolution data, big data, LandScan, activity-based simulation, special event operations

## 5.1 Introduction

Studies from recent evacuation operations associated with wildfires, hurricanes, and other severe events reveal that one of the most important factors for a successful evacuation plan is how to use traffic simulation to reflect evacuees' choices on departure time, shelter locations, and driving route. One-to-one origin/destination trip assignment is widely accepted for evacuation planning purposes. But it assumes all the travelers follow the instructions strictly and do not change their destinations depending on the real-time road conditions. Evacuation operation surveys from several hurricane evacuation studies indicate that evacuees changed or are willing to change their route based on real-time traffic information. Evacuee compliance behavior is one of the key factors for evacuation planning.

In the constraint of high resolution data availability, Traffic Analysis Zones (TAZ) based trip assignment studies have dominated the evacuation simulation domain. It might work for advance-notice evacuation scenarios, such as hurricanes, which give evacuees several days ahead for planning. But it does not perform well in no-notice emergency evacuation scenarios, which causes traffic congestion in a short time. How to represent this congestion is critical for an efficient evacuation plan. The conventional TAZ based models assigned all the trips in the TAZ to the centroid point and then connected it to the nearest node (intersection) on the network. Even activity based traffic simulation, used in Transportation Analysis and Simulation System (TRANSIMS), cannot represent the real-world situation with TAZ models. TRANSIMS assigned all the trips in one TAZ area evenly to all the activity locations in that area. But the trips usually are not generated uniformly for a given area. They depend on the population density in a certain small neighborhood.

A revised framework for evacuation based on uniform high resolution demographic data and activity-based microscopic simulation is proposed, which gives the researchers and practitioners the ability to simulate any area in the world. The aim of this paper is to examine whether the evacuee compliance behavior with population data in different resolutions would impact on the evacuation performance. The simulation results of Alexandria, Virginia indicate that the evacuation performance is not significantly sensitive to evacuee compliance behavior in LPC based assignment compared to TAZ based assignment and TAZ based trip assignment under-estimates the real travel time. Some subsequent discussions on implementing traveler compliance behavior through high resolution data are also suggested.

## 5.2 Literature Reviews

Simulation-based evacuation studies give researchers and practitioners great tools to evaluate evacuation strategies. A detailed review of various emergency evacuation models by Alsnih and Stopher (2004) pointed out the three main procedures to devise emergency evacuation plans, including evacuees' behavior analysis, transportation engineering analysis, and the role of government. The interaction and cooperation among these three aspects are needed to provide better solutions of a mass evacuation on the current transport network. Various scales of evacuation areas were conducted to evaluate the evacuation efficiency and systems performance, from the whole state Tennessee to a corridor in Washington D.C. (Han, Yuan et al. 2006, Liu, Chang et al. 2008).

Microscopic traffic simulation is becoming more popular than conventional macroscopic traffic simulation in evacuation study due to the cheap but fast computing capacity and expectation of detailed performance. Jha, Moore et al. (2004) took advantage of microscopic simulation model (MITSIM) to model the evacuation of a small area -  Los Alamos National Laboratory. Cova and Johnson (2002) presented a method to develop neighborhood evacuation planning with microscopic traffic simulation in the urban – wildland interface. Household-level evacuation planning is implemented in various scenarios. Activity-based traffic simulation provides more realistic simulation at the traffic assignment stage, which is adopted by a population simulation package called Transportation Analysis and Simulation System (TRANSIMS) (Smith, Beckman et al. 1995). Henson and Goulias (2006) reviewed 46 activity-based models and used TRANSIMS to demonstrate their competency for homeland security applications.

Despite these existing research efforts in evacuation simulations, most of these studies assume a one-to-one trip assignment to evacuees' full compliance for predetermined destination and route assignments.

Traveler compliance behavior is a key factor in traffic simulation and evacuation assignments. Many researches focus on the compliance behavior resulting from Advanced Traveler Information Systems (ATIS) with real-time road conditions. Srinivasan and Mahmassani (2000) examined travelers' route choice decisions based on compliance and inertia mechanisms with simulation and empirical data comparison. The results indicate that information quality, network loading, level-of-service, and travelers' prior experience determine route choices. Al-Deek, Khattak et al. (1998) used traveler compliance behavior and traffic system performance to evaluate ATIS. Lu, Han et al. (2013) studied the impact of connected vehicle technology for travelers to exchange real-time traffic information in a car sharing system, which provided another approach to implement an ATIS framework. There are also some papers presenting the evacuee compliance behavior impact in evacuation performance. Pel et al. did a series of studies on evacuation modeling with traveler information and compliance behavior (Pel, Huibregtse et al. 2009, Pel, Hoogendoorn et al. 2010, Pel, Bliemer et al. 2011, Pel, Bliemer et al. 2012). Some results show that traveler compliance behavior affect evacuation efficiency and road capacities have no significant impact on evacuation. The reviewed dynamic traffic simulation models also indicate that some existing models still have some weakness with regard to the choice to evacuate, departure time choice, destination choice, and route choice. Yuan, Han et al. (2007) simulated a Nuclear Plant and Tennessee state with non-compliance route choices in evacuation assignments. The results suggest that there is no significant difference between full compliance and full non-compliance. Revised traffic analysis zones might help to reveal more realistic evacuation performance. All of these studies used TAZ for trip assignment. TAZ is good for planning purpose. But the biggest disadvantage of TAZ is that the data is not usually available to the general audience. Typically, city or county transportation planning agencies are responsible for their TAZ definition. And many of those data are out-of-date and do not represent the current road networks.

Population and demographic data has become more detailed and accurate. Also, high performance computing is much easier to access with faster computation time and lower cost. A revised platform for evacuation modeling built on high resolution demographic data and activity-based microscopic traffic simulation could reveal more realistic and detailed phenomenon in simulation-based evacuation studies.

## 5.3 Evacuation Model

To compare the impact of traveler compliance behavior in evacuation operations with both TAZ and LPC assignments, a real-world evacuation case study with different compliance levels was conducted using the same microscopic simulation software package.

### *5.3.1 Case Study Data Description*

Road networks and trips/population distribution are two major input data for the evacuation case study. To avoid repetitious work on network coding and focus on comparing the performance in different data resolutions, Alexandria, Virginia was chosen for its free but well coded network, provided with TRANSIMS. The road network data is shown in Figure 5.1, which is composed of 3634 links and 2608 nodes. The map source is from OpenStreetMap in Esri ArcGIS 10.1 software package.

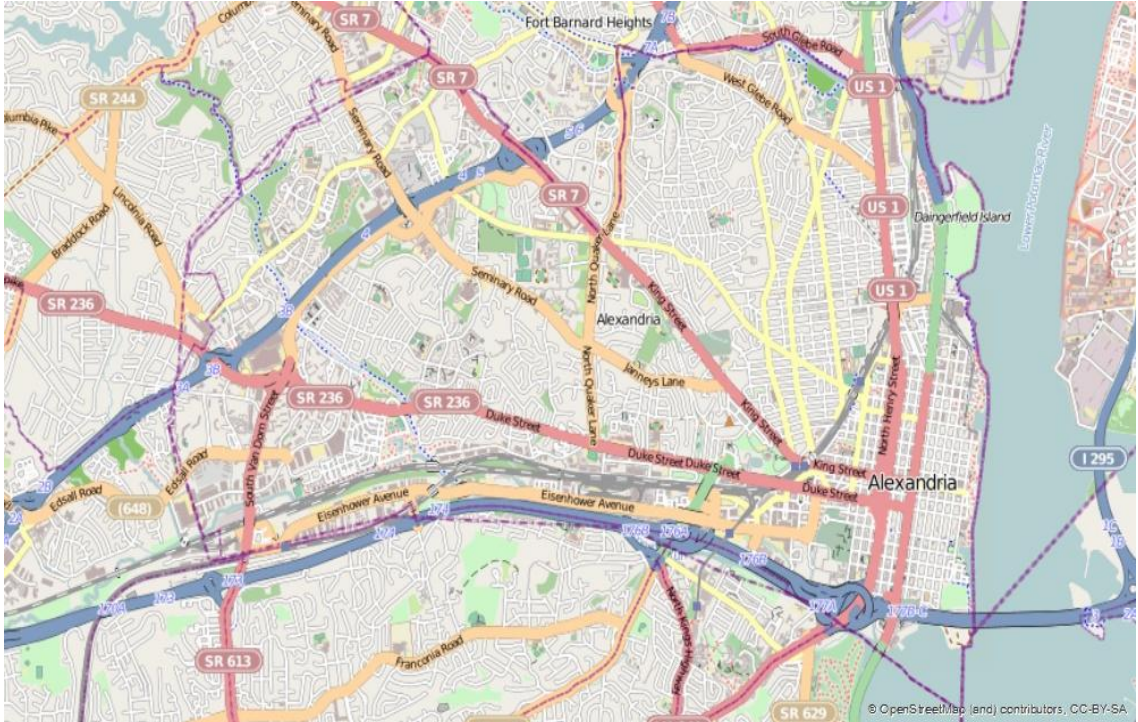

**Figure 5.1. Alexandria City Road Network.**

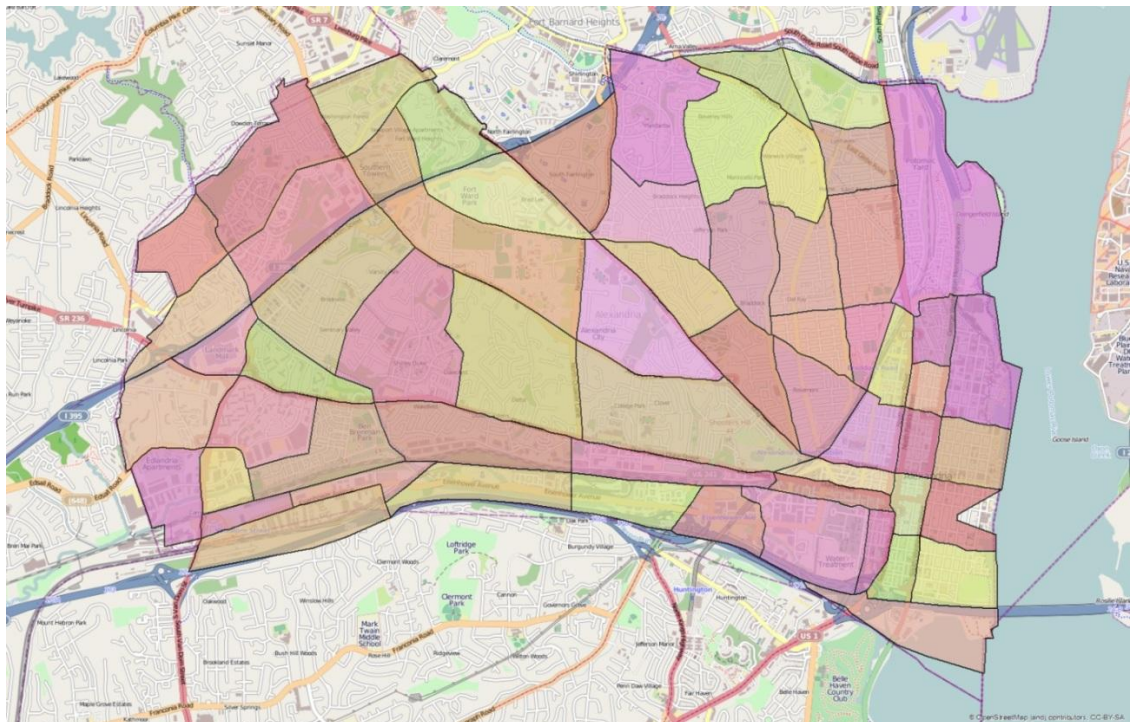

**Figure 5.2. Traffic Analysis Zones in Alexandria.**

For the demographic input for this study, LandScan USA population distribution data was implemented. The LandScan dataset is widely used in the geographic field for population risk analysis (Dobson, Bright et al. 2000, Bhaduri, Bright et al. 2002). LandScan USA population cell (LPC) data is composed of national population distribution in 90m x 90m (3'×3') (Bhaduri, Bright et al. 2007). It is much more accurate than conventional TAZ because some TAZ zones are large in scale or dense in population (You, Nedović-Budić et al. 1998) . Compared to TAZs, LandScan is a challenge for the traditional trip generation method. The trips are generated from cell to cell at very large scale, instead of relatively small scale zone to zone method. LandScan USA 2011 daytime population data is implemented to compare the two resolution input datasets, TAZ and LPC. The TAZ population data is aggregated from LPC data to make the comparison consistent. The total population in Alexandria is 161,519. The Alexandria TAZ distribution map is shown in Figure 5.2. TAZ uses real-world road segment to cut the boundary, which is widely used for traffic planning purpose. Alexandria contains 62 TAZ zones.

In contrast, LPC consists of 5657 non-zero cells (6131 in total), as shown in Figure 5.3. The color represents cells with the number of people per cell. Red means higher population and grey means no population. Each cell is assigned a value for the number of population. Technically, TAZ and LPC have the same zone definition, but LPC size is much smaller. As the vehicle per capita ratio is about 0.8 in U.S., the total predicted trips for evacuation in Alexandria city is 129,215 (that is, 161,519 * 0.8). Only integer values are used in defining the number of trips for both TAZ and LPC. For those LPC cells with only 1 person, 1 trip is generated to ensure every person in the cell has access to evacuate to the safe shelters. Adjustment is amended to make sure TAZ and LPC have the same number of trips. Safe shelters are located near the city boundaries. The objective of the evacuation scenarios is to evacuate all the people to assigned shelters.

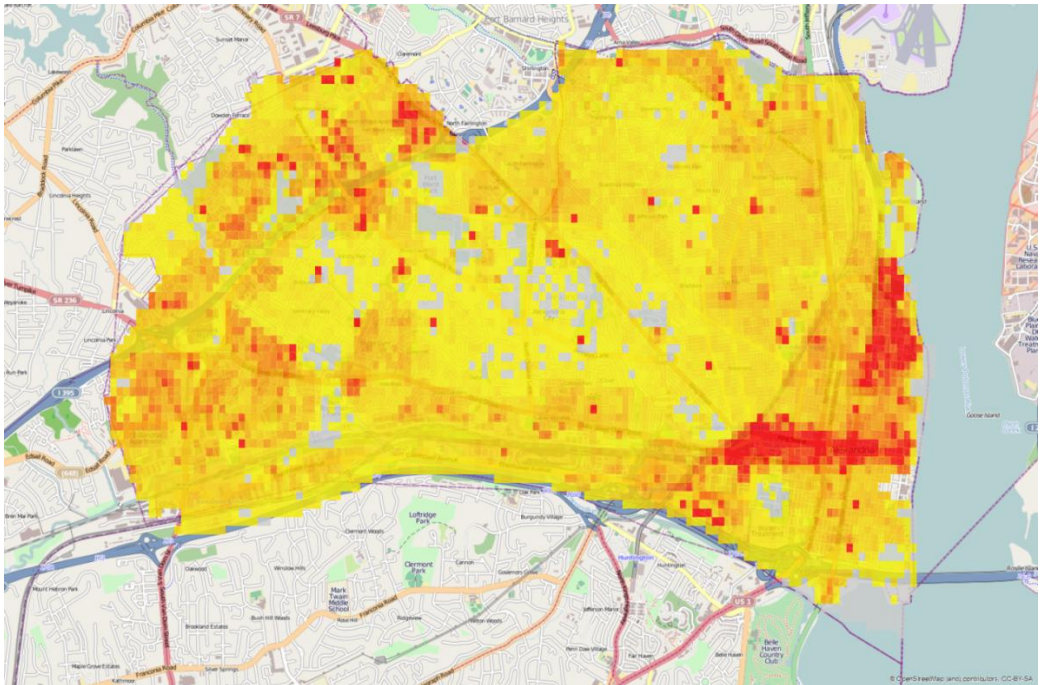

**Figure 5.3. LandScan USA 2011 population data in Alexandria.**

*5.3.2 Simulation Tool*

LandScan population data violates the activity-based traffic assignment with TAZ method. Thus, all existing commercial traffic simulation software is not capable of handling the large number of LandScan USA population cells. TRANSIMS is a widely used open-source traffic modeling tool for transportation planning and operation analysis. It permits traffic model developers to modify the source code according to their specific needs.

There are two major limits to address for applying TRANSIMS to LPC based models. The first is the number of TAZs allowed in trip assignment. TRANSIMS only considers 1000 zones as the maximum for the total number of TAZ. There are 5657 non-zero LPC cells in Alexandria, technically equivalent to 5657 TAZ zones. This limitation has been adjusted. The second limit is the implementation of activity-based traffic simulation in TRANSIMS. It assumes that there are at least two activity locations in a TAZ for trip generation allowing inner-zone trips and at least one activity location for only inter-zone trips. This is understandable in the traditional TAZ concept since the TAZ is usually large enough to include several road segments. But LPC size is much smaller than TAZ and there are many LPC cells with no roads intersecting their areas. In this case, the trips in that zone cannot be assigned to the activity locations. To correct this problem, an internally developed program was written to assign each LPC to their nearest activity location. This meets the real-world situation where travelers prefer to access roads at the nearest spots to their location.

*5.3.3 Routing Modeling with Multiple Destinations*

A full Alexandria city road network, including local roads in neighborhoods, was used to achieve more realistic evacuation performance. The full road network is composed with 2608 nodes, 3634 links, 7718 activity locations, 1062 stop/yield signs, and 261 traffic signals. 129,215 trips, in both 62 TAZ zones and 5657 non-zero LPC cells format, were modeled. In addition, 21 safe shelters were assigned and connected one-to-one with 21 exits in the road network respectively.

The conventional highway-biased shortest network distance assignment patterns in TAZ and LPC formats are shown in Figure 4. The bold green links in Figures 5.4 connects the TAZs central points to their shelters with three kinds of trip routing algorithm. The blue points are the TAZs central points. The red points are the safe shelters as destinations. The LPC cells are not drawn here due to the large amount, but the concept is similar to TAZ. The gold links bridges the LPC central points to their shelters.

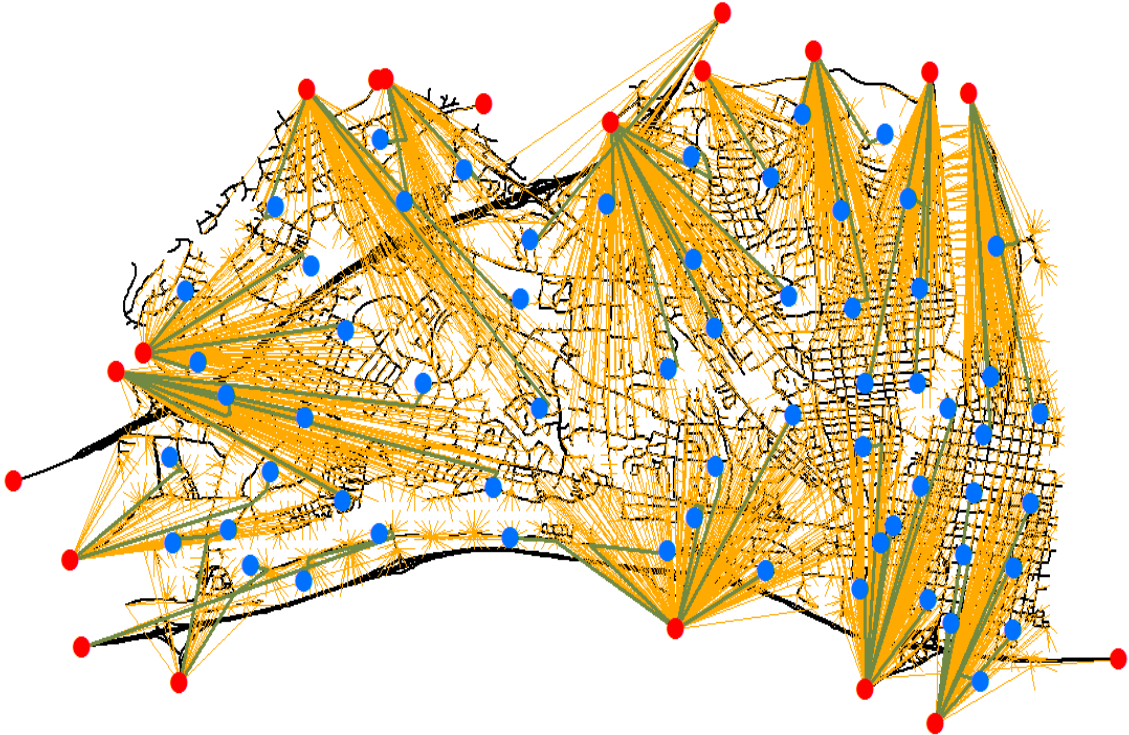

**Figure 5.4. One-to-one destination assignments in TAZ and LPC**

Different from conventional one-to-one origin/destination zone assignment, multiple destinations were implemented to compare traveler compliance behavior impact. When evacuees do not follow the evacuation instructions, they might choose a shelter in a different location. In the Alexandria case study, some of these 21 destinations are near each other. If we decide to use the one adjacent to the assigned shelter, there might be no significant difference for the travel time and cannot reflect travelers' desire to choose another location. In addition to the first route based on highway-biased shortest network distance, the second best and the third best destinations are provided for the travelers and the inter-distance among these three destinations is larger than the threshold. For this study of Alexandria, VA, this threshold was 1 mile. The non-compliant trips are evenly assigned to the second and the third destinations. This multi-destination routing algorithm is both flexible and more representative of actual behavior. Users can define their own destination inter-distance threshold value and non-compliance level based on their road network and shelter distribution. It also can be easily extended to k destinations, where k is equal to 1 to N. Figure 5.5 shows a 3-destination assignment based on TAZ in Alexandria. The zone, pointed by a green arrow, illustrates how three destinations vary for a certain zone. The red line is the first destination. The orange line is the second and the purple is the third. The LPC based assignment has similar pattern.

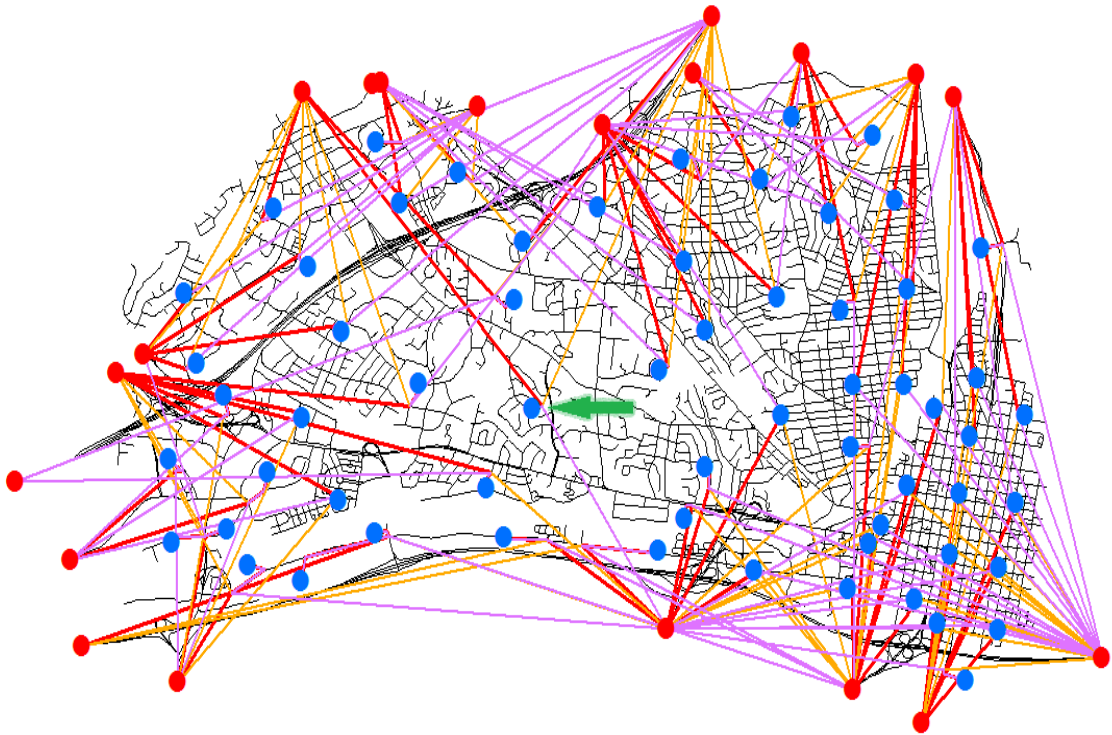

**Figure 5.5. Multiple destinations assignments in TAZ**

The evacuation demand for each TAZ or LPC was converted to 15-minute interval volumes. A typical S-shape loading curve was used in this research. The total Alexandria, VA area is about 15.2 sq. mi. To reflect the real evacuation performance for this area, the evacuation demand was loaded over 6 hours and evacuation simulation time is 7 hours.

*5.3.4 Experimental Scenarios*

For city level evacuation planning such as this conducted by this study, it is very challenging for local transportation agencies to generate an optimal evacuation plan for every evacuee with specific route and destination assignment. Conventionally, transportation planners use traffic analysis zones for trip assignments. This was implemented in simulations as the base case. LPC based trip assignments were also implemented for the comparison study. Five compliance levels were simulated, as 100%, 70%, 50%, 30%, and 0%. In total, 10 evacuation scenarios were simulated in this paper, combing two resolution trip assignment methods and five travelers' compliance levels. TRANSIMS 4.0.8 version package was run on a Window 7 64-bit laptop computer. The configuration of this laptop is 16GB RAM, 2.6GHz Intel(R) Core(TM) i5-3320 CPU, and 500GB hard disk. An 8-hour simulation time was used and the other parameters are set as stated in the preceding sections. After multiple testing of the TRANSIMS simulation, the traffic assignment achieved user equilibrium status after 10 runs of Router and 15 runs of Microsimulator procedures for this case study. Router and Microsimulator are two major programs in TRANIMS for trip assignment and microscopic simulation. To adjust the effect of random number in the Microsimulation procedure, all the simulation results are based on the average of 30 independent runs. To evaluate the evacuation efficiency of these scenarios, various measures of effectiveness (MOEs) are developed. Han et al. (Han, Yuan et al. 2007) suggested a four-tier MOE framework for evacuation, including evacuation time, individual travel time and exposure time, time-based risk and evacuation exposure, and time-space-based risk and evacuation exposure. They can be used in different scenarios in evacuation modeling to satisfy different evacuation planning purpose. Evacuation time is the most significant MOE to be considered in this study. Comparison studies of two resolution data and five compliance levels were analyzed.

## 5.4 Simulation Results

Results from the TRANSIMS simulations on both TAZ and LPC are summarized in Table 5.1. In general, LPC based evacuation performance is not as efficient as TAZ based evacuation. The high resolution data has a better representation on how evacuees travel out of evacuation area. Missing data in the table means that not all the scenarios produce a high percentage evacuation performance. Detailed comparison studies were also presented to reveal the impact of traveler compliance behavior and population distribution in different data resolutions.

**Table 5.1. Evacuation Time for Different Scenarios**

| Population Resolutions | Compliance Level | Percent Evacuated | | | | | | | | |
|---|---|---|---|---|---|---|---|---|---|---|
| | | 25% | 50% | 60% | 70% | 80% | 85% | 90% | 95% | 99%[1] |
| | | Evacuation Times in Minutes | | | | | | | | |
| TAZ | 100% | 114 | 175 | 195 | 220 | 251 | 272 | 295 | 315 | 348 |
| | 70%[2] | 112 | 174 | 197 | 224 | 253 | 277 | 297 | 318 | - |
| | 50%[3] | 114 | 180 | 207 | 241 | 280 | 304 | 329 | 370 | - |
| | 30%[4] | 119 | 189 | 219 | 255 | 301 | 328 | 369 | - | - |
| | 0%[5] | 125 | 204 | 243 | 284 | 342 | 405 | - | - | - |
| LPC | 100%[6] | 120 | 201 | 242 | 290 | 343 | 377 | - | - | - |
| | 70%[7] | 120 | 201 | 241 | 295 | 353 | 396 | - | - | - |
| | 50%[8] | 119 | 207 | 257 | 306 | 353 | 390 | - | - | - |
| | 30%[9] | 121 | 222 | 276 | 324 | 378 | 445 | - | - | - |
| | 0%[10] | 120 | 223 | 275 | 327 | 402 | 465 | - | - | - |

[1] Only 99.26% of the trips arrived at destinations in 7 hours. 99% is used instead of 100%.
[2] Only 98.85% of the trips arrived at destinations in 7 hours.
[3] Only 96.21% of the trips arrived at destinations in 7 hours.
[4] Only 92.48% of the trips arrived at destinations in 7 hours.
[5] Only 86.50% of the trips arrived at destinations in 7 hours.
[6] Only 89.09% of the trips arrived at destinations in 7 hours.
[7] Only 87.94% of the trips arrived at destinations in 7 hours.
[8] Only 88.07% of the trips arrived at destinations in 7 hours.
[9] Only 86.23% of the trips arrived at destinations in 7 hours.
[10] Only 85.87% of the trips arrived at destinations in 7 hours.

For the TAZ based traffic assignment simulations, the evacuation performance from different compliance levels are summarized in Figure 5.6. At the two hour (120 minutes) mark of the simulation, there is no significant difference among different compliance level. After that, the evacuation efficiency decreases while the compliance level decreases. The small non-compliance level, such as 30% non-compliance level, or 70% compliance level, did not impact the system performance significantly. The slight difference between 100% and 70% compliance level might be caused by the computation convergance residuals. The one-mile inter-distance threshold among the three destinations might also have an impact on the simulation results. The exact value of destination distance should also consider local land use information. The 70% scenario even has better performance at the beginning. In other words, a small percent of travelers en-route route decisions based on provided real-time traffic information can help them arrive at destinations efficiently. The full non-compliance scenario, where no traveler uses the assigned best route but 50% travelers choose the second and the third assigned destinations respectively, has the worst performance.

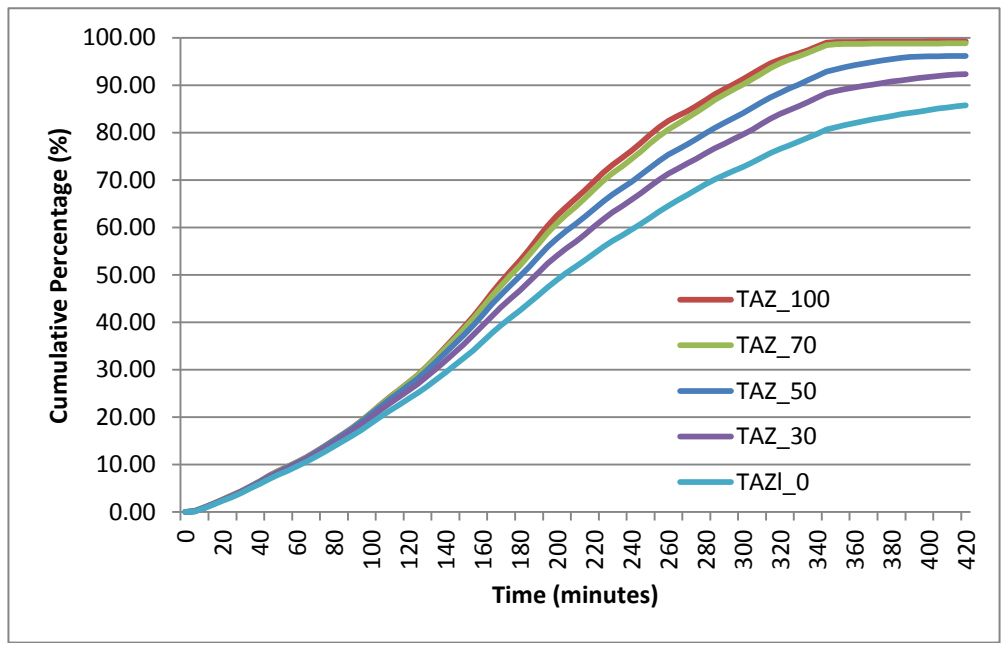

**Figure 5.6. Evacuation curves under TAZ based assignment.**

In contrast, the LPC based trip assignment produced different results for evacuation performance. It has less sensitive to travelers' compliance choices, as shown in Figure 5.7. In the first three hours, evacuation performances are almost the same for different compliance levels. The evacuation efficiency decreases slightly when the compliance level is 30% and 0%. But it does not change the efficiency compared to other compliance levels. The smaller LPC size increases the diversity of each individual's destination choices. The aggregated trips for the 21 destinations with 5657 LPCs do not change significantly the compliance levels, unlike the case with 62 TAZs, where trips can vary largely with small number of zones. The standard deviation of assigned destination population in the TAZ study is much larger than LPC case. The population value of TAZ method ranges in [394, 12497]. But the value of LPC method is [1, 2641]. The evacuee compliance level can change the destination population distribution significantly in TAZ scenario, but not the LPC case.

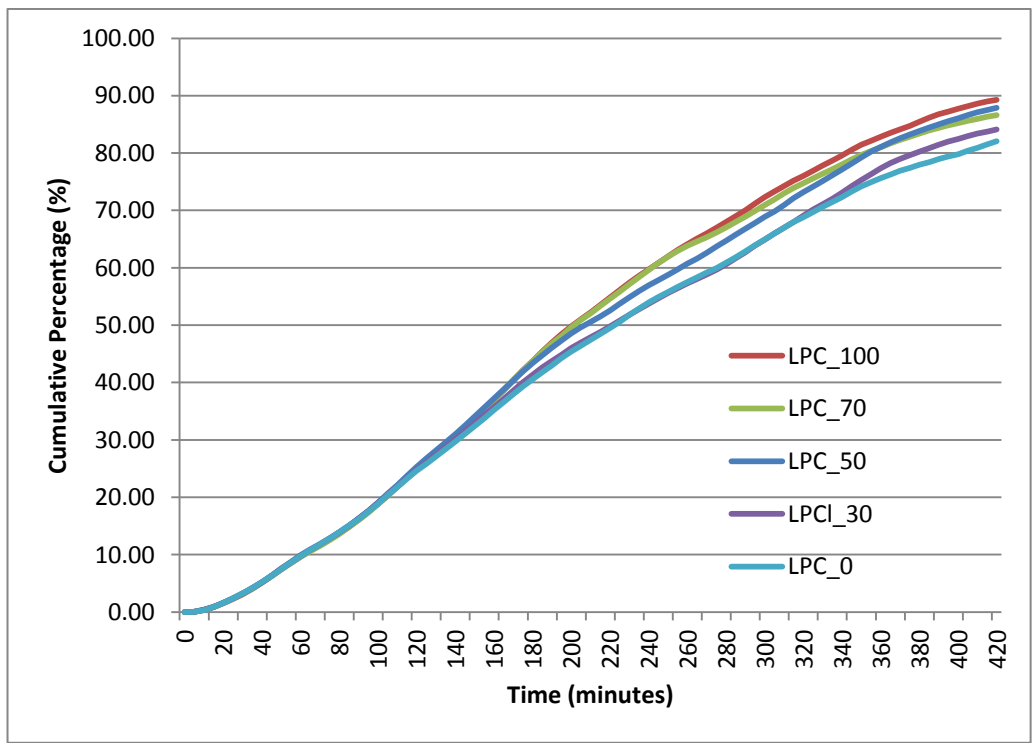

**Figure 5.7. Evacuation curves under LPC based assignment.**

After the 8-hour simulations of 10 scenarios ended, the evacuation efficiencies to compliance level with TAZ and LPC were summarized in figure 5.8. TAZ based trip assignment under-estimate the evacuation times compared to the LPC method. The full compliance evacuation efficiency with LPC assignment is 11.4% more than the conventional TAZ assignment. This is because TRANSIMS evenly assign the trips to all the activity locations in each zone. But the LPC method does not use the default TRANSIMS method. Instead, it assigns each LPC cell to the nearest activity location, which is more realistic. Some LPC cells might have much more trips than others, like downtown areas, which formulates traffic congestion when people are trying to evacuate at the same time. Again, the evacuation performance drops while compliance level decreases in TAZ method, but keeps relatively flat in LPC method.

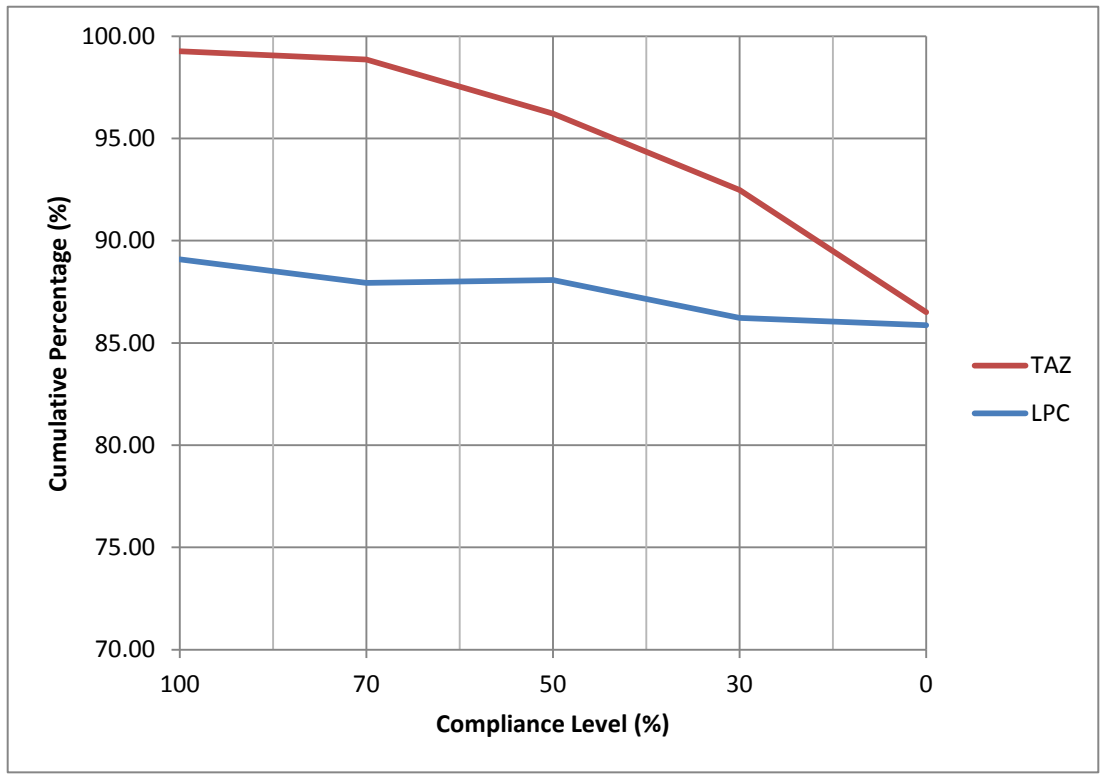

**Figure 5.8. Evacuation efficiency vs. compliance level.**

## 5.5 Conclusions

A comparison study of evacuee compliance behavior impacts on evacuation assignment with both conventional TAZ method and high resolution LandScan USA population data was conducted. A revised framework for evacuation based on uniform high resolution data and activity-based microscopic simulation was proposed, which gives the researchers and practitioners the ability to simulate any area in the world (since LandScan Global has the whole world population data and OpenStreetMap provides world-wide road network.) with more realistic results. Two major findings are concluded: 1) conventional TAZ based trip assignment under-estimates the real travel time, and 2) the evacuation performance is not that sensitive to evacuee compliance behavior in LPC based assignment compared to TAZ based assignment. This first result is caused by TRANSIMS evenly distributing trips to all the activity locations in each traffic analysis zone. But people are using their nearest activity location to access the road network in the real world, where LPC assignment is the case. The second phenomenon is because the standard deviation of assigned destination population in the TAZ study is much larger than the LPC case. The evacuee compliance level can change the destination population distribution significantly in a TAZ scenario, but not the LPC case. This inspires traffic engineers and planners to reconsider the definition of traffic analysis zones. The conventional TAZs might not be efficient in real-world traffic operation, especially evacuation operation cases, where every individual should be counted and treated independently. Beyond that, real-time traveler compliance behavior through advanced traveler information systems should be considered, rather than discussion of predefined compliance situations, since there is no significant impact of compliance level in LPC based assignment.

Current simulations in this paper are based on given well defined Alexandria, VA road network data, which helps us focus on evacuee compliance behavior analysis but limits its application in other area. The purpose of this research is to evacuate any assigned area in the world. A conversion program from OpenStreetMap to TRANSIMS readable network data can make this happen. A realistic and user-friendly open-source traffic simulation package for evacuation purpose can save researchers and practitioners on model building and help them focus on policy studies. Also, the simulations in this paper were implemented with distance-based traffic assignment. A time-based traffic assignment, like dynamic traffic assignment, could improve the performance when we implement intelligent transportation systems (ITS) devices in the city to provide real-time travel information.

# CHAPTER 6
# EVALUATION OF VEHICULAR COMMUNICATION NETWORK IN A CAR SHARING SYSTEM

# Abstract

Yet even with an increase in car sharing programs worldwide, there has been little research on the application of Vehicular Ad hoc Network (VANET) in car sharing systems. This study employed three parameters of communication networks to evaluate VANET performance in car sharing systems. The integration of vehicle mobility generation and vehicular communication networks simulation is crucial to this research. A variety of scenarios on a partial Manhattan grid network were simulated to assess the influence of different parameters in the communication network. Through the performance analysis, it evaluates the feasibility of VANET in car sharing systems and gives some suggestions for future field deployment.

*Keywords*: VANET, car sharing system, NS-2, traffic flow, grid network, traffic simulation

## 6.1 Introduction

Since the 1980's, car-sharing programs have been discussed in Europe. Fellows and Pitfield (2000) evaluated urban car-sharing programs from an economic and operational perspective. There are many car sharing programs, even successful commercial one, like Zipcar (Keegan 2009), are widely researched and used in United States. Communication between shared cars can improve the level of service and promote the effectiveness and efficiency of the car sharing systems. Some communication systems, like in-vehicle radio and/or intercom system, have been implemented in vehicles. Usually, they are used for private companies, such taxi companies. However, such systems are more dependent on human's participation and less time-efficient. Vehicle Communication Networks (VANET) has been widely researched in various fields, both vehicle-to-vehicle and vehicle-to-infrastructure communications. But there is no specific research about how to implement VANET in car sharing systems and the implement strategies. This paper is attempting to understand the feasibility of VANET in a public available car sharing system. The characteristics of shared-cars' movement and the implementation of VANET are discussed. The paper then discusses the integration of vehicle mobility generation and wireless communication network simulation. Three MOEs (Measure of Effectiveness) are used to evaluate the performance of inter-vehicle communication in car sharing systems. A case study is presented using part of the Manhattan map with different vehicle scenarios. Finally, a conclusion is provided.

## 6.2 Literature Review

Car-sharing program has become increasingly popular in urban areas to improve the commuter experience and to reduce the travel cost. Car-sharing, as defined in Transit Cooperative Research Program's report (Millard-Ball, Murray et al. 2005), is a service "that provides members with access to a fleet of vehicles on an hourly basis." Different from car-rental service, car-sharing are mainly used for short distance commuter within a city, rather than long distance travel. It is more like a shuttle system between different hot-spot locations, but the participant is his/her own driver. Shaheen, Sperling et al. (1998) summarized the car-sharing programs in Europe and North America with emphasizing the large range of choices of vehicles, less travel cost, and fewer ownership responsibilities. Especially, the paper talked the car-sharing future with advanced communication technology. Huwer (2004) explored the benefits and effects of combing public transport and car-sharing system. Car-sharing system is more treated as a complimentary service to public transportation system. To better understand the car-sharing usage from users' behaviors, Morency, Trépanier et al. (2007) analyzed the users' temporal patterns, travel distance, and week use variability with transaction datasets in Canada. It provides a good modeling reference for microscopic traffic simulation for car-sharing systems. Car-sharing system is a good case to implement VANET as the most participated vehicles are used within in a city and there is less privacy issue to implement communication equipment.

In recent years, the rapid development of wireless communication and information technologies have enabled the development of vehicular communication systems, especially inter-vehicle communication systems, which can improve the comfort, safety, and operational efficiency of transportation systems (Sichitiu and Kihl 2008). Many such projects have conducted by governments or institutes worldwide.

The eSafety forum (Commission 2002) has led the European government public departments to cooperate with vehicle manufacturers. They focus on various aspects relating to intelligent vehicle safety issues. CarTALK2000 (Reichardt, Miglietta et al. 2002) focuses on the application of inter-vehicle communication in driving assistance systems. The main task is to evaluate the performance of driving assistance system, design suitable data structure and algorithm for software development, and test its function in real scenarios. Hartenstein, Bochow et al. (2001) promote FleetNet as an inter-vehicle communication platform that implements Mobile Ad hoc Network (MANET) in VANET. This method challenges traditional communication technology, including transport layer protocols and information transmission hardware. Car-to-Car Communication Consortium (Kosch and Franz 2005), including six vehicle manufacturers, is an open forum to set up a unique European standard for inter-vehicle communication. It adapts Wireless Local Area Network (WLAN) techniques to implement interactive communication between different vehicles.

Since 1986, the University of California, Berkeley and Caltrans (California Department of Transportation) have cooperated to develop the California Partners for Advanced Transit and Highways (PATH) project. The purpose is to solve the problems of California's transportation systems through interdisciplinary knowledge, such as electrical engineering, information technology, economics and transportation policy studies (Hedrick, Tomizuka et al. 1994). The project contains three program areas: transportation safety research, traffic operations research, and modal applications research. The traffic operations research has a focus on intersections and cooperative systems, which illustrated the communication between vehicles and infrastructures. Also, some detailed communication technology was discussed. Jiang discussed the vehicular safety communication through Dedicated Short-Range Communication (DSRC), which is introduced in 1999 by FCC for vehicular communication systems (Jiang, Taliwal et al. 2006). Connected Vehicle Project (Zhang, Gantt et al. 2009) is a good example of a Vehicle Infrastructure Integration (VII) based on DSRC. A corresponding communication protocol, IEEE 802.11p, is suggested to add wireless access in the vehicular environment (WAVE). Connected Vehicle Program is mainly used for safety and mobility applications, which enable safe and efficient communications among vehicles, infrastructure, and passengers' communications devices. However, V2V systems, which are different from other wireless communication systems, are restricted to the special mobility patterns of communication nodes in traffic networks, especially in dynamic traffic flow scenarios. The combination of a traffic network (based on traffic flow theories) and a wireless communication network is becoming an important part of VANET. It is important to understand the impact of vehicle mobility patterns on VANET to develop more effective and efficient VANET technologies.

Some studies have been done on the analysis of VANET in different traffic flow scenarios. Lu et al. explained the inter-vehicle communication in a unidirectional traffic flow with and without dynamic shockwaves (Lu, Bao et al. 2009, Wang, Lu et al. 2009). They compared the simulation results with the theoretical work by Jin (Jin and Recker 2010). Sun and Lu et al researched on different traffic scenarios on unidirectional traffic streams, bidirectional traffic streams, and multilane traffic streams (Sun, Bao et al. 2009, Sun, Bao et al. 2010, Lu, Bao et al. 2011). Karnadi, Mo et al. (2007) also introduced SUMO for VANET to generate the realistic mobility models. All these work are conducted in a more theoretical traffic condition. A real VANET system, like car-sharing with communication, is needed to promote the feasibility of VANET in real business world.

In 2009, the joint council on Transit Wireless Communications (Communications 2009) was established to discuss the needs of Transit Industry Wireless Communications through information sharing. It mainly focuses on deal with UHF/VHF spectrum, narrow banding, 800MHz rebanding issues. The purpose is to promote and implement standards and rules for transit wireless communication and provide recommendations on commercial purchases on new equipment, system and service. A Car-sharing system, as a complimentary traffic mode to public transportation system, has more random route pattern but with more access points within a city, like the Zipcar location distribution in Figure 6.1. It shows the car access locations in New York / New Jersey area. The distribution is Manhattan is really dense, which provide potential application for vehicle communication. If shared cars can communicate with each other freely, they can provide better service for users. All equipped cars can provide real-time road conditions to the traffic control center, which can help the operation work well. Currently, each vehicle has radio system to access traffic information by radio broadcasting. The radio system cannot provide real-time traffic information because it depends on manually collecting data. How to make these vehicles communicate well with each other becomes an issue.

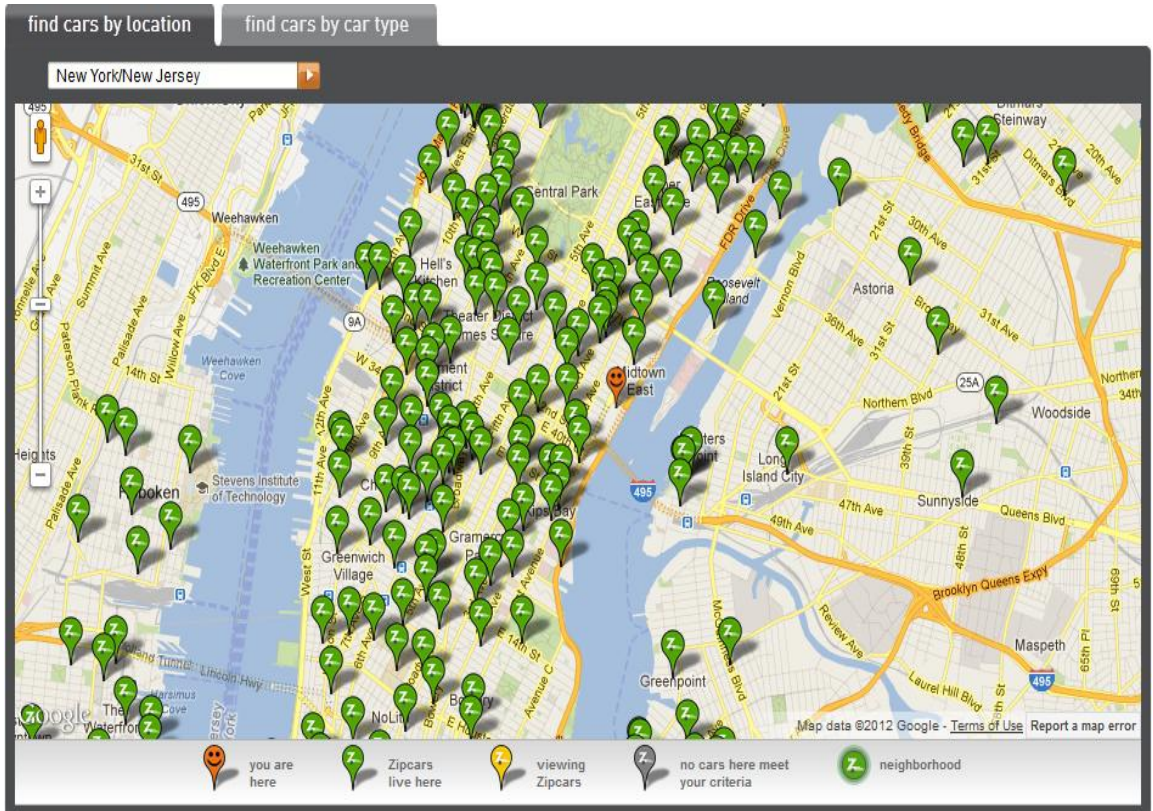

**Figure 6.1. Car-sharing access locations in New York / New Jersey by Zipcar**

Under the condition of wide usage of communication techniques and less contribution to a public system, especially a taxi system, this paper illustrates a simulation method for implementing a vehicular communication network in a taxi system. Through discussion of the feasibility and performance of this VANET system, it proposes some suggestions to implement this system in real world.

## 6.3 Methodology

An important innovation in this paper is to combine a vehicle mobility simulation and vehicular communication networks for car-sharing systems. These tools are used to generate vehicle mobility and simulate communication networks respectively.

*6.3.1 Mobility Model for Taxi Systems*

As mentioned before, the realistic vehicle mobility pattern is essential to research on the feasibility of VANET in car-sharing system. Since the vehicle movement data is not available for free, simulation is the most economic and efficient way to implement it. We adopt Simulation of Urban Mobility (SUMO) as our traffic generation tool, which is open-source, microscopic, multiple-model based traffic simulation software (Behrisch, Bieker et al. 2011). Krajzewicz, Hertkorn et al. (2002) validated SUMO through an example of microscopic car models. Kim, Sridhara et al. (2009) also proved SUMO's effectiveness on realistic mobility simulation of urban mesh networks. As a research tool, SUMO produced fundamental traffic flow characteristics and is suitable for the vehicle movement generation.

To analyze the performance of each vehicle, it is necessary to compute the trajectory of each car and to study the impact of different vehicles' trajectories. The classic microscopic traffic theory, the car following model, can be used to calculate the trajectory of each car. The model is

$$\begin{cases} \frac{d^2}{dt^2} x_{i+1}(t+T) = \lambda \frac{d}{dt} s_i(t) \\ s_i(t) = x_i(t) - x_{i+1}(t) \end{cases} \quad (1)$$

Where $s_i(t)$ is the distance between two adjacent vehicles; $x_i(t)$ is the position of vehicle *i* in the special coordinates; and $\lambda$ is the sensitive coefficient.

The vehicle-driver model is the basic model in SUMO, which is based on the improved car following model by Krauss (Krauss, Wagner et al. 1997). This model can show basic traffic flow characteristics, including free flow and congestion flow. Three kinds of speeds are represented in this vehicle-driver model. The first is the safe speed $v_{safe}$, which is used to satisfy the requirement of collision avoidance. Considering the vehicle's acceleration ability, the second speed is expected vehicle speed, which is the minimum value of all possible maximum values.

There are also random errors for drivers to drive at the expected speed. These can be implemented by subtracting a random error reflecting driving behavior. Also, the vehicle cannot move in the reverse direction, so the stopping speed is 0. The final vehicle speed is

$$v(t) = \max[0, rand[v_{des}(t) - ea, v_{des}(t)]] \tag{2}$$

This model is a kind of collision avoidance model, which performs well in unexpected situations. It makes the traffic simulation more realistic and executable.

### *6.3.2 VANET Configuration in NS-2*

NS-2 (Network Simulator version 2) is a discrete event simulator developed for network research. It is an open source platform written in C++ and Otcl. The latest NS-2 package is version 2.34. It is designed for easy use and study and has been widely used in wireless ad hoc networks. NS-2 provides substantial support for implementing TCP & UDP, communication application such as FTP, CBR, VBR, routing protocols, and multicast protocols over networks (Simulator 2011).

The simulation of the vehicular ad hoc network under NS-2 environment was configured according to the procedure used in Khaled's paper (Toor, Muhlethaler et al. 2008), which emphasizes the application of IEEE 802.11 in NS-2. This paper used constant bit rate (CBR) generators for the application layer. The packet size was 512 bytes. For each scenario, the simulation time was 600s. UDP protocol was used for the transport layer because it presents better than TCP. An Ad hoc On-Demand Distance Vector (AODV) routing protocol was selected for the network layer after we compared AODV, DSR, and DSDV. AODV is more suitable for vehicular communication networks. For the MAC layer, we used IEEE 802.11 DCF because it is easy to implement in NS-2. Two-ray ground model is chosen as the propagation model since it is adequate to the communication in free roadways. The transmission range is 1000 meters, which means that the maximum distance for two vehicles to get connected is 1000 meters.

## 6.4 Performance Evaluation

Because the purpose of researching VANET was to analyze the application performance in car sharing systems in a real road network, a real map is helpful to simulate the more realistic scenarios. Here we used a Manhattan map generated by TIGER (Topologically Integrated Geographic Encoding and Referencing).

### *6.4.1 Simulation Scenario Configuration*

Since we do not have the Zipcar car usage data and our purpose is to research on the feasibility of VANET in a real road map, we assume two different mobility scenarios with a 600-vehicle traffic flow here: the penetration rate for the first scenario is 5% and for the second is 10%. This means there are 30 cars in the first scenario and 60 cars in the second scenario. The cars are equipped to communicate with each other only; the other vehicles are considered to be normal vehicles without communication ability. All moving trajectories are randomly generated. The length of each vehicle is five meters. Their maximum speed is 15.5 m/s, which is approached from 1 m/s in about 10 seconds. We use three parameters to analyze performance: Packet Sending Rate (PSR), Maximum Connectivity Number (MCN), and Number of Vehicles. PSR is the number of packets sent by the source node (vehicle) in one second. MCN is the maximum number of communication links in vehicular communication network, where this means how many pairs of vehicles can set up communication.

To evaluate the communication network performance of car-sharing systems in the Manhattan map scenarios, it mainly focuses on two kinds of scenarios: the impact of different PSR and MCN on the communication performance given the same number of cars (30 cars) and the impact of different numbers of cars when network parameters are held constant (PSR at 10 packets per second).

In addition, three metrics are used to discuss car sharing systems performance. Packet Delivery Fraction (PDF) is the success rate for delivered packets from the source vehicle to the destination vehicle, which can be calculated by the following formula:

$$\text{PDF} = \frac{\text{received packets}}{\text{sent packets}} * 100\% \tag{3}$$

The Average end-to-end Delay of Data Packets contains all possible delay, such as the buffer time to find the communication route, delivery time, and resending time in MAC layer, etc. Formula (4) presents the calculation method,

$$\bar{t} = \frac{t}{N} = \frac{\sum_{1}^{N} t_i}{N} = \frac{\sum_{1}^{N}(t_i^r - t_i^s)}{N} \tag{4}$$

Where $t_i$ is the delay time of each packet; N is the number of packets received by the destination vehicle; $t_i^r$ is the receiving time of each received packet; and $t_i^s$ is the sending time of each received packet. The unit is seconds.

The arrival time of the first packet is the time to build up the routing table, which is used to evaluate the efficiency of routing; a shorter time means the routing table is built more efficiently. The unit is seconds.

### 6.4.2 Simulaiton Results in Scenario with the same number of vehicles but with varying PSR and MCN

When the number of vehicles is 30, it considers two different values of MCN: 5 and 10. The value of PSR ranges from 1 to 20. Figure 6.2 shows that the PDF does not change much when PSR varies. Under the vehicular ad hoc networks, vehicle mobility is not totally chaotic; the taxies have to follow the topology of the road network. In this situation, a communication link will last for a while once it is set up. While the network topology does not change, increasing the PSR results in an increase of received packets, because the vehicles are still within communication range. The communication link is still usable. The PDF is about 11.8% when the MCN is 5; but PDF is about 5.7% when the MCN is 10; increasing MCN results in the decrease of PDF. The Manhattan map is rather large for a taxi system with 30 vehicles. The vehicles are dispersed, and communication links cannot be guaranteed. Many packets are discarded due to unsuccessful communication links, which cause the low PDF problem.

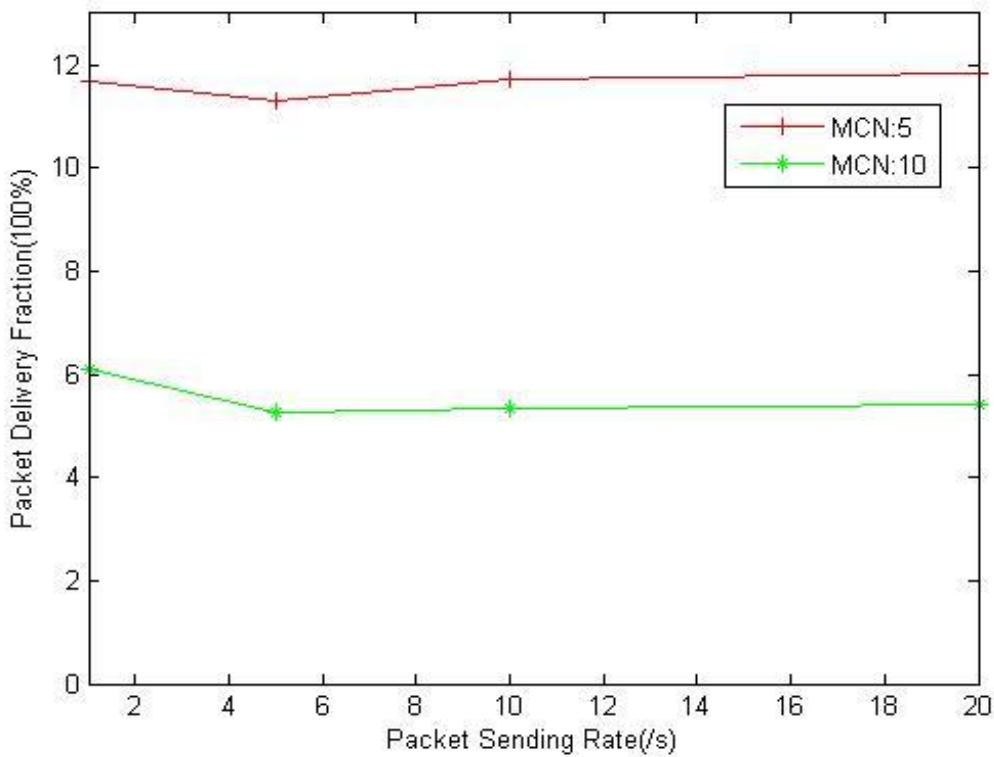

**Figure 6.2. Packet delivery fraction with different values of PSR and MCN.**

The average end-to-end delay for different values of PSR and MCN, shown in Figure 6.3, are nearly the same at 0.09s. The calculation of average end-to-end delay only contains successfully received packets. The numbers of received packets are nearly the same under these two MCN scenarios, so the delays are also similar. It shows that the average end-to-end delay is independent of the PSR due to the mobility pattern of vehicles in vehicular communication networks. The time to deliver packets and maintain communication links will not change too much after the links exist.

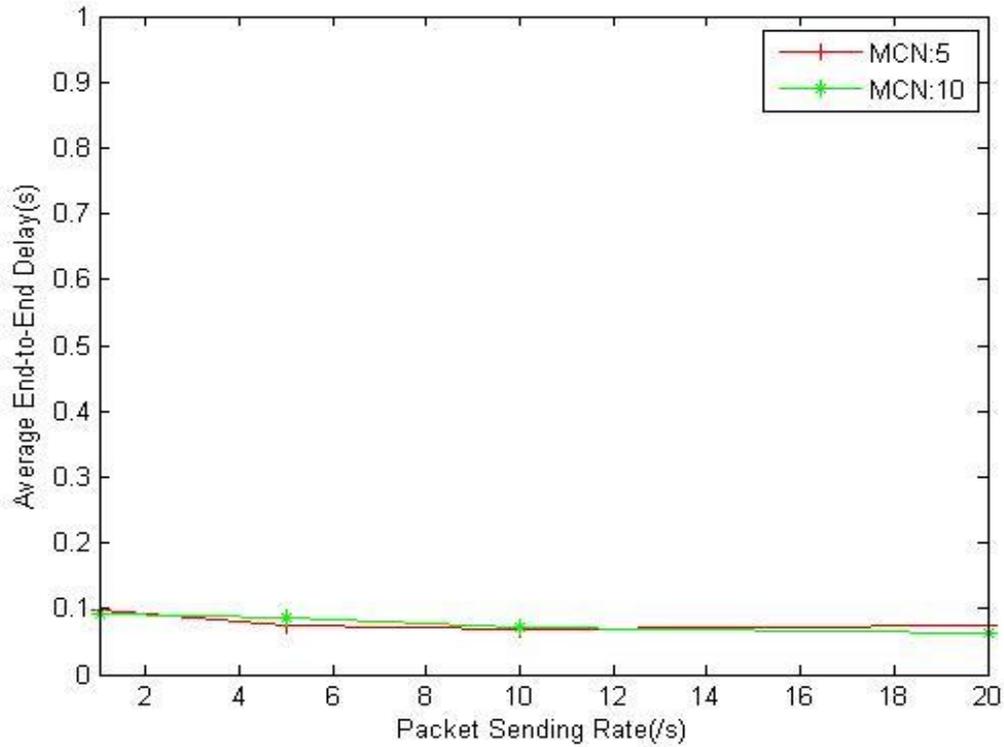

**Figure 6.3. Average end-to-end delay with different values of PSR and MCN**

**Table 6.1. Arrival Time of the First Packet with Different Values of PSR and MCN (seconds)**

| PSR= | 1 | 5 | 10 | 20 |
|---|---|---|---|---|
| **MCN=5** | 58.508 | 58.220 | 58.208 | 58.141 |
| **MCN=10** | 58.509 | 58.962 | 58.247 | 58.187 |

The arrival time of the first packet is shown in table 6.1. Regardless of the values of PSR and MCN, the first packet arrival time is around 58.2 seconds. Under the vehicular communication scenario, the mobility of vehicles follows some rules, which means they follow the roads in particular directions. There is little connection between building the routing table and the number of sent packets.

### *6.4.3 Simulation Results in Scenario with the same PSR and MCN but with different numbers of vehicles*

To examine the impact of different numbers of vehicles, we set the number of vehicles at 30 and 60 respectively. The PSR is 10 and the MCN ranges from 2 to 20. As shown in Figure 6.4, the PDF in 60 vehicles is higher than that in 30 vehicles, but not exactly double value. The increase of numbers of vehicles numbers improves communication possibilities for vehicular ad hoc networks. However, because all the vehicles are distributed randomly, the value of PDF with 60 vehicles is not the double the value with 30 vehicles.

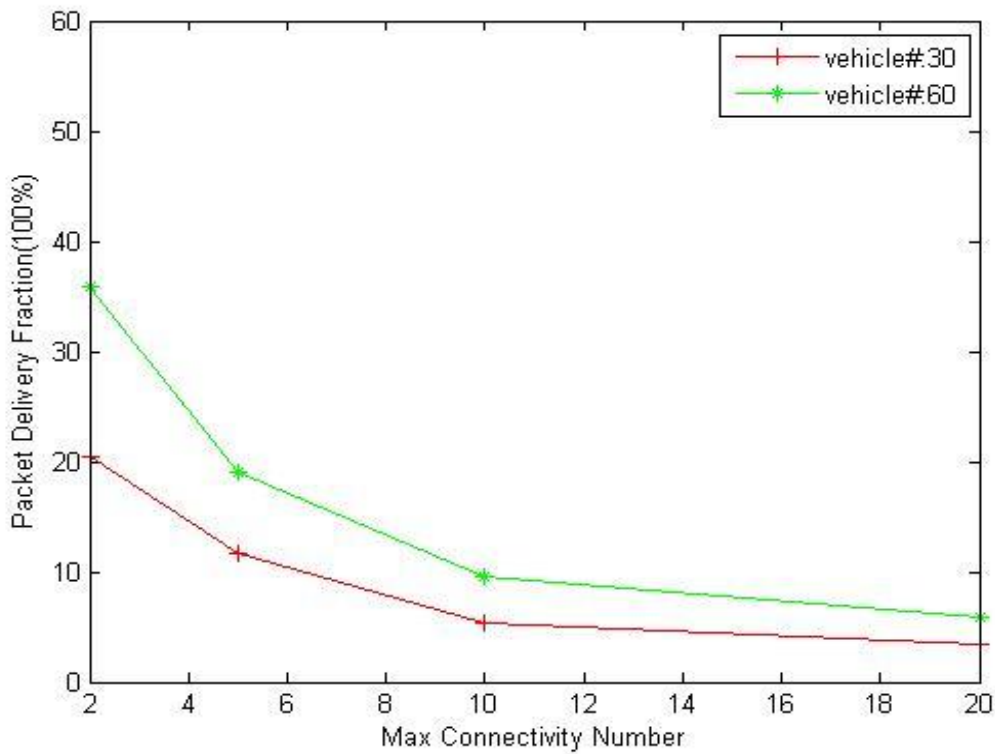

**Figure 6.4. Packet delivery fraction with different numbers of vehicles**

The average end-to-end delay in these two scenarios with different vehicle numbers can be described in Figure 6.5. The delay decreases while the number of vehicles increases. With an increased possibility of setting up communication between a source vehicle and a destination vehicle, the time spent on route finding decreases, explaining the decreased delay.

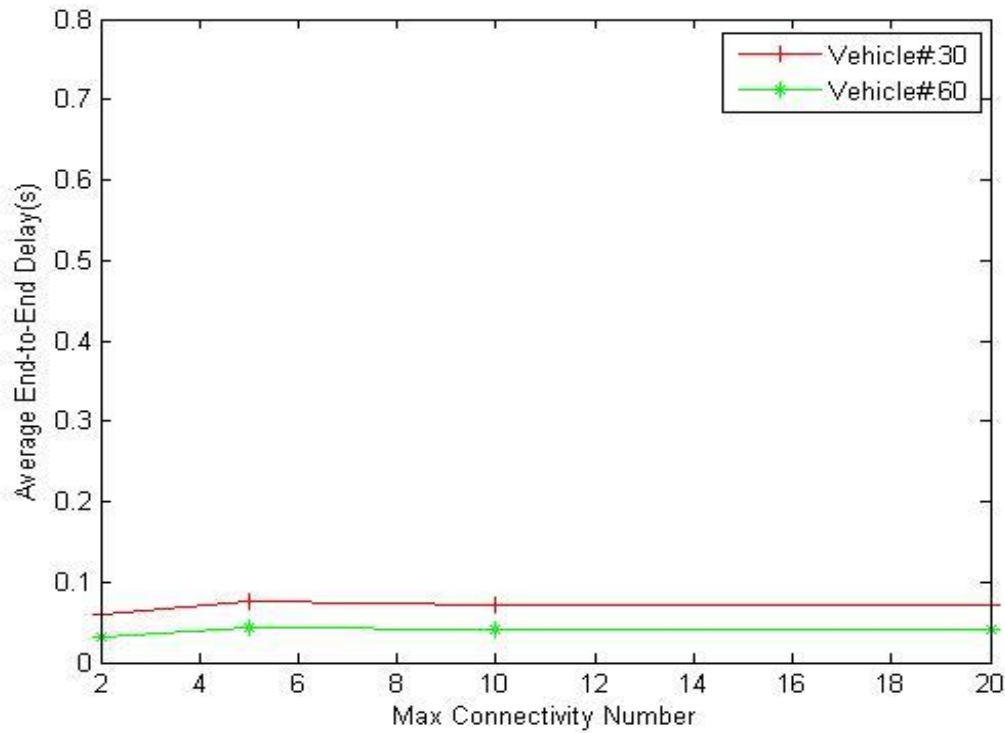

**Figure 6.5. Average end-to-end delay in different vehicle numbers.**

The arrival time of the first packet in this situation is presented in table 6.2. In the 30 vehicles case, the time is almost the same as that in table 1, which is around 58.2 seconds. In the 60 vehicles case, the time decreases to almost the half of that in the 30 vehicles case. After increasing the number of vehicles, the number of communication nodes (vehicles) also increases. There are more opportunities for the packet to deliver, which increase the success rate when delivering the packet and decrease the time to build up the routing table.

**Table 6.2. Arrival Time of the First Packet in Different Vehicle Numbers (seconds)**

| MCN= | 2 | 5 | 10 | 20 |
|---|---|---|---|---|
| **Vehs# 5** | 58.173 | 58.209 | 58.247 | 58.207 |
| **Vehs# 10** | 31.382 | 31.464 | 31.414 | 31.424 |

Based on the performance analysis above, the VANET performance in car-sharing systems is related to the number of vehicles and to the maximum connectivity number, but not to the packet sending rate. The packet delivery fraction increases when the number of vehicle increases, but decreases when the maximum connectivity number increases. The average end-to-end delay decreases when the number of vehicles increases, but does not change very much in relation to the max connectivity number. The arrival time of the first packet decreases when the number of vehicles increases, but has almost no relationship with MCN and PSR. The three parameters indicate the VANET is feasible in car-sharing systems although the communication setup is a little bit longer. After the communication between two vehicles is built, the delay for peer-to-peer communication is not significant. Theoretically, with more vehicles equipped with communication ability, the VANET performance is improved. Meanwhile, the system should control the number of communication links between different pairs of taxies to ensure better performance of the communication networks.

## 6.5 Conclusions

In this paper, an inter-vehicle communication system is simulated in a car-sharing system with different simulation scenarios. Through discussing the packet delivery fraction and average end-to-end delay of the vehicular communication network in different PSR, MCN, and the number of vehicles, we analyze the performance of VANET in car-sharing systems. Not only does this study reveal the impact of different network parameters on VANET systems, but it also is relevant to the feasibility and future development of VANET in these systems. VANET is feasible in car-sharing systems although the communication setup is a little bit longer. After the communication between two vehicles is built, the delay for peer-to-peer communication is not significant. The PSR and the number of vehicles can be treated together as the number of equipped vehicles. With higher number of equipped vehicles and proper control of the maximum connectivity number, it can reach a better performance of VANET in car-sharing systems.

This study is just one extension in our attempts to understand the performance of VANET in car-sharing systems. We mainly focus on how to implement the VANET in a real vehicle movement of car-sharing systems. Due to the lack of real shared cars movement pattern data, we do not compare the influence of before and after VANET in a car-sharing system. Also, only the scenarios with equipped vehicles but not with other equipped vehicles are considered. In the future, more complex traffic scenarios with the participation of more vehicles will be discussed. As an important part of connected vehicle technology, the simulation of inter-vehicle communication with roadside stations in NS-2 is also important. Work on all these simulations can be used as reference to build a better vehicular communication networks in car-sharing systems.

# CHAPTER 7
# CONCLUSIONS

To take advantage of the high resolution demographic data LandScan USA and agent-based microscopic traffic simulation models, many new problems appeared and novel solutions are needed. A series of studies are conducted using LPC data for evacuation assignments with different network configurations, travel demand models, and travelers' compliance behavior.

First, a new MSNDSP problem is defined to represent our research problem in generating the high resolution LPC-based OD matrix for TRANSIMS evacuation simulation. To solve this problem efficiently and effectively, a SNTG algorithm is proposed, which transforms the OD generation from a MSNDSP problem to a normal single source shortest path problem.

Second, an agent-based traffic assignment framework with LPC is proposed for evacuation planning and simulation. TAZ-based traffic assignment underestimates evacuation travel time and simulation computational time due to its even distribution assignment in a TAZ zone. The proposed LPC-based framework provides a more accurate environment to represent how travelers access the road network. It takes more time for evacuees in Alexandria to travel to shelters in daytime under normal traffic conditions than the case in nighttime because of the different temporal demographic patterns. But the travel times with location-based model in both daytime and nighttime are almost the same due to congestions. Location-based departure time choice model improves the evacuation efficiency at the beginning stage of evacuation. But it also causes congestion, which delays the average travel time.

Third, two levels of real-world road networks (major and full) and two scales of trip generation (TAZ and LPC) are modeled and compared using three different route finding algorithms. Two major findings are concluded from the simulation results of 12 scenarios in data resolution and routing algorithm aspects. For the data resolution analysis, the highest resolution data with LPC and full network, which is also the more realistic one, need more travel time for emergency evacuation. For the routing algorithm analysis, the highway-biased trip assignment has relatively better performance on evacuation time than the other two methods. But the performance is not significantly better than HWB method, especially when the network is very congested. The shortest straight-line distance assignment, the naïve user preferred intuitive method, produced the worst results.

Fourth, a comparison study of evacuee compliance behavior impacts on evacuation assignment with both conventional TAZ method and high resolution LandScan USA population data was conducted. The evacuation performance is not that sensitive to evacuee compliance behavior in LPC based assignment compared to TAZ based assignment. This is because the standard deviation of assigned destination population in the TAZ study is much larger than the LPC case. The evacuee compliance level can change the destination population distribution significantly in a TAZ scenario, but not the LPC case.

Finally, an inter-vehicle communication system is simulated in a car-sharing system with different simulation scenarios. The performance of VANET in car-sharing systems is analyzed through discussing the packet delivery fraction and average end-to-end delay of the vehicular communication network in different PSR, MCN, and the number of vehicles. VANET is feasible in car-sharing systems although the communication setup is a little bit longer. After the communication between two vehicles is built, the delay for peer-to-peer communication is not significant. With higher number of equipped vehicles and proper control of the maximum connectivity number, it can reach a better performance.

Michael Robinson, A. K. (2012). <u>Evacuee Route Choice Decisions in a Dynamic Hurricane Evacuation Simulation</u>. Transportation Research Board 2012 Annual Meeting, Washington, D.C., Transportation Research Board of the National Academies.

Millard-Ball, A., G. Murray, J. Burkhardt and J. ter Schure (2005). Car-Sharing: Where and How it Succeeds Final Report, TCRP Project B-26. TRB, National Research Council, Washington DC, Forthcoming.

Morency, C., M. Trépanier, B. Agard, B. Martin and J. Quashie (2007). <u>Car sharing system: what transaction datasets reveal on users' behaviors</u>. Intelligent Transportation Systems Conference, 2007. ITSC 2007. IEEE, IEEE.

Nagel, K., R. J. Beckman and C. L. Barrett (1999). <u>TRANSIMS for urban planning</u>. 6th International Conference on Computers in Urban Planning and Urban Management, Venice, Italy, Citeseer.

Nagel, K. and G. Flötteröd (2009). <u>Agent-based traffic assignment: going from trips to behavioral travelers</u>. 12th International Conference on Travel Behaviour Research (IATBR), Jaipur.

Ng, M., J. Park and S. T. Waller (2010). "A hybrid bilevel model for the optimal shelter assignment in emergency evacuations." <u>Computer-Aided Civil and Infrastructure Engineering</u> **25**(8): 547-556.

Ozbay, K., M. A. Yazici, S. Iyer, J. Li, E. E. Ozguven and J. A. Carnegie (2012). <u>Use of a Regional Transportation Planning Tool for Modeling of Emergency Evacuation: A Case Study of Northern New Jersey</u>. Transportation Research Board 91st Annual Meeting.

Pel, A., M. Bliemer and S. Hoogendoorn (2011). "Modelling traveller behaviour under emergency evacuation conditions." <u>EJTIR</u> **11**(2): 166-193.

Pel, A., M. J. Bliemer and S. Hoogendoorn (2012). "A review on travel behaviour modelling in dynamic traffic simulation models for evacuations." <u>Transportation</u> **39**(1): 97-123.
116

# VITA

Mr. Wei Lu was born in Bengbu, a city in eastern China. He received a Bachelor's degree in automation and electrical engineering from China University of Mining and Technology (CUMT) in 2007. He earned his Master's degree in pattern recognition and intelligent systems in 2010. Then, he attended the University of Tennessee – Knoxville for a doctoral degree in civil engineering with concentration on transportation engineering. Also, he pursued a minor degree in computational science.

Mr. Lu's research interests include traffic simulation and modeling, high performance computing in traffic evacuation, connected vehicle technology, information and communication technology in intelligent transportation systems, geographic information science, and route optimization.

Since June 2012, Mr. Lu has been working in Oak Ridge National Laboratory as a research intern in the geographic information science and technology group. He has developed an interdisciplinary background with traffic engineering, computational science, geographic information science, and statistics.

Mr. Lu is an active member in ITE, ASCE, AAG, and IEEE. He serves as the Editor-in-Chief of the Modern Traffic and Transportation Engineering Research journal. He is also invited to be reviewers on several other journals and international conferences. During his PhD study, he was awarded several paper awards and travel awards.